\documentclass[aps,prd,twocolumn,groupedaddress,nofootinbib]{revtex4-2}

\usepackage{latexsym}
\usepackage{tensor}
\usepackage{amsmath}
\usepackage{amsfonts}
\usepackage{amssymb}
\usepackage{mathrsfs}
\usepackage{wasysym}
\usepackage{textcomp}
\usepackage{xcolor}
\usepackage{url}
\usepackage{booktabs}
\usepackage{hyperref}

 \hypersetup{
           breaklinks=true,   % splits links across lines
           colorlinks=true,   % displays links as colored text instead of blocks
           citecolor = [rgb]{0,0,0.6},  %navy blue
           linkcolor = [rgb]{0.8,0.52,0}, %dark orange
           pdfusetitle=true,  % \title and \author values into pdf metadata
        }

\newcommand{\Vin}{\text{\textcent} } 
\def\a{\alpha}
\def\b{\beta}

\def\m{\mu}

\def\l{\lambda}
\def\n{\nu}

\def\a{\alpha}
\def\b{\beta}

\def\G{\Gamma}

\def\rd{{\rm d}}
\def\rmm{{\rm m}}
\def\rn{{\rm n}}
\def\rmp{{\rm p}}

\def\tC{{\tilde{C}}}

\def\cD{{\mathcal{D}}}
\def\cP{{\mathcal{P}}}
\def\cK{{\mathcal{K}}}
\def\cR{{\mathcal{R}}}

\def\tG{ {\tilde \G}}

\def\hatGamma{{\hat{\Gamma}}}

\def\bcP{{\bar{\cP}}}
\def\bnabla{{\bar{\nabla}}}
\def\hnabla{{\hat{\nabla}}}
\def\bGamma{{\bar{\Gamma}}}
\def\bR{{\bar{R}}}

\usepackage{appendix} 

\def\varbG{\delta\bGamma}

\def\vartC{\delta\tC}

\def\varhG{\delta\hatGamma}
\def\varg{\delta g}

\def\varbGL{\delta_{\bGamma}\mathcal{L}}

\def\vartCL{\delta_{\tC}\mathcal{L}}

\def\varhGL{\delta_{\hatGamma}\mathcal{L}}

\def\bbD{\mathbb{D}}
\def\bbE{\mathbb{E}}

\def\tbbE{\tilde{\mathbb{E}}}

\def\hmone{\hspace*{-1 pt}}
\def\hmtwo{\hspace*{-2 pt}}
\def\hmthree{\hspace*{-3 pt}}
\def\hmfour{\hspace*{-4 pt}}
\def\hmfive{\hspace*{-5 pt}}
\def\hmsix{\hspace*{-6 pt}}
\def\hmseven{\hspace*{-7 pt}}

\def\hmten{\hspace*{-10 pt}}
\def\hmtwenty{\hspace*{-20 pt}}

\begin{document}

%Title of paper
\title{Covariant and Manifestly Projective Invariant \texorpdfstring{\\}{} Formulation of Thomas-Whitehead Gravity}

%Authors
\author{Tyler Grover}
\email[]{tyler-grover@uiowa.edu}

\author{Kory Stiffler}
\email[]{kory-stiffler@uiowa.edu}
\affiliation{Department of Physics and Astronomy, 
The University of Iowa, Iowa City, IA 52242, USA}

\author{Patrick Vecera}
\email[]{pvecera@ucsb.edu}
\affiliation{Department of Mathematics,
University of California, Santa Barbara, CA 93106-3080, USA}

\date{\today}

\begin{abstract}
Thomas-Whitehead (TW) gravity is a recently formulated projectively invariant extension of Einstein-Hilbert gravity. Projective geometry was used long ago by Thomas et. al. to succinctly package equivalent paths encoded by the geodesic equation. Projective invariance in gravity has further origins in string theory through a geometric action constructed from the method of coadjoint orbits using the Virasoro Algebra.  A projectively invariant connection arises from this construction, a part of which is known as the diffeomorphism field. 
TW gravity exploits projective Gauss-Bonnet terms in the action functional to endow the diffeomorphism field with dynamics, while allowing the theory to collapse to general relativity in the limit that  the diffeomorphism field vanishes and the connection becomes Levi-Civita. In the original formulation of TW gravity, the diffeomorphism field is projectively invariant but not tensorial and the connection is projectively invariant but not affine. In this paper we reformulate TW gravity in terms of projectively invariant tensor fields and a projectively invariant covariant derivative, derive field equations respecting these symmetries, and show that the field equations obtained are classically equivalent across formulations. This provides a 'Rosetta Stone' between this newly constructed covariant and projective invariant formulation of TW gravity and the original formulation that was manifestly projective invariant, but not covariant.
\end{abstract}

%\maketitle must follow title, authors, abstract, and keywords
\maketitle

\section{Introduction}
Nearly a century ago~\cite{Thomas:1925a,Thomas:1925b}, Thomas used projective geometry to package the set of equivalent paths encoded by the geodesic equation into a projectively invariant framework. In these seminal works, the connection is extended to be projectively invariant via the addition of a projective direction to make projective invariance manifest. This extra projective dimension was subsequently utilized by Veblin and Hoffman~\cite{Veblen1930} in an attempt at unification of gravity with electromagnetism $\grave{\text{a}}$ la Kaluza and Klein~\cite{Kaluza:1921tu,Klein:1926tv}. Whitehead further generalized the work of Thomas by allowing certain components of the projective connection to be arbitrary~\cite{Whitehead}. 

In the past few decades Whitehead's generalization has been revisited~\cite{BaileyT.N.1994TSBf,Roberts,Eastwood,Eastwood2,Crampin} and projective geometry has been investigated in regards to the equivalence principle and cosmology~\cite{Hall:2007wp,Hall:2008,Hall:2009zza,Hall:2011zza} and Einstein metrics in general~\cite{GoverA.Rod2012DEgE,CapA.2014Emip}. In making projective invariance manifest, the aforementioned projective direction extends the $\rd$-dimensional base manifold $\mathcal{M}$ to a $\rd+1$-dimensional  manifold $\mathcal{N}$ now known as the Thomas cone.   Thomas-Whitehead (TW) gravity is a recently formulated model that extends Einstein-Hilbert gravity to the Thomas cone~\cite{Brensinger:2017gtb,Brensinger:2019mnx,Brensinger:2020gcv,SamBrensinger}. Additional dynamical fields are introduced to perform this extension. The resulting TW gravity is projectively invariant and reduces to Einstein-Hilbert gravity upon setting the additional dynamical fields to zero. 

In TW gravity, the additional projective direction is integrated out. This leads back to a model on the $\rd$-dimensional manifold $\mathcal{M}$, where the additional fields take the form of a rank-two object $\mathcal{D}_{ab}$ known as the diffeomorphism field and a projective connection $\Pi^c{}_{ab}$. The connection and the metric are independent in the spirit of the Palatini formalism \cite{palatini, Borunda:2008}, allowing the field equations to collapse naturally to the Einstein equations when the diffeomorphism field  vanishes and the connection is Levi-Civita. In this way projective geometry can influence Riemannian geometry by acting as sources in the energy-momentum tensor, providing an avenue for  geometric explanations of dark energy, dark matter, inflation, and other physical phenomenon. Indeed, the diffeomorphism field has recently been shown to be a possible source of dark energy~\cite{Brensinger:2019mnx} and inflation~\cite{InflationNewPaper:2022}. All stemming from the principle of projective invariance. 

The roots of TW gravity go deeper still, to string theory and the Virasoro algebra. The notion of the diffeomorphism field as a \emph{dynamical} field arises from geometric actions built via the method of coadjoint orbits~\cite{Kirillov:1962,Kirillov:1982kav,Zaccaria:1981fi,Balachandran:1983oit,Balachandran:1986hv,Balachandran:1987st,Alekseev:1988ce,Alekseev:1988vx,Delius:1990pt}. In~\cite{Rai:1989js}, the method of coadjoint orbits was used to produce the two-dimensional Wess-Zumino-Witten action~\cite{DiVecchia:1984ksr,Witten:1983ar} and the Polyakov 2D quantum gravity action~\cite{Polyakov:1987zb,Polyakov:1981rd}. More specifically, a vector field arises from the Kac-Moody algebra and is interpreted as a Yang-Mills field $A_a$, and a rank-two field $D_{ab}$ arises from the Virasoro algebra and is interpreted as the diffeomorphism field of TW gravity~\cite{Rai:1989js,Rodgers:1994ck,Lano:1994gx,Branson:1996pe,Branson:1998bc,Brensinger:2017gtb,Brensinger:2019mnx,Brensinger:2020gcv,SamBrensinger,Rodgers:2022rvo}. This interpretation becomes clear from the transformation properties of $A_a$ and $D_{ab}$~\cite{Rai:1989js,InflationNewPaper:2022,Rodgers:2022rvo}: $A_a$ indeed transforms as a Yang-Mills field and $D_{ab}
$ transforms as part of a projective connection of Thomas~\cite{Thomas:1925a,Thomas:1925b}.

In its original formulation~\cite{Brensinger:2017gtb,Brensinger:2019mnx,Brensinger:2020gcv,SamBrensinger} TW gravity was written in terms of the non-tensorial, but projectively invariant diffeomorphism field $\cD_{ab}$, the non-affine but projectively invariant connection $\Pi^c{}_{ab}$, and the metric $g_{ab}$. The projective Schouten tensor $\cP_{ab}$ was introduced in lieu of $\cD_{ab}$, though this fails to be projectively invariant. Furthermore, the non-affine connection $\Pi^c{}_{ab}$ can be expanded in terms of the affine connection  $\Gamma^c{}_{ab}$, though this connection is not by itself projectively invariant.  Can we have the best of both worlds: a projectively invariant tensor field and a projectively invariant affine connection? The answer is yes, and the purpose of this paper is to demonstrate precisely how this is done. Expressing TW gravity in the resulting covariant and manifestly projective invariant form is analogous to writing the Maxwell Lagrangian in a manifestly Lorentz invariant form using $F_{\mu\nu}$ rather than $\vec{E}$ and $\vec{B}$ fields. More to the point, a covariant and manifestly projective invariant formulation of TW gravity is analogous to the formulation of general relativity (GR) itself in terms of tensors and covariant derivatives rather than connections and partial derivatives. 

As this new formulation of TW gravity is covariant, the usual tricks of GR become at our disposal: partial integration with respect to the covariant derivative is now available for instance. This was not available in the original formulation of TW gravity, and so deriving the field equations was quite cumbersome~\cite{Brensinger:2020gcv,SamBrensinger}. Now, with a covariant formulation, deriving the equations of motion becomes much more straightforward. Since the new formulation is manifestly projective invariant, this simplifies how the action and field equations are expressed, much like, again, how GR is simpler to write in terms of the Ricci tensor and metric, rather than expanding fully in terms of derivatives of the metric.

We note a key distinction between the additional, projective direction of TW gravity and higher dimensions. In TW gravity, the projective direction can be thought of as a gauge dimension more so than an extra spacetime dimension. Upon integrating out this gauge dimension, TW gravity maintains its projective invariance on the $\rd$-dimensional manifold $\mathcal{M}$, though it is no longer manifestly so when expressed in terms of the $\rd$-dimensional projective Schouten tensor $\cP_{ab}$, affine connection $\Gamma^c{}_{ab}$, and metric $g_{ab}$. The goal of this paper is to make projective invariance manifest on the $\rd$-dimensional manifold $\mathcal{M}$ by introducing the \emph{projectively invariant Schouten tensor}  $\bcP_{ab}$  and \emph{projectively invariant affine connection} $\bar{\Gamma}^c{}_{ab}$ that is used to construct the \emph{projectively invariant covariant derivative} $\bar{\nabla}_a$. Thus TW gravity is effectively a $\rd$-dimensional (set $\rd = 4$ to describe our Universe), projectively invariant model of gravity and \emph{not} a higher dimensional gravitational model.

This paper is structured as follows. Section~\ref{s:Review} is a review of TW gravity. Here the equations of motion are seen to be either manifestly projective invariant or covariant, but not both simultaneously. Section~\ref{s:Meat} contains the main results of the paper: the revelation of the covariant and manifestly projective invariant formulation of TW gravity. Here we present a covariant and manifestly projective invariant derivation of the field equations, utilizing the familiar tensorial tricks of Einstein-Hilbert gravity such as integrating by parts with a covariant derivative. The independent fields we vary are the metric $g^{mn}$, the projectively invariant Schouten tensor $\bcP_{ab}$, and the projectively invariant trace-removed Palatini tensor $\tC^c{}_{ab}$. The tensor $\tC^c{}_{ab}$ is the difference between the projectively invariant affine connection $\bar{\Gamma}^c{}_{ab}$ and the Levi-Civita connection $\hat{\Gamma}^c{}_{ab}$. We demonstrate the on-shell (with equations of motion enforced) equivalence of the equations of motion across the different formulations, summarize the equations of motion, and discuss solutions in Section~\ref{s:Summary}.

%%%%%%%%%%%%%%%%%%%%%%%%%%%%%%%%%%%%
%%%%%%%%%%%%%%%%%%%%%%%%%%%%%%%%%%%%
%%%%%%%%%%%%%%%%%%%%%%%%%%%%%%%%%%%%
%%%%%%%%%%%%%%%%%%%%%%%%%%%%%%%%%%%%

\section{\label{s:Review}Review of TW Gravity}
The foundations of TW gravity are with respect to a connection $\Gamma^a{}_{bc}$ that is \emph{not connected} to a metric. In fact, to start there is no metric at all as TW gravity is based on the notion of equivalence of paths, regardless of any metric, as first investigated by Thomas~\cite{Thomas:1925a,Thomas:1925b}. Transforming from one path to another path is known as a projective transformation in this context. This notion of projections is a generalization of the more familiar notion of projections in Euclidean geometry. In TW gravity, a projection indeed identifies points on one path with points on another path, but these points generally are not connected by straight lines as in a Euclidean projection. We define a projective transformation of a connection $\Gamma^{a}{}_{bc}$ as
\begin{align}\label{e:ProjTrans}
    \Gamma^a{}_{bc} \to \Gamma^a{}_{bc} + \delta^a{}_{(b} v_{c)}
\end{align}
with $v_b$ an arbitrary one form. Here and throughout the paper, parenthesis denote symmetrization of enclosed indices as described in Eq.~\eqref{e:SymmetrizationAndAntisymmetrization} of Appendix~\ref{s:Conventions}.
\subsection{\label{s:TWConnections}TW Connections}
As originally presented in \cite{Brensinger:2017gtb,Brensinger:2019mnx,Brensinger:2020gcv,SamBrensinger}, the TW connection $\tilde{\Gamma}^\alpha_{\ \beta\gamma}$ on the $\rd + 1$-dimensional Thomas cone $\mathcal{N}$ is projectively invariant with respect to projective transformations as in Eq.~\eqref{e:ProjTrans}. The TW connection is expressed in terms of the fields $\mathcal{D}_{ab}$ and $\Pi^a{}_{bc}$ on the $\rd$-dimensional base manifold\footnote{Here we are using the labels $\mathcal{M}$ and $\mathcal{N}$ as in~\cite{Brensinger:2019mnx,Brensinger:2020gcv}. Since the Thomas cone is a bundle over $\mathcal{M}$, the works~\cite{Crampin,Brensinger:2017gtb,SamBrensinger,Brensinger:2024udu} refer to these manifolds as $M$ and $VM$.} $\mathcal{M}$
\begin{equation}
    \label{eq: Gammatilde}
    \tilde{\Gamma}^{\alpha}{}_{ \beta\gamma} = \begin{cases}
        \tilde{\Gamma}^a_{\ bc} = \tensor{\Pi}{^a_{bc}} \\
        \tensor{\tilde{\Gamma}}{^{\lambda}_{bc}} = \Upsilon^\lambda \mathcal{D}_{bc} \\ 
        \tensor{\tilde{\Gamma}}{^a_{b \lambda}} = \tensor{\tilde{\Gamma}}{^a_{ \lambda b}} = \alpha_\lambda \tensor{\delta}{^a_b} \\ 
        \tensor{\tilde{\Gamma}}{^{\lambda}_{b \lambda}} = \tensor{\tilde{\Gamma}}{^{\lambda}_{ \lambda b}} = \tensor{\tilde{\Gamma}}{^{\lambda}_{ \lambda \lambda}} = 0 \\ 
    \end{cases}\, 
\end{equation}
where 
\begin{align}
\label{e:Pi}
{\Pi}^{a}_{\,\,\,\,b c} =& { \G}^{a}_{\,\,\,\,b c} + \delta^a{}_{(c}~ \a_{b)}~~~,~~~\a_b = -\tfrac{1}{\rd +1} \Gamma^c{}_{bc}\\
        \Upsilon^\alpha =& (0,0,\dots,0,\lambda)~~~,~~~\a_\b = \left( \a_c, \lambda^{-1} \right)~. \label{UpsilonOmega}
\end{align}
Latin indices $a,b,c,\dots$ are over the manifold $\mathcal{M}$ and Greek indices $\alpha, \beta, \gamma, \dots$ excluding $\lambda$ are over the manifold $\mathcal{N}$. The coordinates $x^a$ over $\mathcal{M}$ are extended to the coordinates $x^\alpha$ over $\mathcal{N}$ by appending the volume coordinate $\lambda$ as
\begin{align}
    x^\alpha = & (x^a , \lambda)~.
\end{align}
We note here that the connection is torsion-free, so symmetric in its lower two indices, and written in the conventions of Roberts~\cite{Roberts,Crampin,Brensinger:2017gtb,Brensinger:2019mnx,Brensinger:2020gcv,SamBrensinger}. A more general treatment, including a revival of Kaluza-Klein-like solutions, can be found in the recent work~\cite{Brensinger:2024udu}.

The connection $\Gamma^a{}_{mn}$ transforms as an affine connection under general coordinate transformations on $\mathcal{M}$ 
\begin{align}\label{e:GTrans}
        {\G}'^{a}{}_{mn} =& \frac{\partial x'^a}{\partial x^b}\frac{\partial x^p}{\partial x'^m}\frac{\partial x^q}{\partial x'^n}{\G}^{b}{}_{pq} + \frac{\partial^2 x^b}{\partial x'^m \partial'^n}\frac{\partial x'^a}{\partial x^b}  ~.
\end{align}
Its normalized trace $\alpha_b$ therefore transforms on the manifold $\mathcal{M}$ as
\begin{align}
	\label{e:alphaTransformOnM}
	\alpha'_b =&   \frac{\partial x^m}{\partial x'^b} \alpha_m  - \frac{1}{\rd+1}  \frac{\partial^2 x^m}{\partial x'^a \partial x'^b}\frac{\partial x'^a}{\partial x^m}~.
\end{align}
Under what is known as a Thomas cone (TC$\mathcal{N}$)-transformation
\begin{align}
        \label{e:TcTrans}
        \hspace*{-4 pt}x'^\a =& ( x'^0(x^m), x'^1(x^m), \dots , x'^{\rd -1}(x^m) , \lambda' = \lambda J^{\tfrac{1}{\rd+1}} ) \\
         J \equiv & |\partial x^m / \partial x'^n|
\end{align}
the connection $\Gamma^\alpha{}_{\mu\nu}$ likewise transforms as an affine connection on $\mathcal{N}$
\begin{align}
        \label{e:tGTrans}
        \tG'^{\alpha}{}_{\mu\nu} =& \frac{\partial x'^\a}{\partial x^\rho}\frac{\partial x^\sigma}{\partial x'^\m}\frac{\partial x^\beta}{\partial x'^\n}\tG^{\rho}{}_{\sigma\beta} + \frac{\partial^2 x^\beta}{\partial x'^\mu \partial'^\nu}\frac{\partial x'^\a}{\partial x^\b}~.
\end{align}
Interestingly, the object $\alpha_\beta$ transforms on the manifold $\mathcal{N}$ as a rank one tensor under TC$\mathcal{N}$-transformations as shown in the first equality below
\begin{align}\label{e:alphatransform}
   &\alpha'_\beta = \frac{\partial x^\mu}{\partial x'^\beta} \alpha_\mu \cr
    &= \frac{-1}{\rd+1} \delta^b{}_\beta \hspace*{-3 pt}\left[\frac{\partial x^m}{\partial x'^b} \Gamma^n{}_{mn}  + \frac{\partial x'^a}{\partial x^m} \frac{\partial^2 x^m}{\partial x'^a \partial x'^b}\right]\hspace*{-3 pt}  + \delta^\lambda{}_{\beta} \frac{1}{\lambda'}.~~~~~
\end{align}
This is consistent with $\alpha_b$ transforming as the normalized trace of an affine connection on the manifold $\mathcal{M}$, Eq.~\eqref{e:alphaTransformOnM}, as demonstrated by the expansion following the second equality.

The fields $\mathcal{D}_{ab}$ and $\Pi^a{}_{bc}$ have the advantage of being projectively invariant under a projective transformation as defined in Eq.~\eqref{e:ProjTrans}. The drawback of $\mathcal{D}_{ab}$ is that it is not a tensor and the drawback of $\Pi^a{}_{bc}$ is that it is not an affine connection. Instead, these transform under general coordinate transformations on $\mathcal{M}$ as
\begin{align}\label{e:Dtrans}
    \mathcal{D}'_{ab} =& \frac{\partial x^m}{\partial x'^a}\frac{\partial x^n}{\partial x'^b} \left[\mathcal{D}_{mn} - \partial_m j_n - j_m j_n + j_c \Pi^c{}_{mn}\right]  \\
    {\Pi'}^a_{\ bc} = &\frac{\partial x'^a}{\partial x^p} \frac{\partial x^m}{\partial x'^b}\frac{\partial x^n}{\partial x'^c}\Pi^p_{\ mn} + \frac{\partial x'^a}{\partial x^m}\frac{\partial^2 x^m}{\partial x'^b \ \partial x'^c} \cr
    & -   \delta^a_{\ (b}\frac{\partial x^m}{\partial x'^{c)}}j_m\\
    j_m \equiv & \frac{1}{\rd+1}\partial_m  \log J = \frac{1}{\rd+1} \frac{\partial x'^q}{\partial x^p} \frac{\partial^2 x^p}{\partial x^m \partial x'^q}
\end{align}
where we see the last three terms are the non-tensorial part of the  $\mathcal{D}_{ab}$ transformation and the last term is the non-affine part of the $\Pi^a{}_{bc}$ transformation.

In particular, this means that the resulting gravitational model written in terms of $\mathcal{D}_{ab}$ and $\Pi^a{}_{bc}$ known as Thomas-Whitehead gravity~\cite{Brensinger:2017gtb,Brensinger:2019mnx,Brensinger:2020gcv,SamBrensinger} though manifestly projective invariant does not have field equations that are covariant, or even tensorial at all in the case of many of the field equations. The main goal of this paper is to express Thomas-Whitehead gravity in terms of a set of fields that are both covariant and manifestly projective invariant on the manifold $\mathcal{M}$. This will lead to field equations that are likewise covariant and manifestly projective invariant on the manifold $\mathcal{M}$, while at the same time being dynamically equivalent to the original set of field equations derived in~\cite{Brensinger:2020gcv,SamBrensinger} at least classically. 

Derivative operators associated with each connection are defined as
\begin{align}
\label{e:tnablaaction}
    \hspace*{-5 pt}\tilde{\nabla}_\alpha v_\beta =& \partial_\alpha v_\beta - \tilde{\Gamma}^\mu{}_{\a\b} v_\m ~,~
    \tilde{\nabla}_\alpha v^\beta = \partial_\alpha v_\m + \tilde{\Gamma}^\b{}_{\a\m} v^\m \\
    \label{e:nablaaction}
    %%%%%%%%%%%%%%%%%%%%%%%%%%%%%%%%%%%%%%%%%%%%%%%%
    \nabla_a v_b =& \partial_a v_b - \Gamma^m{}_{ab} v_m ~,~
    \nabla_a v^b = \partial_\alpha v_\m + \Gamma^b{}_{am} v^m \\
    %%%%%%%%%%%%%%%%%%%%%%%%%%%%%%%%%%%%%%%%%%%%%%%%
    \label{e:bnablaaction}
    \breve{\nabla}_a v_b =& \partial_a v_b - \Pi^m{}_{ab} v_m ~,~
    \breve{\nabla}_a v^b = \partial_\alpha v_\m + \Pi^b{}_{am} v^m
\end{align}
and similarly when acting on higher rank objects. The derivatives $\tilde{\nabla}_\alpha$ and $\nabla_a$ will be covariant derivatives when acting on tensors associated with their respective manifolds, as their associated connections are each affine on their respective manifolds. The derivative operator $\breve{\nabla}_a$, however, will not generally be a covariant derivative regardless on what it acts as its associated connection $\Pi^m{}_{ab}$ is not affine. 

The derivative $\tilde{\nabla}_\alpha$ is tensorial under TC$\mathcal{N}$ transformations, Eq.~\eqref{e:TcTrans}, as well as projectively invariant under Eq.~\eqref{e:ProjTrans}. In contrast, neither $\nabla_a$ 
nor $\breve{\nabla}_a$ satisfy both of these conditions on the manifold $\mathcal{M}$. The derivative $\nabla_a$ is tensorial under general coordinate transformations on the manifold $\mathcal{M}$, but it is not projectively invariant under Eq.~\eqref{e:ProjTrans}. Conversely, the derivative $\breve{\nabla}_a$ is not tensorial under general coordinate transformations on the manifold $\mathcal{M}$, but is projectively invariant under  Eq.~\eqref{e:ProjTrans}. Similarly, the field $\mathcal{D}_{ab}$ is projectively invariant, but not tensorial on $\mathcal{M}$. Combining $\mathcal{D}_{ab}$ with appropriate factors of the connection $\Gamma^c{}_{ab}$ forms the tensorial combination $\mathcal{P}_{ab}$
\begin{align}
    \label{e:Pdef}
    \mathcal{P}_{ab} & = \mathcal{D}_{ab} - \partial_a \alpha_b + \Gamma^c_{\ ab} \alpha_c + \alpha_a \alpha_b ~.
\end{align} 
The tensor $\cP_{ab}$, however, is not projectively invariant owing to its dependence on the connection $\Gamma^c{}_{ab}$. 

In this paper, we will introduce the projectively invariant covariant derivative $\bar{\nabla}_a$ (with connection $\bar{\Gamma}^{c}{}_{ab}$) and projectively invariant Schouten tensor $\bcP_{ab}$,  both being tensorial on $\mathcal{M}$ more specifically as well as projectively invariant. We then demonstrate that TW-gravity can be written in terms of these tensorial and projectively invariant $\bar{\nabla}_a$ and $\bcP_{ab}$. With this we will accomplish our goal of expressing TW-gravity in a covariant and manifestly projective invariant form in terms of $\rd$-dimensional objects. At the same time, we will find the dynamics to be classically equivalent to the dynamics of TW-gravity written in terms of the original $\breve{\nabla}_a$ and $\mathcal{D}_{ab}$.

%%%%%%%%%%%%%%%%%%%%%%%%%%%%%%%%%%
%%%%%%%%%%%%%%%%%%%%%%%%%%%%%%%%%%%
%%%%%%%%%%%%%%%%%%%%%%%%%%%%%%%%%%%%
%%%%%%%%%%%%%%%%%%%%%%%%%%%%%%%%%%%%%

\subsection{\label{s:TWCurvatureTensors}TW Curvature Tensors}

Curvature on the Thomas cone is defined in the usual way, through the commutator
\begin{align}\label{e:CurvatureContravariant}
    [\tilde{\nabla}_\mu,\tilde{\nabla}_\nu ] \mathcal{V}^\alpha \equiv & \mathcal{K}^\alpha{}_{\beta\mu\nu}\mathcal{V}^\beta \\
    \label{e:CurvatureCovariant}
     [\tilde{\nabla}_\mu,\tilde{\nabla}_\nu ] \mathcal{V}_\beta \equiv & -\mathcal{K}^\alpha{}_{\beta\mu\nu}\mathcal{V}_\alpha 
\end{align}
with $\mathcal{V}^\alpha$ an arbitrary vector.
This leads to the usual formula in terms of the connection
\begin{align}
    \mathcal{K}^\alpha_{\ \beta\mu\nu} =& \partial_{\mu} \tilde{\Gamma}^\alpha_{\ \nu\beta} - \partial_\nu \tilde{\Gamma}^\alpha_{\ \beta\mu} +  \tilde{\Gamma}^\rho_{\ \beta\nu}\tilde{\Gamma}^\alpha_{\ \mu\rho} - \tilde{\Gamma}^\rho_{\ \beta\mu}\tilde{\Gamma}^\alpha_{\ \nu\rho}~. 
\end{align}

We define an analogue of the Riemann tensor known as the \emph{equi-projective curvature tensor} using $\Pi^c{}_{ab}$ as 
\begin{align}
    \mathcal{R}^a_{\ bcd} = \partial_c \Pi^a_{\ bd} - \partial_d \Pi^a_{\ bc} +  \Pi^e_{\ bd}\Pi^a_{\ ce} - \Pi^e_{\ bc}\Pi^a_{\ de} 
\end{align}
which is projectively invariant by construction, but not tensorial. We can also define the Riemann curvature tensor, which will not generally be built from the Levi-Civita connection, and is given by 
\begin{align}\label{e:RiemannGen}
    R^a_{\ bcd} = \partial_c \Gamma^a_{\ bd} - \partial_d \Gamma^a_{\ bc} + \Gamma^e_{\ bd}\Gamma^a_{\ ce}  -  \Gamma^e_{\ bc}\Gamma^a_{\ de}~.   
\end{align}
Each of $\cK^a{}_{bcd}$, $R^a{}_{bcd}$, and $\cR^a{}_{bcd}$ are antisymmetric in their last two indices by construction
\begin{align}
	\cK^a_{\ bcd} =& - \cK^a_{\ bdc} \\
	R^a_{\ bcd} =& - R^a_{\ bdc} \\
	\mathcal{R}^a_{\ bcd} =& - \cR^a_{\ bdc}~.
\end{align}

We form the Ricci tensor and its projectively invariant analogue on both manifold $\mathcal{M}$ and $\mathcal{N}$ as
\begin{align}
    \mathcal{R}_{bd} \equiv & \mathcal{R}^a_{\ bad} \\
    R_{bd} \equiv & R^{a}_{\ bad} \\
	\mathcal{K}_{\beta\nu} =& \mathcal{K}^\alpha_{\ \beta\alpha\nu} = \cK^a_{\ \beta a \nu}~. 
\end{align}
Notice that it matters not whether the contraction forming $\cK_{\beta\nu}$ includes the $\lambda$-direction or not as $\cK^\alpha_{\ \beta\mu\nu}$ vanishes when any of its lower indices equals $\lambda$. Furthermore, the only non-vanishing components of $\mathcal{K}_{\beta\nu}$ are those on the manifold $\mathcal{M}$, namely $\mathcal{K}_{bn}$. It should be noted that both $\mathcal{R}_{bd}$ and $\mathcal{K}_{\beta\nu}$ are symmetric while $R_{bd}$ is not.  The antisymmetric part of $R_{bd}$ is proportional to the curl of $\alpha_b$ as well as the antisymmetric part of $\mathcal{P}_{bd}$
\begin{align}
	\label{eq:AntiSymP-R}
 \mathcal{P}_{[bd]} = \frac{1}{d+1} R_{[bd]} = -\partial_{[b} \alpha_{d]}~  
\end{align}
where here and throughout the paper, square brackets denote antisymmetrization of enclosed indices as described in Eq.~\eqref{e:SymmetrizationAndAntisymmetrization} of Appendix~\ref{s:Conventions}.
To form scalars such as the Ricci tensor requires a metric, to which we turn to next.

%%%%%%%%%%%%%%%%%%%%%%%%%%%%%%%%%%
%%%%%%%%%%%%%%%%%%%%%%%%%%%%%%%%%%%
%%%%%%%%%%%%%%%%%%%%%%%%%%%%%%%%%%%%
%%%%%%%%%%%%%%%%%%%%%%%%%%%%%%%%%%%%%
\subsection{The TW Action}
Much of this section is a review of~\cite{Brensinger:2020gcv}. To construct an action from $\mathcal{K}^\alpha{}_{\beta\mu\nu}$, we need the metric on the Thomas cone
\begin{align}\label{e:BigGSuccinct}
        G_{\alpha\beta} &= \delta^a_{\,\,\alpha} \delta^b_{\,\,\beta} \,g_{ab} - \lambda_0^2 g_\alpha g_\beta \\
        \label{e:BigGInverseSuccinct}
        G^{\alpha\beta} &= g^{ab} (\delta^\alpha_{\ a} - g_a \Upsilon^\alpha)(\delta^\beta_{\ b} - g_b \Upsilon^\alpha) - \lambda_0^{-2} \Upsilon^\alpha \Upsilon^\beta~. 
\end{align} 
The metric $g_{ab}$ is the metric on the $\rd$-dimensional manifold $M$. The object $g_\alpha$ is defined as
\begin{align}
    g_\alpha = (g_a, \lambda^{-1})
\end{align}
with $g_a$ defined as
\begin{align}\label{e:ga}
    g_a =& - \tfrac{1}{d+1} \partial_a \log \sqrt{|g|} =  - \tfrac{1}{\rd + 1} \hat{\Gamma}^c{}_{ac} \\
    \label{e:Gammahat}
    \hat{\Gamma}^c{}_{ab} =& \tfrac{1}{2}g^{cm}\left(\partial_{(a}g_{b)m} - \partial_m g_{ab}\right)~.
\end{align} 
The object $g_\alpha$ transforms on the manifold $\mathcal{N}$ as a rank one tensor under TC$\mathcal{N}$-transformations, and as the normalized trace of an affine connection on the manifold $\mathcal{M}$: it has the same transformations as $\alpha_b$ and $\alpha_\beta$ in Eqs.~\eqref{e:alphaTransformOnM} and~\eqref{e:alphatransform}, respectively. This will be useful in forming invariants on the Thomas cone. The symbol $|g|$ is the absolute value of the determinate of the metric $g_{ab}$.
The Levi Civita connection $\hat{\Gamma}^c{}_{ab}$ differs from the full connection $\Gamma^c{}_{ab}$ by a tensor $C^c{}_{ab}$ known as the Palatini field~\cite{Brensinger:2020gcv}
\begin{align}\label{e:GtoLCGPlusC}
    \Gamma^c{}_{ab} =& \hat{\Gamma}^c{}_{ab} + C^c{}_{ab}~.
\end{align}
These relations to the Levi Civita connection will facilitate integration by parts when deriving the equations of motion.

Using Eq.~\eqref{e:Gammahat} we can expand the Riemann curvature tensor as
\begin{align}
    \label{e:RiemannExpandedLCQ}
    R^a_{\ bcd} =& \hat{R}^a_{\ bcd} + Q^a_{\ bcd} \\
    \label{e:QRiemann}
    Q^a_{\ bcd} =& \hat{\nabla}_c C^a_{\ bd} - \hat{\nabla}_d C^a_{\ bc} + C^e_{\ bd}C^a_{\ ce}  -  C^e_{\ bc}C^a_{\ de}
\end{align}
where $\hat{R}^a_{\ bcd}$ is the usual Levi-Civita Riemann tensor, built from $\hat{\Gamma}^a{}_{bc}$ rather than $\Gamma^a{}_{bc}$ as in Eq.~\eqref{e:RiemannGen}, and $\hat{\nabla}_a$ the Levi-Cita covariant derivative for the connection $\hat{\Gamma}^c{}_{ab}$. The Levi-Civita Ricci curvature and scalar along with the analogous versions involving $Q^a{}_{bcd}$ are defined in the usual way
\begin{align}\label{e:Riccihat}
\hat{R}_{bd} =& \hat{R}^a{}_{bad}~~~,~~~\hat{R} = g^{bd}\hat{R}_{bd}~~~\\
    \label{e:QRicci}
    Q_{bd} =& Q^a{}_{bad}~~~,~~~Q = g^{bd}Q_{bd}~.
\end{align}

 With the metric $G_{\alpha\beta}$ now in hand from Eq.~\eqref{e:BigGSuccinct}, we can construct invariants on the Thomas cone such as
\begin{align}
    \label{e:KKretschmann}
    \mathcal{K}^\alpha{}_{\beta\mu\nu}\mathcal{K}_\alpha{}^{\beta\mu\nu} \equiv& \mathcal{K}^\alpha{}_{\beta\mu\nu}\mathcal{K}^\rho{}_{\sigma\kappa\tau}G_{\alpha\rho}G^{\beta\sigma}G^{\mu\kappa}G^{\nu\tau} \\
    \label{e:KKRicci}
    \mathcal{K}_{\alpha\beta}\mathcal{K}^{\alpha\beta} \equiv &\mathcal{K}^\rho{}_{\alpha\rho\beta}\mathcal{K}^\sigma{}_{\mu\sigma\nu}G^{\alpha\mu}G^{\beta\nu} \\
    \label{e:KScalar}
    \mathcal{K} \equiv & \mathcal{K}_{\alpha\beta}G^{\alpha\beta} = \cK_{ab} g^{ab} ~. 
\end{align}
These all are projectively invariant under Eq.~\eqref{e:ProjTrans} as well as invariant with respect to TC$\mathcal{N}$-transformations, Eq.~\eqref{e:TcTrans}.  Note that in forming $\cK$ it matters not which metric is used as $\mathcal{K}_{\mu\nu}$ vanishes when either of its indices indices equals $\lambda$ and $G^{ab} = g^{ab}$ as seen in Eq.~\eqref{e:BigGInverseSuccinct}. 

A portion of the expansion of Eq.~\eqref{e:KKretschmann} produces the projective Cotton-York tensor
\begin{align}
    \label{e:CottonYorkDef}
    K_{bcd} \equiv & g_\alpha \mathcal{K}^\alpha{}_{bcd} \cr
    =& \breve{\nabla}_{[c} \mathcal{D}_{d]b} + g_a \mathcal{K}^a{}_{bcd} \cr
    =& \nabla_{[c} \mathcal{P}_{d]b} + \Vin_a \mathcal{K}^a{}_{bcd}~~~
\end{align}
where $\breve{\nabla}_c \cD_{db}$ is defined as in Eq.~\eqref{e:bnablaaction} and $\Vin_a$ is defined as
\begin{align}
    \label{e:Vincent}
    \Vin_a \equiv & g_a - \alpha_a  = \tfrac{1}{\rd+1} C^c{}_{ac}~.
\end{align}
The object $\Vin_a$ is in fact a tensor on $\mathcal{M}$, as the non-tensorial parts of the transformations of $g_a$ and $\alpha_a$ cancel in the net transformation law for $\Vin_a$. 

The projective Cotton-York tensor is antisymmetric in its last two indices by construction and inheritance from the projective Riemann curvature tensor
\begin{align}
	K_{bcd} = - K_{bdc}~.
\end{align}
The projective Cotton-York tensor $K_{bcd}$ is clearly a tensor on both $\mathcal{M}$ and $\mathcal{N}$ as well as projectively invariant. In Eq.~\eqref{e:CottonYorkDef}, the first line demonstrates the tensor character on $\mathcal{N}$ as recall both $g_\alpha$ and $\mathcal{K}^\alpha{}_{\beta\mu\nu}$ are tensors on $\mathcal{N}$. The second line demonstrates projective invariance even when restricted to the manifold $\mathcal{M}$ as each of $\mathcal{D}_{ab}$, $g_a$, and $\mathcal{K}^a{}_{bcd}$ are projectively invariant. The third line demonstrates the tensor character on the manifold $\mathcal{M}$ as each of $\mathcal{P}_{ab}$, $\Vin_a$, and $\mathcal{K}^a{}_{bcd}$ are tensors on $\mathcal{M}$. None of the lines alone demonstrate both tensorality and projective invariance on the manifold $\mathcal{M}$, once again showing the importance of defining the covariant and projectively invariant $\bcP_{ab}$ and $\bar{\nabla}_a$ that we will introduce later.

The TW action is the sum of a projective Einstein-Hilbert action and a projective Gauss-Bonnet action
\begin{align} 
    \label{e:TWAction} 
    S & = -\frac{1}{2 \tilde{\kappa}_0 \lambda_0} \int  d^{\rd} x d \lambda  \sqrt{|G|} (\mathcal{K} + 2 \Lambda_0) \cr
    -&\frac{\tilde{J}_0 c}{\lambda_0} \hspace*{-3 pt}\int \hspace*{-3 pt} d^{\rd} x d \lambda \sqrt{|G|}  \left[\mathcal{K}^\alpha_{\ \beta\mu\nu}\mathcal{K}_\alpha^{\ \beta\mu\nu}- 4 \mathcal{K}_{\alpha\beta}\mathcal{K}^{\alpha\beta}  + \mathcal{K}^2 \right] ~~~~~
\end{align}
where we have included a cosmological constant $\Lambda_0$ that was not included in the original works~\cite{Brensinger:2017gtb,Brensinger:2019mnx,Brensinger:2020gcv,SamBrensinger} but was included in the more recent work~\cite{InflationNewPaper:2022}. This is the only substantive change to this review of~\cite{Brensinger:2020gcv}. In the action Eq.~\eqref{e:TWAction}, the first line is the projective Einstein-Hilbert action that includes a coupling of the metric to $\cD_{ab}$ and the second line is the projective Gauss-Bonnet action that endows $\mathcal{D}_{ab}$ with dynamics while avoiding higher derivatives \cite{Lovelock:1971yv}. In four dimensions, taking the limit where $\Gamma^c{}_{ab}$ is Levi-Civita and $\mathcal{D}_{ab} = 0$, the TW action collapses to the Einstein-Hilbert action. This follows from the fact that the functional variation of the Gauss-Bonnet action with respect to the metric is identically zero~\cite{Lanczos:1938sf}. We see the TW action is clearly both invariant under TC$\mathcal{N}$ transformations as well as projectively invariant as it is constructed entirely from such objects $\mathcal{K}^\alpha{}_{\beta\mu\nu}$, $G_{\alpha\beta}$ and physical constants. Both invariances are manifest in this $\rd+1$-dimensional form as $\mathcal{K}^{\alpha}{}_{\beta\mu\nu}$ is built from the connection $\tilde{\Gamma}^\alpha{}_{\mu\nu}$ which is affine on $\mathcal{N}$ as well as projectively invariant. 

Our goal in this paper is to maintain this covariance and manifest  projective invariance after reducing to the $\rd$-dimensional manifold $\mathcal{M}$\footnote{Here and throughout, by manifest projective invariance we mean that each object, such as $\cK_{abcd}$ in the action, is projectively invariant.}. To illuminate the behavior on the spacetime manifold $\mathcal{M}$ we can expand $S$ and integrate out $\lambda$ 
\begin{align} 
\label{e:TWActionExpanded}
    S = & \int d^\rd x \sqrt{|g|} \mathcal{L} \\
    \label{e:LTW}
    \mathcal{L} =& -\frac{1}{2 \kappa_0}(\mathcal{K}+ 2 \Lambda_0)   + \mathbb{L} \\
    \label{e:LPGB}
    \mathbb{L} =& J_0 c \left[\lambda_0^2  K_{bcd} K^{bcd} \hspace*{-3 pt} -  \hspace*{-2 pt}\mathcal{K}_{abcd} \mathcal{K}^{abcd} \hspace*{-2 pt}+ \hspace*{-2 pt}4 \mathcal{K}_{bd} \mathcal{K}^{bd}\hspace*{-3 pt} -\hspace*{-2 pt} \mathcal{K}^2 \right]\hspace*{-2 pt}.
\end{align}
 The full TW Lagrangian is denoted by $\mathcal{L}$ that includes the projective Gauss-Bonnet (PGB) Lagrangian denoted by $\mathbb{L}$. The rescaled coupling constants are
\begin{align}
    \label{e:J0k0Renorm}
    \kappa_0^{-1} =& \tilde{\kappa}_0^{-1} \int d\lambda \ \lambda^{-1} ,~~~J_0 = \tilde{J}_0  \int d\lambda \ \lambda^{-1} ~. 
\end{align}
A possible relationship between these scalings and renormalization in a quantum field theory is under investigation. 

On $\mathcal{M}$ we use the notation that indices are raised and lowered with $g^{ab}$ and $g_{ab}$, respectively
\begin{align}
\label{e:cKraiselower1}
    \cK_{abcd} \equiv & g_{ma} \cK^m{}_{bcd}\\
\label{e:cKraiselower2}
    \cK^{abcd} = & g^{bp} g^{cm} g^{dn} \cK^a{}_{pmn}\\
\label{e:cKraiselower3}
    K^{bcd} \equiv & g^{bp}g^{cm}g^{dn}K_{pmn}~.
\end{align}
On $\mathcal{N}$ whether indices are raised with $g^{ab}$ or $G^{\alpha\beta}$ is unambiguous for the projective Cotton-York tensor and the last three indices of the projective Riemann tensor $\mathcal{K}^a{}_{bcd}$. This is because $G^{ab} = g^{ab}$ and the lambda components of these tensor vanish aside from $\mathcal{K}^\lambda{}_{bcd}$. Care must be taken, however, to be clear what is meant by lowering the first index of $\mathcal{K}^a{}_{bcd}$. On $\mathcal{M}$ we no longer have the full $G_{\alpha\beta}$ as its $\lambda$ components have been integrated out, so raising and lowering is done via $g^{ab}$ and $g_{ab}$, respectively.

Quadratic dynamics for the fields can be seen in the projective Cotton-York tensor term when expanded using Eq.~\eqref{e:CottonYorkDef}. The curvature tensors contain interactions as can be seen upon expanding in terms of the $\rd$-dimensional $\mathcal{D}_{ab}$ and $\Pi^c{}_{ab}$ or $\cP_{ab}$ and $\Gamma^c{}_{ab}$ as
\begin{align}
\label{e:KRiemannExpand}
    \mathcal{K}^a_{\ bcd} 
    = &\mathcal{R}^a_{\ bcd} + \delta^a_{\ [c} \mathcal{D}_{d]b} \cr
     =& R^a_{\ bcd} + \delta^a_{\ [c} \mathcal{P}_{d]b} - \delta^a_{\ b} \mathcal{P}_{[cd]}  \\
\label{e:KRicciExpand}
    \mathcal{K}_{mn} 
        = & \mathcal{R}_{mn} + (\rd-1) \mathcal{D}_{mn} \cr
        =&R_{mn} + (\rd-1) \mathcal{P}_{mn} - \rd \mathcal{P}_{[mn]} \\ 
    \label{e:KscalarExpand}
    \mathcal{K} = &\mathcal{R} + (\rd-1) \mathcal{D} = R + (\rd-1) \mathcal{P}~~~ \\
    \label{e:RDtraces}
    \mathcal{R} \equiv & g^{ab} \mathcal{R}_{ab}~,~\mathcal{D} \equiv g^{ab} \mathcal{D}_{ab} \\
    \label{e:RPtraces}
    R \equiv & g^{ab} R_{ab}~,~\cP \equiv g^{ab} \cP_{ab}~.
\end{align}
All other curvature components vanish aside from $\mathcal{K}^\lambda{}_{bcd}$ which can be solved for using the expressions of the projective Cotton-York tensor in Eq.~\eqref{e:CottonYorkDef}. We see explicitly each object in TW gravity is not covariant and manifestly projective invariant on $\mathcal{M}$ when written in terms of these expansions. If the curvature tensors are expanded through the first equality in each of the above, they are manifestly projective invariant, but not covariant. If the second equality is used in each, they are no longer manifestly projective invariant, but they are covariant. The same is true for the projective Cotton-York tensor as was shown in Eq.~\eqref{e:CottonYorkDef}. In the next section, we introduce the covariant and manifestly projective invariant $\bcP_{ab}$ and $\bar{\nabla}_a$ and expand the above tensors in terms of these. 

We conclude this section with a brief aside on the values of the physical constants of TW gravity that are consistent with physical phenomena, specifically early Universe inflation.  For $\rd = 4$ the units are
\begin{align}\label{e:Units}
\begin{split}
        [J_0] = \frac{M L^2}{T}~~~,~~~ [\mathcal{D}_{ab}] &= [R_{ab}] = L^{-2}\cr
         \left[\kappa_0 \right] =\frac{T^2}{ML} ~.  
        \end{split}
\end{align}
In this manner we see that $J_0$ has units of angular momentum, while $[\kappa_0] \propto \left[\frac{8 \pi G}{c^4}\right]$ in accordance with general relativity. Also, $\lambda$ and $\lambda_0$ both have units of length so their quotient is dimensionless. The parameter $\lambda_0$ sets the characteristic projective length scale appearing in the model.  Using natural units $\hbar = c = 1$ and defining the reduced Planck mass $M_p$ as
\begin{align}
    M_p = \sqrt{\frac{\hbar c}{8 \pi G}}
\end{align} a recent analysis in~\cite{InflationNewPaper:2022} suggests that if $\lambda_0 \lesssim 6\cdot10^6 M_p^{-1}$ the inflaton $\phi \propto \cP$ appearing naturally in TW gravity would have provided early Universe inflation matching current inflationary cosmology data within the $95\%$ confidence level. Specifically, it was found that to produce $N=60$ \emph{e}-folds, the parameters would have to satisfy the constraints $\lambda_0^2 \lesssim 10^3 J_0 \kappa_0$ and $10^8 \lesssim J_0 \lesssim 4\cdot10^{10}$ all within the $95\%$ confidence level. Furthermore, a value of $\kappa_0 \sim M_{p}^{-2}$ would result in Einstein-Hilbert gravity + cosmological constant at the end of inflation within the $95 \%$ confidence level \footnote{The value of $\kappa_0$ was not discussed in this way in \cite{InflationNewPaper:2022}. However, it is easy to see that the TW action reduces to $(-2\kappa_0)^{-1} (1 + 8 J_0 \kappa_0 \lambda_0^{-1} \phi)R$ plus cosmological constant at the end of inflation if $\phi \propto \cP$ settles to a constant value, with the traceless parts of $\mathcal{P}_{ab}$ equal to zero and assuming a Levi-Civita connection. The analysis in \cite{InflationNewPaper:2022} leads to $8J_0 \kappa_0  \lambda_0^{-1} \phi <<1$ at the end of $N=60$ \emph{e}-folds of inflation within the $95\%$ confidence level. Thus the TW action reduces to approximately $(-2\kappa_0)^{-1} R$ plus cosmological constant at the end of inflation, where we identify $\kappa_0 \sim M_p^{-2}$ to match general relativity.}. Putting this all together results in the above limit on $\lambda_0$. This suggest that the characteristic projective scale defined by $\lambda_0$ is less than a few million Planck lengths if it is TW gravity itself that is responsible for inflation. The importance of these results is that it appears \emph{inflation is automatically is built into the geometry of TW gravity with no additional field needed}. The above analysis for the limit on $\lambda_0$ relies on the assumption that TW gravity must precisely match Einstein-Hilbert gravity + cosmological constant at the end of inflation to then transition to radiation dominance. Whether or not that is true depends on the specific interactions of TW gravity with matter because matter acts like radiation in the early Universe. These interactions are an ongoing avenue of research~\cite{Brensinger:2017gtb,Brensinger:2019mnx,SamBrensinger,Brensinger:2020gcv}.

%%%%%%%%%%%%%%%%%%%%%%%%%%%%%%%%%%
%%%%%%%%%%%%%%%%%%%%%%%%%%%%%%%%%%%
%%%%%%%%%%%%%%%%%%%%%%%%%%%%%%%%%%%%
%%%%%%%%%%%%%%%%%%%%%%%%%%%%%%%%%%%%%
\subsection{\label{s:MPIEOM}Manifestly Projective Invariant form of the TW Field Equations}
We note that from here on, we work in units where the speed of light $c = 1$, but keep the constants $J_0$, $\kappa_0$, and $\lambda_0$ owing to the discussion in the previous section. The equations of motion with respect to the independent fields $\cD_{ab}$, $\Pi^{c}{}_{ab}$, and $g^{mn}$ were originally derived in~\cite{Brensinger:2017gtb} in constant-volume coordinates. Here we review the more general, non-constant volume coordinate version of the field equations originally derived in~\cite{Brensinger:2020gcv,SamBrensinger}.
The field equations for $\mathcal{D}_{ab}$ are~\footnote{We have divided by a factor of $2 \sqrt{|g|}$ from the original version presented in~\cite{Brensinger:2020gcv,SamBrensinger}. This makes the equation tensorial, though it is not manifest here. }
\begin{align}\label{e:DEOM}
    0= \mathbb{D}^{ab} \equiv & - \tfrac{(\rd - 1)}{4 J_0\kappa_0}  g^{ab} + \tfrac{1}{2}\hat{\mathcal{K}}_{c}^{\ (ab)c} \cr
    & +  \tfrac{\lambda_0^2}{\sqrt{|g|}} \breve{\nabla}_c ( \sqrt{|g|} K^{(ab)c}) -  \lambda_0^2 g_c \breve{\mathcal{K}}^{(ab)c} 
\end{align}
where the following were introduced for convenience
\begin{align} 
    \label{khat equation}
    &\hmseven\hat{\mathcal{K}}_d^{\  a b c} =  \mathcal{K}^{ \bar d}_{\ \bar a \bar b \bar c} \mathcal{G}_{d \bar d}^{\ \ a \bar a [b |\bar b| c] \bar c} \\
    &\hmseven\mathcal{G}_{d \bar{d}}^{\ \ \ a \bar{a} b \bar{b} c \bar{c}} \nonumber\\
    &\quad= G_{d \bar{d}} g^{a \bar{a}} g^{b \bar{b}} g^{c \bar{c}} - 4 \delta^b_{\ d} \delta^{\bar{b}}_{\ \bar{d}} g^{a \bar{a}} g^{c \bar{c}} + \delta^b_{\ d} \delta^{\bar{b}}_{\ \bar{d}} g^{ac} g^{\bar{a} \bar{c}}\\
    &\hmseven\breve{\mathcal{K}}^{abc} = g^{ap}g^{bm}g^{cn}\breve{\nabla}_{[m}\cD_{n]p} ~.
\end{align}
Note that $\hat{\cK}_d^{\ abc}$, that includes $G_{d\bar{d}}$ in its definition, should \emph{not} be confused with $\cK_d^{\ abc} = g_{dm} \cK^{mabc}$ that,  here on $\mathcal{M}$, has its indices raised or lowered via $g^{dm}$ or $g_{dm}$  as in Eq.~\eqref{e:cKraiselower2}.

The field equations for $\Pi^c{}_{ab}$ are~\footnote{We have divided by a factor of $2 \sqrt{|g|}$ from the original version presented in~\cite{Brensinger:2020gcv,SamBrensinger}. }
\begin{align}\label{e:EOMGamma}
0=\tilde{\mathcal{E}}_c{}^{ab} \equiv & \mathcal{E}_c^{\ ab} - \tfrac{1}{d+1} \delta_c^{\ ( a}\mathcal{E}_{d}^{\ b)d} \\
\label{e:CurlyE}
\mathcal{E}_c^{\ ab}  \equiv & \tfrac{1}{4 J_0\kappa_0 \sqrt{|g|}} \breve{\nabla}_c( \sqrt{|g|}g^{ab})\hspace*{-2 pt} - \hspace*{-2 pt} \tfrac{1}{2\sqrt{|g|}}\breve{\nabla}_d (\sqrt{|g|}\hat{\mathcal{K}}_{c}^{\ (ab) d}) \nonumber\\
    &\hmthree + \tfrac{\lambda_0^2}{\sqrt{|g|}} \breve{\nabla}_d (\sqrt{|g|}g_c \breve{\mathcal{K}}^{(ab)d})  \hmtwo - \hmtwo \lambda_0^2  K^{(ab)d} \mathcal{D}_{dc}.
\end{align}
The field equations for $g^{mn}$ without any external matter sources are~\footnote{This equation is equivalent, up to the addition of the cosmological constant $\Lambda_0$,  to that originally presented in~\cite{Brensinger:2020gcv,SamBrensinger}, though written in a slightly different way such that each individual term is projectively invariant.}
\begin{align}\label{e:cHmn}
	&0 = \mathcal{H}_{mn} \equiv   -\mathcal{K}_{mn} +  \kappa_0 \left( 2 \mathbb{L}_{mn} -  g_{mn} \mathcal{L} \right)   \cr 
	& \qquad\qquad\quad -  g_{mn} \tfrac{2 J_0 \kappa_0 \l_0^2}{(\rd+1)\sqrt{|g|}} \breve{\nabla}_a (\sqrt{|g|}\cK^{a}{}_{bcd}K^{bcd}) \\
       \label{e:PGBLmn}
      & \mathbb{L}_{mn} = J_0 \lambda_0^2  (K_{mcd}K_n^{\ cd} + 2 K_{bcm}K^{bc}_{\ \ n}) \cr
    & \qquad \quad + J_0   (\cK_{mbcd}\cK_n^{\ bcd} -\cK^a_{\ mcd}\cK_{an}^{\ \ cd} - 2 \cK^{abc}_{\ \ \ m} \cK_{abcn}) \cr
    & \qquad \quad + J_0  \left( 8 \cK_{mb}\cK^b_{\ n} - 2 \cK \cK_{mn}\right)
\end{align}
where the TW Lagrangian $\mathcal{L}$ and PGB Lagrangian $\mathbb{L}$ were defined in Eqs.~\eqref{e:LTW} and~\eqref{e:LPGB}, respectively. Recall all indices are raised or lowered via $g^{dm}$ or $g_{dm}$ here on $\mathcal{M}$ as explained in Eqs.~\eqref{e:cKraiselower1}, \eqref{e:cKraiselower2}, and~\eqref{e:cKraiselower3}.

All equations of motion are manifestly projective invariant as they are composed entirely of projectively invariant objects $\mathcal{K}^a{}_{bcd}$, $\breve{\nabla}_a$, $g_{ab}$, $g_a$, and $\mathcal{D}_{ab}$. None are covariant, and in fact $\mathcal{E}_c{}^{ab}$ is not tensorial, though $\mathbb{D}^{ab}$ and $\mathcal{H}_{mn}$ are tensorial as illustrated in the next section.

%%%%%%%%%%%%%%%%%%%%%%%%%%%%%%%%%%%%%%

\subsection{Covariant Form of the TW Field Equations}
Rewriting the equations in terms of $\mathcal{P}_{ab}$, $\nabla_a$, and $\Vin_a$ makes manifest the tensorial nature of $\mathbb{D}^{ab}$ and $\mathcal{H}_{mn}$ and illuminates how $\mathcal{E}_c{}^{ab}$ fails to be tensorial. The details of these recastings are found in appendix~\ref{a:EOMPItoTdetails}. For $\mathbb{D}^{ab}$ we have
\begin{align}\label{e:DEOMtensorial}
    & \hmsix 0 \hmtwo =\hmtwo\mathbb{D}^{ab}\hmtwo = 
    \hmtwo \lambda_0^2 (\nabla_c \hmtwo- \hmtwo(\rd\hmtwo+\hmtwo 2)\Vin_c) K^{(ab)c} \hmtwo -\hmtwo(\rd \hmtwo -\hmtwo 1) \mathbb{K}^{ab}\hmtwo -\hmtwo 2 \mathcal{K}^{ab}\\
    \label{e:bbK}
    &\hspace*{6 pt} \mathbb{K}^{ab} \equiv g^{ab}   \left(\cK + \tfrac{1}{4 J_0 \kappa_0}\right) - 4 \mathcal{K}^{ab}
\end{align}
which is covariant for the reasons described above.
To illustrate the tensorial nature of $\mathcal{H}_{mn}$, we express it in the familiar form of the Einstein Equations  
\begin{align}\label{e:EE}
	\hat{R}_{mn} - & \tfrac{1}{2} g_{mn} \hat{R} = \kappa_0 \hat{T}_{mn}  + g_{mn} \Lambda_0\\
	\label{e:Thatmn}
 \hat{T}_{mn} =&
   - \kappa_0^{-1}\left( Q_{mn} - \tfrac{1}{2}  g_{mn} Q \right) + 2 \mathbb{L}_{mn} - g_{mn} \mathbb{L} \cr 
   & - \kappa_0^{-1} \left((\rd -1) \mathcal{P}_{mn} - \rd \cP_{[mn]} - \tfrac{\rd - 1}{2} g_{mn} \mathcal{P} \right)\cr 
  & - g_{mn} \tfrac{2 J_0 \lambda_0^2}{\rd+1} (\nabla_a -(\rd+1) \Vin_a)(K_{bcd} \mathcal{K}^{abcd} )
\end{align}
where recall $\cP_{mn}$, $\cP$, $Q_{mn}$, $Q$, $\mathbb{L}_{mn}$, and $\mathbb{L}$ were defined in Eqs.~\eqref{e:Pdef}, \eqref{e:QRiemann}, \eqref{e:QRicci}, \eqref{e:RDtraces}, \eqref{e:RPtraces}, \eqref{e:PGBLmn}, and~\eqref{e:LPGB}.

That $\mathcal{E}_c{}^{ab}$ is non-tensorial is easily seen when cast into the following form 
\begin{align}\label{e:GammaEOMNablaP}
    \mathcal{E}_c{}^{ab} =&\mathbb{E}_c{}^{ab} + g_c \mathbb{D}^{ab} + \delta^{(a}{}_{c}\mathfrak{g}^{b)}
\end{align}
where for brevity we have defined the tensor $\mathbb{E}_c^{\ ab}$ and the non-tensorial object $\mathfrak{g}^b$ as
\begin{align}\label{e:GammaEOMEscrK}
    \mathbb{E}_c{}^{ab} \equiv& (\nabla_c - 2 \Vin_c) \mathbb{K}^{ab} - \delta^{(a}{}_{c} \Vin_d \mathbb{K}^{b)d} \cr
    &- [(\nabla_d - (\rd+1)\Vin_d) \mathcal{K}_c{}^{(ab)d} - 2 \Vin_c \mathcal{K}^{ab}] \cr
    &- \lambda_0^2 (\mathcal{P}_{cd} - \nabla_c \Vin_d - \Vin_c \Vin_d) K^{(ab)d}   \\
    \label{e:mathfrakg}
    \mathfrak{g}^b \equiv & \tfrac{1}{4 J_0 \kappa_0}g_d g^{bd} \hspace*{-2 pt} - \hspace*{-2 pt}  \tfrac{1}{2}(\nabla_d \hspace*{-2 pt} -\hspace*{-2 pt} (\rd+3) \Vin_d)(g^{bd} \mathcal{K}\hspace*{-2 pt} -\hspace*{-2 pt} 4 \mathcal{K}^{bd}).~~~~~~
\end{align}
Recall here on $\mathcal{M}$ the tensor $\cK_c^{\ abd}$ has its indices raised or lowered via $g^{cm}$ or $g_{cm}$ as in Eqs.~\eqref{e:cKraiselower1} and~\eqref{e:cKraiselower2}.

The non-tensorality of $\mathcal{E}_c{}^{ab}$ is explicitly seen in the second and third terms of Eq.~\eqref{e:GammaEOMNablaP} along with the first term of Eq.~\eqref{e:mathfrakg}: these take the form of the non-tensorial $g_c$ multiplying a tensor, the product of which is necessarily non-tensorial. An obvious question is: does the actual equation of motion $\tilde{E}_c{}^{ab}$, formed by subtracting from $\mathcal{E}_c{}^{ab}$ its trace, turn out to be tensorial? The answer is no as seen below
\begin{align}\label{e:cEtilde}
    0=\tilde{\mathcal{E}}_c^{\ ab} =& \tbbE_c{}^{ab}+\hspace*{-2 pt} g_c \mathbb{D}^{ab} \hspace*{-2 pt} - \hspace*{-2 pt} \tfrac{1}{\rd+1} \delta^{(a}_{\ \  c}g_m \mathbb{D}^{b)m}  \\
    \label{e:tbbE}
    \tbbE_c{}^{ab} \equiv & \bbE_c{}^{ab} - \tfrac{1}{\rd+1} \delta_c{}^{(a} \bbE_m{}^{b)m}
\end{align}
where we still see the presence of the non-tensorial $g_c$ times the tensor $\mathbb{D}^{ab}$.  Since $\mathbb{E}_c{}^{ab}$ is itself a tensor, we do however see that $\tilde{\mathcal{E}}_c{}^{ab}$ becomes tensorial upon enforcing the equation of motion $\mathbb{D}^{ab} = 0$.

Regardless, making the tensorial or non-tensorial nature of the equations manifest has now completely hidden their projective invariance. Now written in terms of the covariant $\cP_{ab}$, $\nabla_a$, $\Vin_a$, and $Q_{ab}$, we have to expose the projective transformations on each of these, and show that they cancel, to demonstrate that the equations are indeed still projectively invariant. Under a projective transformation, Eq.~\eqref{e:ProjTrans}, these transform as
\begin{align}
    \mathcal{P}_{ab} \to & \mathcal{P}_{ab} + \nabla_a v_b - v_a v_b\\
    \Vin_a \to & \Vin_a + v_b\\
    Q_{mn} \to & Q_{mn} + \nabla_m v_n - \rd \nabla_n v_m + (\rd-1)v_m v_n ~. 
\end{align}
In this venture, some helpful findings are that the following terms are all projectively invariant, though clearly not manifestly so
\begin{align}
\label{e:PQ}
	&\mathcal{P}_{cd} - \nabla_c \Vin_d - \Vin_c \Vin_d~~~ \\
	&Q_{mn} \hmtwo  - \hmtwo \tfrac{1}{2}  g_{mn} Q \hmtwo + \hmtwo (\rd -1) \mathcal{P}_{mn} \hmtwo - \hmtwo \rd \cP_{[mn]} \hmtwo - \hmtwo \tfrac{\rd - 1}{2} g_{mn} \mathcal{P} \\
	\label{e:bnabla1}
	& (-\tfrac{1}{\rd+1}\nabla_a + \Vin_a)K_{bcd} \mathcal{K}^{abcd}~~~\\
	&(\nabla_c - 2 \Vin_c) \mathbb{K}^{ab} - \delta^{(a}{}_{c} \Vin_d \mathbb{K}^{b)d}\\
	\label{e:bnabla2}
	&(\nabla_d - (\rd+1)\Vin_d) \mathcal{K}_c{}^{(ab)d} - 2 \Vin_c \mathcal{K}^{ab}~~~\\
	&(\nabla_c - (\rd+2)\Vin_c) K^{(ab)c}\\
	\label{e:bnabla3}
	&(\nabla_d - (\rd+3) \Vin_d)(g^{bd} \mathcal{K} - 4 \mathcal{K}^{bd})~.
\end{align}
This section and the previous section have served to illustrate the main driving force for the introduction of covariant and projectively invariant fields to which we now turn.

%%%%%%%%%%%%%%%%%%%%%%%%%%%%%%%%%
%%%%%%%%%%%%%%%%%%%%%%%%%%%%%%%%%%%
%%%%%%%%%%%%%%%%%%%%%%%%%%%%%%%%%%%
%%%%%%%%%%%%%%%%%%%%%%%%%%%%%%%%%%%%

\section{\label{s:Meat}Covariant and Manifestly Projective  Invariant  Formulation of TW Gravity}

\subsection{Projectively Invariant Connection and Schouten Tensor}

We introduce a set of fields, denoted with overbars, that are both tensorial and projectively invariant. We begin by introducing a modification of the connection on the manifold $\mathcal{M}$ and define 
\begin{align}\label{e:Gammabar}
	\bar{\Gamma}^a{}_{bc} & = \Gamma^a{}_{bc} - \delta^a{}_{(b} \Vin_{c)} = \Pi^a_{\ bc} - \delta^a_{\ (b}g_{c)}  
\end{align}
where $\Vin_c$ and $g_c$ were defined in Eqs.~\eqref{e:Vincent} and~\eqref{e:ga}. Modifying $\Pi^a{}_{bc}$ by subtracting the $g_c$ term has promoted the connection to affine. At the same time, changing $\Gamma^a{}_{bc}$ by subtracting the tensor $\Vin_c$ has not changed its affine character. So $\bar{\Gamma}^a{}_{bc}$ indeed transforms as an affine connection, as in Eq.~\eqref{e:GTrans}.

From $\bar{\Gamma}^a{}_{bc}$ we can construct a covariant derivative operator $\bar{\nabla}_a$, which is covariant from the first equality in Eq~\eqref{e:Gammabar} and manifestly projective invariant from the second. The operation of $\bar{\nabla}_a$ is the usual rule for covariant derivatives on a rank-m  covariant, rank-n contravariant arbitrary tensor $T_{a_1 \dots a_{\rmm}}^{\ \ \ \ \ \ b_1\dots b_{\rn}} $
\begin{widetext}
\begin{align}\label{e:bnablaGenAction}
    \bar{\nabla}_c T_{a_1 \dots a_{\rmm}}^{\ \ \ \ \ \ b_1\dots b_{\rn}} =& \partial_c T_{a_1 \dots a_{\rmm}}^{\ \ \ \ \ \ b_1\dots b_{\rn}}  - \sum_{\rmp=1}^{\rmm} \bar{\Gamma}^{d}_{\ ca_{\rmp}} T_{a_1 \dots a_{\rmp-1}d a_{\rmp+1} \dots a_{\rmm}}^{\ \ \ \ \ \ \ \ \ \ \ \ \ \ \ \ \ \ \ \ b_1 \dots b_{\rn}}  +\sum_{\rmp=1}^{\rn} \bar{\Gamma}^{b_{\rmp}}_{\ cd} T_{a_1 \dots a_{\rmm}}^{\ \ \ \ \ \ b_1 \dots b_{\rmp-1} d b_{\rmp+1}\dots b_{\rn}}~.
\end{align} 
Thus, substituting the defining Eq.~\eqref{e:Gammabar}, we have the following relationship between the three derivatives
\begin{align}\label{e:bnablatoOthernablas}
	\bar{\nabla}_c T_{a_1 \dots a_{\rmm}}^{\ \ \ \ \ \ b_1\dots b_{\rn}} =& (\nabla_c \hmtwo + \hmtwo  (\rmm \hmtwo - \hmtwo \rn) \Vin_c) T_{a_1 \dots a_{\rmm}}^{\ \ \ \ \ \ b_1\dots b_{\rn}}  \hmtwo + \hmtwo \sum_{\rmp=1}^{\rmm} \Vin_{a_{\rmp}} T_{a_1 \dots a_{\rmp-1}c a_{\rmp+1} \dots a_{\rmm}}^{\ \ \ \ \ \ \ \ \ \ \ \ \ \ \ \ \ \ \ \  b_1 \dots b_{\rn}} \hmtwo - \hmtwo\sum_{\rmp=1}^{\rn} \delta^{b_{\rmp}}_{\ \ c} ~\Vin_d T_{a_1 \dots a_{\rmm}}^{\ \ \ \ \ \ b_1 \dots b_{\rmp-1} d b_{\rmp+1}\dots b_{\rn}} \cr
    =& (\breve{\nabla}_c \hmtwo + \hmtwo (\rmm \hmtwo - \hmtwo \rn) g_c) T_{a_1 \dots a_{\rmm}}^{\ \ \ \ \ \ b_1\dots b_{\rn}} \hmtwo + \hmtwo \sum_{\rmp=1}^{\rmm} g_{a_{\rmp}} T_{a_1 \dots a_{\rmp-1}c a_{\rmp+1} \dots a_{\rmm}}^{\ \ \ \ \ \ \ \ \ \ \ \ \ \ \ \ \ \ \ \  b_1 \dots b_{\rn}}  -\sum_{\rmp=1}^{\rn} \delta^{b_{\rmp}}_{\ \ c} ~g_d T_{a_1 \dots a_{\rmm}}^{\ \ \ \ \ \ b_1 \dots b_{\rmp-1} d b_{\rmp+1}\dots b_{\rn}}.
\end{align}
\end{widetext}
With this, it is straightforward to show that the projectively invariant terms listed in \eqref{e:bnabla1}, \eqref{e:bnabla2}, and~\eqref{e:bnabla3} can all be written manifestly so as total derivatives with $\bar{\nabla}_a$.

In terms of the Levi-Civita connection $\hat{\Gamma}^a{}_{bc}$, $\bar{\Gamma}^a{}_{bc}$ can be expressed as
\begin{align}\label{e:GammabarC}
    \bar{\Gamma}^a_{\ bc} =& \hat{\Gamma}^a_{\ bc} + \tilde{C}^{a}_{\ bc} \\
    \label{e:Ctilde}
    \tilde{C}^{a}_{\ bc} \equiv & C^{a}_{\ bc} - \tfrac{1}{\rd+1} \delta^a_{\ (b} C^{m}_{\ \ c)m} \cr
    =& C^{a}{}_{bc} - \delta^a_{\ (b} \Vin_{c)}  ~.
\end{align}
This will be an important identity in the coming sections when we derive the field equations as it allows us to use partial integration with $\bar{\nabla}_a$ equivalently to $\hat{\nabla}_a$ their partial integrations differ by a factor of $\tilde{C}^c{}_{bc} = 0$. This feature forces the traces of the Levi-Civita connection and the projectively invariant connection to be identical. Indeed, tracing Eq.~\eqref{e:GammabarC} and substituting Eq.~\eqref{e:ga} we find that $g_a$ is proportional to both the trace of the Levi-Civita connection and the trace of the projectively invariant connection
\begin{align}
    g_a& = - \tfrac{1}{\rd+1} \hat{\Gamma}^c{}_{ac} = - \tfrac{1}{\rd+1} \bar{\Gamma}^c{}_{ac} ~.
\end{align}

We define the tensorial and manifestly projective invariant tensor $\bar{\mathcal{P}}_{mn}$ by 
\begin{align}
    \label{e:PBarDef}
    \bar{\mathcal{P}}_{mn} =& \mathcal{P}_{mn} - \nabla_m \Vin_n - \Vin_m \Vin_n \cr
    =& \mathcal{D}_{mn} - \breve{\nabla}_m g_n - g_m g_n 
 \end{align}
which is tensorial from the first equality and projectively invariant from the second. Comparing this with the first of the terms listed in~\eqref{e:PQ} we see it is indeed nothing more than $\bcP_{cd}$. We construct the trace in the usual manner, namely 
\begin{align} 
	\bar{\mathcal{P}} =   & g^{mn} \bar{\mathcal{P}}_{mn} = \mathcal{P} - \nabla^m \Vin_m - \Vin_m \Vin^m~.  
	\end{align}
The new field $\bar{\mathcal{P}}_{mn} = \bar{\mathcal{P}}_{nm}$ is symmetric by construction and contains the same degrees of freedom as $\mathcal{D}_{bc}$ and $\mathcal{P}_{bc}$.  

%%%%%%%%%%%%%%%%%%%%%%%%%%%%%%%%%%%%%%%%%%%%%%%
%%%%%%%%%%%%%%%%%%%%%%%%%%%%%%%%%%%%%%%%%%%%%%%%%
%%%%%%%%%%%%%%%%%%%%%%%%%%%%%%%%%%%%%%%%%%%%%%%
%%%%%%%%%%%%%%%%%%%%%%%%%%%%%%%%%%%%%%%%%%%%%%%%%

\subsection{Projectively Invariant Curvature Tensors}
From the projectively invariant connection, we build a projectively invariant Riemann curvature tensor
\begin{align}
	\bar{R}^a_{\ bcd} = & \partial_c \bar{\Gamma}^a{}_{bd} - \partial_d \bar{\Gamma}^a{}_{bc} + \bar{\Gamma}^m{}_{bd} \bar{\Gamma}^a{}_{cm} - \bar{\Gamma}^m{}_{bc} \bar{\Gamma}^a{}_{dm} \cr
	= & R^a{}_{bcd} + \delta^a{}_{[c} \nabla_{d]} \Vin_b + \delta^a{}_{[c} \Vin_{d]} \Vin_b - \delta^a{}_b \nabla_{[c} \Vin_{d]} \cr
    = & \mathcal{R}^a{}_{bcd} + \delta^a{}_{[c} \breve{\nabla}_{d]} g_b + \delta^a{}_{[c} g_{d]} g_b 
    %- \delta^a{}_b \breve{\nabla}_{[c} g_{d]}  %this last term vanishes owing to the fact that g_{d} is a total derivative
\end{align} 
which is covariant and manifestly projectively invariant when expressed as in the first line. In contrast, it is only covariant in each of its terms when expressed as in the second line and only manifestly projective invariant when expressed as in the third line. We will express the other curvature tensors in this section in the same way: covariant and manifestly projectively invariant in their first lines, manifestly only one or the other in each of their other lines. The derivative $\breve{\nabla}_d g_b$ is defined as in Eq.~\eqref{e:bnablaaction}. It is worth noting that all three of the curvature tensors are commutators of derivative operators, and so are antisymmetric in their last two indices. However, the lack of metric compatibility means they are not generally antisymmetric in the first two indices (i.e. $R_{abcd} \neq - R_{bacd}$). Proceeding in the usual manner, we also construct the associated Ricci tensors by contracting over the first and third indices
\begin{align}
	\bar{R}_{bd} = \bar{R}_{db} =& \bar{R}^a{}_{bad} \nonumber \\ 
        \qquad = & \tfrac{1}{2}R_{(bd)} + \tfrac{\rd-1}{2} \nabla_{(b} \Vin_{d)} + (\rd-1) \Vin_b \Vin_d \nonumber \\
        = & \mathcal{R}_{bd} + (\rd-1) \breve{\nabla}_b g_d + (\rd - 1) g_b g_d~.
\end{align}   
We then take the trace to construct the Ricci scalars  
\begin{align}
	\bar{R} =  g^{bd} \bar{R}_{bd} =& R + (\rd -1) \nabla^b \Vin_b + (\rd-1) \Vin^b \Vin_b  \nonumber \\
 = & \mathcal{R} + (\rd -1) \breve{\nabla}^b g_b + (\rd-1) g^b g_b 
\end{align}
where raising/lowering is defined as occurring \emph{outside} of the derivative and the raised $g^b$ is defined as follows
\begin{equation}
\nabla^b \equiv g^{bd}\nabla_d~~~,~~~\breve{\nabla}^b \equiv g^{bd}\breve{\nabla}_d~~~,~~~g^b \equiv g^{bd} g_d~.
\end{equation}

With the above considerations in mind, we can  express the projective Riemann tensor in terms of projectively invariant fields
\begin{align}
	\label{e:KRiemannAllForms}
    \mathcal{K}^a{}_{bcd} = & \bar{R}^a{}_{bcd} + \delta^a{}_{[c} \bar{\mathcal{P}}_{d]b} \nonumber \\
    = & R^a{}_{bcd} + \delta^a{}_{[c} \mathcal{P}_{d]b} - \delta^a{}_b \mathcal{P}_{[cd]} \nonumber \\ 
    = & \mathcal{R}^a{}_{bcd} + \delta^a{}_{[c} \mathcal{D}_{d]b}~.  
\end{align}
Contraction over the first and third indices as before yields the projective Ricci tensor. In terms of the three curvatures this is
\begin{align}
\label{e:KRicciAllForms}
    \mathcal{K}_{mn}  = & \bar{R}_{mn} + (\rd -1) \bar{\mathcal{P}}_{mn} \nonumber\\ 
        = & R_{mn} + (\rd-1) \mathcal{P}_{mn} - \rd \mathcal{P}_{[mn]} \nonumber \\  
        = & \mathcal{R}_{mn} + (\rd-1) \mathcal{D}_{mn}
\end{align}
which is necessarily symmetric.  Contraction with the metric again yields the projective Ricci scalar. In terms of the three curvatures this is
\begin{align}
\label{e:KScalarAllForms}
    \mathcal{K}  = & \bar{R} + (\rd -1) \bar{\mathcal{P}} \nonumber\\
   = & R + (\rd-1) \mathcal{P} \nonumber  \\
   = & \mathcal{R} + (\rd-1) \mathcal{D}~ .
\end{align}

Finally, the projective Cotton-York tensor is now expressed simply as
\begin{align}
	K_{bcd} = & \bar{\nabla}_{[c} \bar{\mathcal{P}}_{d]b} \nonumber  \\  
 = & \nabla_{[c} \mathcal{P}_{d]b} + \Vin_a \mathcal{K}^a{}_{bcd} \nonumber \\
 =&\breve{\nabla}_{[c} \mathcal{D}_{d]b} + g_a \mathcal{K}^a{}_{bcd}~. 
\end{align}
In summary, we see each of the above curvature tensors are covariant and manifestly projective invariant in their first lines, where they are each written entirely in terms of the barred objects.

It will become useful to expand these tensors in terms of the Levi-Civita connection as
\begin{align}
	\label{e:bRiemannQt}
	\bar{R}^a_{\ b c d} =& \hat{R}^a_{\ bcd} + \tilde{Q}^a_{\ bcd} \\
	\label{e:bRicciQt}
	\bar{R}_{mn} =& \hat{R}_{mn} + \tilde{Q}_{mn} \\
	\label{e:bRScalarQt}
	\bar{R} =& \hat{R} + \tilde{Q}
\end{align}
where we have introduced
\begin{align}
	\label{e:QtRiemann}
	\tilde{Q}^a_{\ bcd} \equiv &\hnabla_{[c} \tC^a_{\ d]b} - \tC^e_{\ b[c}\tC^a_{\ d]e} \\
	\label{e:QtRicci}
	\tilde{Q}_{mn} \equiv \tilde{Q}^a_{\ man} = &\hnabla_{a} \tC^a_{\ mn} - \tC^b_{\ ma}\tC^a_{\ nb} \\
	\label{e:QtScalar}
	\tilde{Q} \equiv g^{mn}\tilde{Q}_{mn} = &\hnabla_{a} \tC^{am}_{\ \ \ m} - \tC^b_{\ ma}\tC^{am}_{\ \ \ b}~.
\end{align}
In terms of $\bnabla_a$, these take the following form
\begin{align}
	\label{e:QtRiemannbnabla}
	\tilde{Q}^a_{\ bcd} = &\bnabla_{[c} \tC^a_{\ d]b} + \tC^e_{\ b[c}\tC^a_{\ d]e} \\
	\label{e:QtRiccibnabla}
	\tilde{Q}_{mn}  = &\bnabla_{a} \tC^a_{\ mn} + \tC^b_{\ ma}\tC^a_{\ nb} \\
	\label{e:QtScalarbnabla}
	\tilde{Q} = &\bnabla_{a} \tC^{am}_{\ \ \ m} - \tC^b_{\ ma}\tC^{am}_{\ \ \ b}~.
\end{align}
Note that $\tilde{Q}$ can be written with $\hnabla_a$ and $\bnabla_a$ interchangeable with no other modification and no change in signs. This is because the divergence of an arbitrary vector is the same in terms of either $\hnabla_a$ and $\bnabla_a$ as we will prove in the next section. This feature will allow us to integrate by parts Lagrangians with either $\hnabla_a$ or $\bnabla_a$ as we will see shortly.

In terms of $\tilde{Q}^a_{\ bcd}$ and $Q^a_{\ bcd}$ and their associated Ricci tensors and Ricci scalars, Eqs.~\eqref{e:KRiemannAllForms}, \eqref{e:KRicciAllForms}, and \eqref{e:KScalarAllForms} take the form
\begin{align}
\label{e:KRiemannQExpandAllForms}
\mathcal{K}^a{}_{bcd} = & \hat{R}^a_{\ bcd} + \tilde{Q}^a_{\ bcd}  + \delta^a{}_{[c} \bar{\mathcal{P}}_{d]b} \cr
    = & \hat{R}^a_{\ bcd} + Q^a_{\ bcd} + \delta^a{}_{[c} \mathcal{P}_{d]b} - \delta^a{}_b \mathcal{P}_{[cd]} \\
%%%%%%%%%%%%%%%%%%%%%%%
    \label{e:KRicciQExpandAllForms}
\mathcal{K}_{mn} = &  \hat{R}_{mn} + \tilde{Q}_{mn} + (\rd -1) \bar{\mathcal{P}}_{mn} \cr
        = & \hat{R}_{mn} + Q_{mn} + (\rd-1) \mathcal{P}_{mn} - \rd \mathcal{P}_{[mn]} \\
    \label{e:KScalarQExpandAllForms}
        \mathcal{K}  = & \hat{R} + \tilde{Q} + (\rd -1) \bar{\mathcal{P}} \cr
   = & \hat{R} + Q + (\rd-1) \mathcal{P} ~~~.
\end{align}
These relationships will be useful when we compare the forms of the equations of motion across different formulations.

%%%%%%%%%%%%%%%%%%%%%%%%%%%%%%%%%%%%
%%%%%%%%%%%%%%%%%%%%%%%%%%%%%%%%%%%%
%%%%%%%%%%%%%%%%%%%%%%%%%%%%%%%%%%%%
%%%%%%%%%%%%%%%%%%%%%%%%%%%%%%%%%%%%%
\subsection{Covariant and Manifestly Projective Invariant TW Field Equations}
Here we derive the covariant and manifestly projective invariant field equations by varying the action with respect to $\bar{\cP}_{ab}$, $\tC^c{}_{ab}$, and $g^{mn}$. First, we point out two helpful tricks. One is that the divergence of an arbitrary vector $A^a$ is the same whether it is taken with $\bnabla_a$ or $\hnabla_a$. This owes to the vanishing of the trace of $\tC^a{}_{bc}$ as follows 
\begin{align}\label{e:divbnablaProof}
    \bnabla_a A^a =& \partial_a A^a + \bGamma^{a}{}_{ba}A^b \cr
                 =& \partial_a A^a + \hat{\Gamma}^{a}{}_{ba}A^b + \tC^{a}{}_{ba}A^b \cr
                 =& \partial_a A^a + \hat{\Gamma}^{a}{}_{ba}A^b \cr
                 =& \hat{\nabla}_a A^a ~. 
\end{align}
Second, we have that a total derivative $\bar{\nabla}_a$ of an arbitrary vector $A^a$ is a surface term
\begin{align}
    \int d^\rd x \sqrt{|g|} \bar{\nabla}_a A^a =& \int d^\rd x \sqrt{|g|} \hat{\nabla}_a A^a \cr
    =& \int d^{\rm}x \sqrt{|g|} (\partial_a A^a + \hat{\Gamma}^a{}_{ba}A^b) \cr
    =& \int d^{\rm}x(\sqrt{|g|} \partial_a A^a + A^a \partial_a \sqrt{|g|} ) \cr
    =&  \int d^\rd x \partial_a (\sqrt{|g|} A^a) \cr
    =&~ \text{surface term} 
\end{align}
where we have used Eq.~\eqref{e:ga} in going from the second to third lines.  This means we can use partial integration with $\bar{\nabla}_a$ at the Lagrangian level at no cost to the field equations. For instance given an arbitrary rank three tensor $T^{abc}$
\begin{align}
    T^{a bc} \bar{\nabla}_a \delta\bar{\cP}_{bc} =& - (\bar{\nabla}_a T^{abc}) \delta \bar{\cP}_{bc} 
\end{align}
up to a surface term. Thus, we can work entirely with the TW Lagrangian that we copy from Eqs.~\eqref{e:LTW} and~\eqref{e:LPGB} to here for convenience
\begin{align} 
    \mathcal{L} =& -\frac{1}{2 \kappa_0}(\mathcal{K}+ 2 \Lambda_0)   + \mathbb{L}\tag{\ref{e:LTW}} \\
    \mathbb{L} =& J_0 c \left[\lambda_0^2  K_{bcd} K^{bcd} \hspace*{-3 pt} -  \hspace*{-2 pt}\mathcal{K}_{abcd} \mathcal{K}^{abcd} \hspace*{-2 pt}+ \hspace*{-2 pt}4 \mathcal{K}_{bd} \mathcal{K}^{bd}\hspace*{-3 pt} -\hspace*{-2 pt} \mathcal{K}^2 \right]\hspace*{-2 pt}  \tag{\ref{e:LPGB}}
\end{align}
and need not work with the full action, Eq.~\eqref{e:TWActionExpanded}.

The connection $\bar{\nabla}_a$ provides as much calculational functionality in TW gravity as the Levi-Civita connection $\hat{\nabla}_a$ does in conventional general relativity. This is a powerful new feature that the projectively invariant and tensorial $\bar{\nabla}_a$ and $\bcP_{ab}$ provide for TW gravity, moving the results of the previous work~\cite{Brensinger:2020gcv,SamBrensinger} closer to the language of more conventional language in general relativity. In the following, we will demonstrate the utility of this new calculation tool.

%%%%%%%%%%%%%%%%%%%%%%%%%%%%%%%%%%%%
%%%%%%%%%%%%%%%%%%%%%%%%%%%%%%%%%%%%
%%%%%%%%%%%%%%%%%%%%%%%%%%%%%%%%%%%%
%%%%%%%%%%%%%%%%%%%%%%%%%%%%%%%%%%%%%
\subsubsection{Field Equations for \texorpdfstring{$\bcP_{ab}$}{Pab}}
Varying the Lagrangian with respect to $\bcP_{ab}$ depends only on the following two variations
\begin{align}
    \delta_{\bcP}\cK^a{}_{bcd} = & \delta^a{}_{[c} \delta\bcP_{d]b} ~~~,~~~\delta_{\bcP} K_{bcd} = \bnabla_{[c} \delta\bcP_{d]b}
\end{align}
along with contractions. Considering only these variations, we have
\begin{align}
    &\delta_{\bcP} \mathcal{L} =  - \tfrac{1}{2\kappa_0} g^{ab} \delta_{\bcP} \cK_{ab} + 2 J_0 \lambda_0^2 K^{bcd} \delta_{\bcP} K_{bcd} \cr
    &\qquad\quad- 2 J_0 \cK_a^{\ bcd} \delta_{\bcP}\cK^a{}_{bcd}  - 2 J_0 ( -4 \cK^{ab} + \cK g^{ab})\delta_{\bcP} \cK_{ab} \cr
   &= - \tfrac{1}{2\kappa_0} g^{ab} (\rd -1) \delta\bcP_{ab} + 2 J_0 \lambda_0^2 K^{bcd} \bnabla_{[c} \delta\bcP_{d]b} \cr
    &- \hmtwo  2 J_0 \cK_a^{\ bcd} \delta^a_{\ [c} \delta\bcP_{d]b} \hmtwo -\hmtwo 2 J_0  (- 4 \cK^{ab}\hmtwo + \hmtwo \cK g^{ab})(\rd \hmtwo - \hmtwo 1)\delta\bcP_{ab}   ~~~~~~
\end{align}
where recall here on $\mathcal{M}$ indices are raised and lowered via $g^{ma}$ and $g_{am}$ as in Eqs.~\eqref{e:cKraiselower1}, \eqref{e:cKraiselower2}, and~\eqref{e:cKraiselower3}. Integrating by parts in the second term, contracting indices in the third, and renaming and reordering indices yields
\begin{align}
   \hmfive \delta_{\bcP} \mathcal{L} =& \big(\hmtwo - \hmtwo \tfrac{\rd -1}{2 \kappa_0} g^{ab} \hmtwo  + \hmtwo  2 J_0 \lambda_0^2 \bnabla_c K^{(ab)c}  \hmtwo  - \hmtwo  2 J_0 (\rd \hmtwo  - \hmtwo 1) \cK g^{ab}\cr
    & \quad + 8 J_0 (\rd -1) \cK^{ab}  - 4 J_0 \cK^{ab}\big) \delta\bcP_{ab} 
\end{align}
where we have used the antisymmetry of the last two indices of $K^{abc}$ to reorder the indices. 

Simplifying we have
\begin{align}
    \delta_{\bcP} \mathcal{L} =& 2 J_0 \mathbb{D}^{ab} \delta \bcP_{ab} 
\end{align}
where the equation of motion for $\bcP_{ab}$ is 
\begin{align}\label{e:PbarEOM}
               0=\mathbb{D}^{ab} =& \lambda_0^2 \bar{\nabla}_c K^{(ab)c} -(\rd - 1) \mathbb{K}^{ab} -2\mathcal{K}^{ab} ~.
\end{align}
Recall the projectively invariant tensor $\mathbb{K}^{ab}$ was defined in Eq.~\eqref{e:bbK}. Using Eq.~\eqref{e:bnablatoOthernablas} to expand the first term, it is straightforward to show that this is the \emph{exact} same as Eq.~\eqref{e:DEOMtensorial} that was derived via variation with respect to $\cD_{ab}$. In the form of Eq.~\eqref{e:PbarEOM}, however, we see this equation of motion is covariant and manifestly projectively invariant as it is written entirely in terms of such objects. 

%%%%%%%%%%%%%%%%%%%%%%%%%%%%%%%%%%%%
%%%%%%%%%%%%%%%%%%%%%%%%%%%%%%%%%%%%
%%%%%%%%%%%%%%%%%%%%%%%%%%%%%%%%%%%%
%%%%%%%%%%%%%%%%%%%%%%%%%%%%%%%%%%%%%
\newpage
\subsubsection{Field Equations for \texorpdfstring{$\tC^c{}_{ab}$}{Ccab}}
Similar to $\bcP_{ab}$, the only variations that contribute to the field equations for $\tC^c{}_{ab}$ are
\begin{align}
    \delta_{\bGamma}\cK^a{}_{bcd} = & \bnabla_{[c} \delta\bGamma^a{}_{d]b}~~~,~~~\delta_{\bGamma} K_{bcd} =  -\delta\bGamma^m{}_{b[c} \bcP_{d]m}
\end{align}
we will work with $\bGamma^c{}_{ab}$ for most of this calculation, expanding in terms of $\tC^c{}_{ab}$ at the very end, so that we can borrow this calculation when we derive the metric field equations. Our first step parallels that for $\bcP_{ab}$
\begin{align}
   & \delta_{\bGamma} \mathcal{L} =  - \tfrac{1}{2\kappa_0} g^{ab} \delta_{\bGamma} \cK_{ab} + 2 J_0 \lambda_0^2 K^{bcd} \delta_{\bGamma} K_{bcd} \cr
    &\qquad\quad- 2 J_0  \cK_a^{\ bcd} \delta_{\bGamma}\cK^a{}_{bcd} - 2 J_0  ( -4 \cK^{ab} + \cK g^{ab})\delta_{\bGamma} \cK_{ab}  \cr
    &= - \tfrac{1}{2\kappa_0} g^{ab} \bnabla_{[c} \delta\bGamma^c{}_{b]a} + 2 J_0 \lambda_0^2 K^{bcd} (-\delta\bGamma^m{}_{b[c} \bcP_{d]m}) \cr
   & - \hmtwo 2 J_0 \cK_a^{\ bcd} \bnabla_{[c} \delta\bGamma^a_{\ d]b}\hmtwo -\hmtwo 2 J_0 (- 4 \cK^{ab} \hmtwo+ \hmtwo\cK g^{ab})\bnabla_{[c} \delta\bGamma^c{}_{b]a} 
\end{align}
where recall here on $\mathcal{M}$ indices are raised and lowered via $g^{ma}$ and $g_{am}$ as in Eqs.~\eqref{e:cKraiselower1}, \eqref{e:cKraiselower2}, and~\eqref{e:cKraiselower3}.
Integrating by parts, relabeling some indices, and collecting like terms, such as the first and last terms that form a $\mathbb{K}^{ab}$ term (Eq.~\eqref{e:bbK}), and simplifying results in
\begin{align}
  & \delta_{\bGamma}  \mathcal{L} = 2 J_0 \delta\bGamma^c_{\ a[b} \bnabla_{c]} \mathbb{K}^{ab}  \cr
     & \qquad + 2 J_0 \left[- \lambda_0^2 K^{(ab)d} \bcP_{dc} +\bnabla_d \cK_c^{\ (b|d|a)} \right] \delta\bGamma^c_{\ ab}~~~~~~~
\end{align}
where vertical lines denote exclusion from the symmetrization parenthesis.
Introducing a Kronecker delta symbols in the first term to move the antisymmetrization of indices off of $\delta\bGamma^c{}_{ab}$ and $\bnabla_{c}$, using the antisymmetry property $\cK^{mbda} = -\cK^{mbad}$ in the second term on the second line, and simplifying results in 
\begin{align}\label{e:varbGL}
    &\delta_{\bGamma}\mathcal{L}\hmtwo= \hmtwo2 J_0[ \delta^a_{\ [m}\bnabla_{c]} \mathbb{K}^{bm} \hmtwo  - \hmtwo\lambda_0^2 \bcP_{dc}K^{(ab)d}\hmtwo  -\hmtwo \bnabla_d \cK_c^{\ (ab)d} ]\delta\bGamma^{c}_{\ ab}. 
\end{align}

As a quick aside, we note that the tensor $\mathbb{E}_c{}^{ab}$ defined in Eq.~\eqref{e:GammaEOMEscrK} can be expressed as 
\begin{align}\label{e:bbE}
    \mathbb{E}_c{}^{ab} =& \bnabla_c \mathbb{K}^{ab} - \bnabla_d \cK_c^{\ (ab)d} - \lambda_0^2 \bcP_{dc} K^{(ab)d}~.
\end{align}
Utilizing Eq.~\eqref{e:bnablatoOthernablas} to expand the $\bar{\nabla}_c$ derivatives and with the definition of $\bcP_{dc}$, Eq.~\eqref{e:PBarDef}, it is straightforward to show that Eq.~\eqref{e:bbE} indeed reduces to Eq.~\eqref{e:GammaEOMEscrK}. With our newfound expression for  $\mathbb{E}_c{}^{ab}$ in Eq.~\eqref{e:bbE}, the variation Eq.~\eqref{e:varbGL} can be cast in the following form
\begin{align}\label{e:varbGL1}
     \varbGL =&2 J_0 \Big[ \bbE_c{}^{ab}   - \delta^a_{ \ c}\bnabla_m \mathbb{K}^{bm}\Big]\varbG^c_{\ ab}  ~. 
\end{align}
To facilitate the goal of simultaneously isolating the equations of motion for $\tC^c{}_{ab}$ whilst keeping track of the contributions to the equations of motion for the metric $g^{mn}$, we substitute in the traceless combination $\tbbE_c{}^{ab}$ defined in Eq.~\eqref{e:tbbE}
to express Eq.~\eqref{e:varbGL1} as
\begin{align}\label{e:varbGL2}
    &\hmfive \varbGL = 2 J_0  \left\{ \tbbE_c^{\ ab} \hmtwo + \hmtwo \tfrac{2}{\rd \hmtwo+ \hmtwo 1} \delta_c^{\ a} \bbE_m^{\ bm} \hmtwo - \hmtwo \delta_c^{\ a} \bnabla_m \mathbb{K}^{bm} \right\} \varbG^c_{\ ab} . 
\end{align}
Collecting the rightmost two terms, this expression simplifies to
\begin{align}\label{e:varbGL2b}
     \varbGL =&  2J_0 (\tbbE_c{}^{ab}\varbG^c{}_{ab} +  V^a \varbG^m{}_{am})\\
     \label{e:Va}
     V^a \equiv &  \tfrac{2}{\rd+1} \bbE_m{}^{am} - \bnabla_m \mathbb{K}^{am}\cr
      = & \tfrac{1}{\rd+1}\bnabla_b \bbD^{ab} - \tfrac{\lambda_0^2}{\rd+1}\left(\bnabla_b \bnabla_c K^{(ab)c} + 2 \bcP_{db}K^{bad}\right)~~~~~~
\end{align}
where the equality between the definition of $V^a$ and the expression its second line is proved in appendix~\ref{a:Va}.
Using Eq.~\eqref{e:GammabarC}  we split the variation Eq.~\eqref{e:varbGL2b} into two parts 
\begin{align}
    \label{e:varbGL3}
    &\varbGL = \varhGL + \vartCL  \\
    \label{e:varhGL0}
    &\varhGL =  2J_0 (\tbbE_c{}^{ab}\varhG^c{}_{ab} +  V^a \varhG^m{}_{am}) \\
    \label{e:vartCL}
    &\vartCL = 2 J_0 \tbbE_c{}^{ab} \vartC^c{}_{ab}
\end{align}
where we have exploited the  tracelessness of $\tC^c{}_{ab}$ to simplify $\delta_{\tilde{C}} \mathcal{L}$ to a single term involving $\tilde{\mathbb{E}}_c{}^{ab}$. Thus Eq.~\eqref{e:vartCL} yields the equations of motion for $\tC^c{}_{ab}$ and  Eq.~\eqref{e:varhGL0} leads to a contribution to the metric $g^{mn}$ field equations. Summarizing, the equations of motion for $\tC^c{}_{ab}$ are
\begin{align}
    \label{e:EOMC}
     0 = \tbbE_c{}^{ab} \equiv&  \bbE_c{}^{ab} - \tfrac{1}{\rd+1} \delta_c{}^{(a} \bbE_m{}^{b)m} \\\label{e:CEOMbbE}
     \mathbb{E}_c^{\ ab} = &  \bnabla_c \mathbb{K}^{ab}  -  \bnabla_d  \cK_c^{\ (ab)d}  - \lambda_0^2 \bcP_{dc} K^{(ab)d}   
\end{align}
where recall the projectively invariant tensor $\mathbb{K}^{ab}$ was defined in Eq.~\eqref{e:bbK}.
The equations of motion for $\tC^c{}_{ab}$ are covariant and manifestly projective invariant as they are composed entirely of such fields. Where the diffeomorphism field equations for $\cD_{ab}$ are identical to those for $\bcP_{ab}$, the equations of motion for $\Pi^c{}_{ab}$ are different from those for $\tC^c{}_{ab}$. This is clearly seen in Eq.~\eqref{e:cEtilde} in that 
$\tilde{\mathbb{E}}_c^{\ ab}$  and $\tilde{\mathcal{E}}_c^{\ ab}$ differ by terms involving the diffeomorphism field equations tensor $\mathbb{D}^{ab}$.
We rewrite this below in terms of $\tilde{\mathbb{E}}_c^{\ ab}$ vanishing instead of $\tilde{\mathcal{E}}_c^{\ ab}$, as this enforces the equations of motion for $\tilde{C}^c_{\ ab}$
\begin{align}\label{e:bbEtomcE}
    0=\tilde{\mathbb{E}}_c{}^{ab} =& \tilde{\mathcal{E}}_c{}^{ab} - \left(g_c \mathbb{D}^{ab} - \tfrac{1}{\rd+1} \delta_c{}^{(a}g_{m} \mathbb{D}^{b)m} \right)~. 
\end{align}
We see here that upon enforcing the diffeomorphism field equations (going on-shell)
\begin{align}
\mathbb{D}^{ab}=0 \Rightarrow \tilde{\mathbb{E}}_c^{\ ab} = \tilde{\mathcal{E}}_c^{\ ab} =0 
\end{align}
thus the equations of motion for $\tC^c{}_{ab}$ and $\Pi^a{}_{ab}$ are identical on-shell.

%%%%%%%%%%%%%%%%%%%%%%%%%%%%%%%%%%%%
%%%%%%%%%%%%%%%%%%%%%%%%%%%%%%%%%%%%
%%%%%%%%%%%%%%%%%%%%%%%%%%%%%%%%%%%%
%%%%%%%%%%%%%%%%%%%%%%%%%%%%%%%%%%%%%
\newpage
\subsubsection{Field Equations for \texorpdfstring{$g^{mn}$}{gmn}}
To derive the field equations for the metric, we must vary the product of $\sqrt{|g|}$ and the Lagrangian, Eq.~\eqref{e:LTW}
\begin{align}\label{e:vargLTW0}
    \delta_g(\sqrt{|g|} \mathcal{L}) =& \tfrac{\sqrt{|g|}}{2}  g^{mn} \delta g_{mn} \mathcal{L} - \tfrac{\sqrt{|g|}}{2\kappa_0} \delta g^{mn} \cK_{mn}  \cr
    & + \sqrt{|g|}\delta_g \mathbb{L} |_{\delta \hat{\Gamma}^c_{\ ab} = 0} + \sqrt{|g|} \delta_{\hat{\Gamma}} \mathcal{L} \cr
    =& \tfrac{\sqrt{|g|}}{2 \kappa_0} \delta g^{mn} ( -  \cK_{mn} - \kappa_0 g_{mn} \mathcal{L} ) \cr
    & + \sqrt{|g|}\delta_g \mathbb{L} |_{\delta \hat{\Gamma}^c_{\ ab} = 0} + \sqrt{|g|} \delta_{\hat{\Gamma}} \mathcal{L} 
\end{align}
where in going from the first to the second equality, we have used the relationship
\begin{align}\label{e:deltagdowntoup}
	\delta g_{mn} = - g_{ma}g_{nb}\delta g^{ab}~.
\end{align} 
The term involving $\delta_g \mathbb{L}$ is evaluated setting variations of the Levi-Civita connection $\delta \hat{\Gamma}^c_{\ ab}$ to zero to avoid double counting of such variations present in the last term involving $\delta_{\hat{\Gamma}}\mathcal{L}$. 

We next focus on this last term, continuing with Eq.~\eqref{e:varhGL0} working to rewrite it in terms of a variation with respect to $g^{mn}$. As proved in appendix~\ref{a:delGammatodelg}, a variation with respect to the Levi-Civita connection can be expressed as a variation with respect to the metric as 
\begin{align}\label{e:delGammatodelg}
          \tbbE_c{}^{ab} \delta\hat{\Gamma}^c{}_{ab} =&  \tfrac{1}{2} \delta g^{mn} \hat{\nabla}_a \tbbE^{*\ \ a}_{mn}\\
          \label{e:Stardef}
        \tbbE^{*\ \ a}_{mn} \equiv & \tbbE_{(mn)}^{\ \ \ \ \ a} - \tbbE^a_{\ mn}\\
            \label{e:delGammaTracetodelg}
            V^a \delta\hat{\Gamma}^m{}_{am} =&  \tfrac{1}{2}\delta g^{mn} g_{mn} \bar{\nabla}_a V^a  
\end{align}
up to total derivatives.
Using this, Eq.~\eqref{e:varhGL0} becomes
\begin{align}
   \varhGL =&  \varg^{mn}J_0 \left( g_{mn} \bnabla_a V^a + \hnabla_a \tbbE^{* \ \ a}_{mn} \right)~. 
\end{align}
Using Eqs.~\eqref{e:bnablaGenAction}~ and \eqref{e:GammabarC} to switch to $\bnabla_a$ we have 
\begin{align}
  \hmfive \varhGL & \hmtwo = \hmtwo  \varg^{mn} \hmtwo J_0 \hmtwo \left[ g_{mn} \bnabla_a V^a \hmtwo  + \hmtwo  \bnabla_a \tbbE^{* \ \ a}_{mn} \hmtwo  + \hmtwo  \tC^p_{\ a(m}\tbbE^{* \ \  a}_{n)p} \right].
\end{align}

Finally, using the identity below that is proved in appendix~\ref{a:DivV}
\begin{align}\label{e:DivV}
    \bnabla_a V^a = \tfrac{1}{\rd+1} \bnabla_a\bnabla_b \bbD^{ab} - \tfrac{\lambda_0^2}{\rd+1} \bnabla_a\left(\cK^a{}_{bcd} K^{bcd}\right)
\end{align}
we have
\begin{align}\label{e:varhGL}
    \hmtwenty \varhGL & =  \varg^{mn}J_0\Big[ -  \tfrac{\lambda_0^2}{\rd+1} g_{mn} \bnabla_a\left(\cK^a{}_{bcd} K^{bcd}\right) \cr
    &  + \tfrac{1}{\rd+1}g_{mn}\bnabla_a\bnabla_b \bbD^{ab} + \bnabla_a \tbbE^{* \ \ a}_{mn} \hmtwo  + \hmtwo  \tC^p_{\ a(m}\tbbE^{* \ \  a}_{n)p} \Big].
\end{align}

Our final task is to expand the third term of Eq.~\eqref{e:vargLTW0}, $\sqrt{|g|}\delta_g \mathbb{L} |_{\delta \hat{\Gamma}^c_{\ ab} = 0}$ . We start by expanding the PGB Lagrangian itself, Eq.~\eqref{e:LPGB}, to expose the factors of the metric $g_{mn}$ and its $g^{mn}$
\begin{align}
    \mathbb{L} =&  J_0 (\lambda_0^2  g^{ab} \hmone  g^{cm} \hmone g^{dn} \hmtwo K_{bcd} K_{amn} \hmtwo - \hmtwo g_{am} g^{bn}\hmone g^{cp}\hmone g^{dq}\hmone \mathcal{K}^{a}_{\ bcd} \mathcal{K}^{m}_{\ npq}) \nonumber\\
    & + J_0 (4 g^{am}g^{bn}\mathcal{K}_{ab} \mathcal{K}_{mn}-  g^{ab}g^{mn}\mathcal{K}_{ab}\cK_{mn}) . 
\end{align}

Variation of $\mathbb{L}$ with respect to the metric (but not Levi-Civita connection to prevent the aforementioned double counting) proceeds via several product rules resulting in 
\begin{align}
    &\delta_g \mathbb{L} |_{\delta \hat{\Gamma}^c_{\ ab} = 0} =
    J_0 \lambda_0^2 (\delta g^{ab} K_{acd}K_b^{\ cd} + 2 \delta g^{dn} K_{bcd}K^{bc}_{\ \ n}) \nonumber\\
    &\hmtwo - \hmtwo  J_0 ( \delta g_{am} \cK^a_{\ bcd} \cK^{mbcd} \hmtwo + \hmtwo \delta g^{bn} \hmone \cK^a_{\ bcd} \cK_{an}^{\ \ cd} \hmtwo + \hmtwo 2 \delta g^{dq} \hmone \cK^a_{\ bcd}\cK_{a}^{\ bc}{}_{q}) \nonumber\\
    &\hmtwo + \hmtwo J_0 (8 \delta g^{an}\cK_{ab}\cK^b_{\ n} \hmtwo - \hmtwo 2 \delta g^{mn} \cK_{mn} \cK) ~. 
\end{align}
Recall here on the manifold $\mathcal{M}$, we raise and lower indices with $g^{mn}$ and $g_{mn}$ as for example $\cK_{a}{}^{bc}{}_q = g_{am}g^{bn}g^{cp}\cK^m{}_{npq}$. Using Eq.~\eqref{e:deltagdowntoup}, collecting like terms, and relabeling indices we have
\begin{align}\label{e:vargLmanifest}
&\delta_g \mathbb{L} |_{\delta \hat{\Gamma}^c_{\ ab} = 0} = \delta g^{mn}   \left[ J_0 \lambda_0^2\left(K_{mcd} K_n{}^{cd} + 2 K_{bcm} K^{bc}{}_n\right) \right. \nonumber\\
    &\quad +J_0 \left( \cK_{mbcd}\cK_n{}^{bcd} - \cK^a{}_{mcd}\cK_{an}{}^{cd} - 2 \cK^{abc}{}_{m}\cK_a{}_{bcn} \right)  \nonumber\\
    &\quad \left. + J_0 \left(8\cK_{mb} \cK^b_{\ n}  - 2 \cK \cK_{mn}\right) \right]~. 
\end{align}
Identifying the term in brackets as $\mathbb{L}_{mn}$ from Eq.~\eqref{e:PGBLmn} results in
\begin{align}\label{e:varbbL}
&\delta_g \mathbb{L} |_{\delta \hat{\Gamma}^c_{\ ab} = 0} = \delta g^{mn} \mathbb{L}_{mn}~.
\end{align}

We insert Eqs.~\eqref{e:varhGL} and~\eqref{e:varbbL} into Eq.~\eqref{e:vargLTW0} and collect all results together into the equation of motion tensor $\mathbb{H}_{mn}$
\begin{align}
      \delta_g(\sqrt{|g|} \mathcal{L}) =& \tfrac{\sqrt{|g|}}{2\kappa_0} \delta g^{mn} \mathbb{H}_{mn}
\end{align}
where the associated equation of motion is
\begin{align}\label{e:mtapiH}
   \hmfive 0=\mathbb{H}_{mn} =& \mathcal{H}_{mn} + \tfrac{2 J_0\kappa_0 }{\rd+1}g_{mn}\bnabla_a\bnabla_b \bbD^{ab}  
   \cr
    & + 2 J_0 \kappa_0  \left[ \bnabla_a \tbbE^{* \ \ a}_{mn} \hmtwo  + \hmtwo  \tC^p_{\ a(m}\tbbE^{* \ \  a}_{n)p}  \right]~~~~~~~~ \\
 %%%%%%%%%%%%%%%%%%%%%%%%%%%%%%%
 %%%%%%%%%%%%%%%%%%%%%%%%%%%%%%%
 	\label{e:cHbar}
    \mathcal{H}_{mn} =& -\mathcal{K}_{mn} +  \kappa_0 \left( 2 \mathbb{L}_{mn} -  g_{mn} \mathcal{L} \right) \cr
    & -  g_{mn} \tfrac{2 J_0  \kappa_0  \lambda_0^2}{\rd + 1}\bnabla_a \left(\cK^a{}_{bcd}K^{bcd}\right)
\end{align}
To show that $\mathcal{H}_{mn}$ is equivalent to the form in Eq.~\eqref{e:cHmn}, Eq.~\eqref{e:bnablatoOthernablas} is used to relate the $\bnabla_a$ term to a $\breve{\nabla}_a$ term as
\begin{align}\label{e:nablahatKK}
    \bnabla_a(\cK^a{}_{bcd} K^{bcd}) =& (\breve{\nabla}_a - (\rd + 1) g_a)(\cK^a{}_{bcd} K^{bcd}) \cr
    =& \tfrac{1}{\sqrt{|g|}} \breve{\nabla}_a (\sqrt{|g|}\cK^a{}_{bcd} K^{bcd})
\end{align}
where the last equality follows from the definition of $g_a$ in Eq.~\eqref{e:ga} along with the facts that $\breve{\nabla}_a$ obeys a Leibniz rule and acts on $|g|$ as merely a partial derivative. 

Finally, utilizing Eqs.~\eqref{e:LTW}, \eqref{e:KRicciQExpandAllForms}, and \eqref{e:KScalarQExpandAllForms}, we cast Eq.~\eqref{e:mtapiH} in the form of Einstein equations
\begin{align}\label{e:EECPIFull}
	&\hat{R}_{mn} - \tfrac{1}{2} g_{mn} \hat{R} = \kappa_0 \mathbb{T}_{mn}  + g_{mn} \Lambda_0\\
	\label{e:EECPITmnFull}
    &\mathbb{T}_{mn} \hmtwo = \hmtwo
   -\kappa_0^{-1} \hmtwo\left[\tilde{Q}_{mn}\hmtwo -\hmtwo \tfrac{1}{2}g_{mn} \tilde{Q}\hmtwo + \hmtwo(\rd-1) \left(\bcP_{mn} \hmtwo  -\hmtwo \tfrac{1}{2}g_{mn} \bcP \right)\right]\nonumber \\ 
  & \quad  - g_{mn} \tfrac{2J_0 \lambda_0^2 }{\rd+1}\bnabla_a (K_{bcd} \mathcal{K}^{abcd})+ 2 \mathbb{L}_{mn} -  g_{mn} \mathbb{L} \cr
  &+\hmtwo 2 J_0 \hmtwo \left[  \tfrac{1}{\rd+1}g_{mn}\bnabla_a\bnabla_b \bbD^{ab} \hmtwo  + \hmtwo \bnabla_a \tbbE^{* \ \ a}_{mn} \hmtwo  + \hmtwo  \tC^p_{\ a(m}\tbbE^{* \ \  a}_{n)p}  \right]\hmtwo.
\end{align}

%%%%%%%%%%%%%%%%%%%%%%%%%%%%%%
%%%%%%%%%%%%%%%%%%%%%%%%%%%%%%
\section{\label{s:Summary}Classical Equivalence of Field Equations Across Formulations}
We have seen that the equations of motion derived in terms of the old fields $\mathcal{D}_{ab}$, $\Pi^a{}_{bc}$, and $g^{mn}$ are equivalent to those derived in terms of the new fields $\bcP_{ab}$, $\tC^a{}_{bc}$, and $g^{mn}$ upon enforcement of the equations of motion (going on-shell). They describe the same classical system, regardless of field choice, as expected. Equation~\eqref{e:bbEtomcE} is a particularly interesting relationship between $\tilde{\mathbb{E}}_c{}^{ab}$ and  $\tilde{\mathcal{E}}_c{}^{ab}$ and  $g_c \mathbb{D}^{ab}$ 
as just the right combination to remove the non-tensorial part of $\tilde{\mathcal{E}}_c{}^{ab}$ to create the tensorial  $\tilde{\mathbb{E}}_c{}^{ab}$. Furthermore, we see that the tensorial result $\tilde{\mathbb{E}}_c{}^{ab}$ is also manifestly projective invariant as it is composed entirely of such objects $\tilde{\mathcal{E}}_c{}^{ab}$ and $g_c\mathbb{D}^{ab}$. 

In the following, we investigate some nuances in the variation of the Lagrangian and summarize the field equations and their classical equivalence across formulations. We start by investigating the tensorial or non-tensorial character of the fields and their variations.  We suspect that this will have interesting consequences at the quantum level in terms of Faddeev-Popov-like projective gauge fixing conditions. It is therefore important to work out the details at the classical level here first, leaving the quantum calculation for a future endeavor. 

\subsection{Equivalence of the Variation of the Action}
Let us start by summarizing the variation in terms of the old fields
\begin{align}\label{e:varSoldfields}
    & \delta S \hmtwo = \hmtwo 2 J_0 \hmfour \int \hmfour  d^\rd x \sqrt{|g|}\hmtwo \left[\mathbb{D}^{ab} \hmone\delta\cD_{ab}\hmthree  + \hmthree \tilde{\mathcal{E}}_c^{\ ab} \hmone \delta \Pi^c_{\ ab}\hmthree  +\hmthree  \tfrac{1}{4 J_0\kappa_0} \mathcal{H}_{mn} \delta g^{mn} \hmtwo \right]\hmtwo. 
\end{align}
Using Eqs.~\eqref{e:PBarDef}, \eqref{e:Gammabar}, and~\eqref{e:GammabarC} we can solve for the old field variations $\delta\cD_{ab}$ and $\delta\Pi^c{}_{ab}$ in terms of the new field variations $\delta\bcP_{ab}$ and $\delta\tC^c{}_{ab}$
\begin{align}\label{e:varD}
     \delta\mathcal{D}_{ab} =& \delta\bar{\mathcal{P}}_{ab}   - \tfrac{1}{\rd+1} \bnabla_a \delta\hat{\Gamma}^{m}{}_{bm}  \cr
     & - g_c ( \delta \hat{\Gamma}^c{}_{ab} - \tfrac{1}{\rd+1} \delta^c{}_{(a} \delta\hat{\Gamma}^m{}_{b)m})- g_c \delta\tC^c{}_{ab}  \\
     \label{e:varPi}
   \delta\Pi^c{}_{ab}   =& \delta\hat{\Gamma}^c{}_{ab} + \delta \tC^{c}{}_{ab}+\delta^c{}_{(a} \delta g_{b)} ~. 
\end{align}
With this substitution, the variation becomes
\begin{align}\label{e:SOldToNew1}
    &\delta S = 2 J_0 \int d^\rd x \sqrt{|g|}\big(\mathbb{D}^{ab} \delta\bcP_{ab} + \tilde{\mathcal{E}}_c{}^{ab} \delta \tC^c{}_{ab}\cr
    &  - g_c\mathbb{D}^{ab}\delta\tC^c{}_{ab} + \tfrac{1}{4 J_0\kappa_0} \mathcal{H}_{mn} \delta g^{mn} -\tfrac{1}{\rd+1} \mathbb{D}^{ab} \bnabla_a \delta \hat{\Gamma}^m{}_{bm} \cr
    &  - g_c \mathbb{D}^{ab} ( \delta \hat{\Gamma}^c{}_{ab}  - \tfrac{1}{\rd+1} \delta^c{}_{(a} \delta\hat{\Gamma}^m{}_{b)m}) + \tilde{\mathcal{E}}_c{}^{ab} \delta\hat{\Gamma}^c{}_{ab}\big).
\end{align}
The third term on the first line and second term on the second line can be rewritten as follows 
\begin{align}\label{e:TracelessTransfer1}
&    g_c \mathbb{D}^{ab} \delta\tC^c{}_{ab} = (g_c \mathbb{D}^{ab} - \tfrac{1}{\rd+1} \delta^{(a}{}_c g_m \mathbb{D}^{b)m}) \delta \tC^c{}_{ab} \\
    \label{e:TracelessTransfer2}
&    g_c \mathbb{D}^{ab} ( \delta \hat{\Gamma}^c{}_{ab} - \tfrac{1}{\rd+1} \delta^c{}_{(a} \delta\hat{\Gamma}^m{}_{b)m}) \cr
&\qquad\qquad\quad = (g_c \mathbb{D}^{ab} - \tfrac{1}{\rd+1} \delta^{(a}{}_c g_m \mathbb{D}^{b)m}) \delta \hat{\Gamma}^c{}_{ab} ~. 
\end{align}
Essentially, fully contracting a traceless object with another enforces tracelessness on both. 

Substituting Eqs.~\eqref{e:TracelessTransfer1} and~\eqref{e:TracelessTransfer2} into Eq.~\eqref{e:SOldToNew1}, the second and third terms on the first line can be combined using Eq.~~\eqref{e:bbEtomcE}, as can the second and third terms on the second line. Furthermore, in Eq.~\eqref{e:SOldToNew1} the first term on the second line can be integrated by parts. All this leaves us with
\begin{align}
    &\hmten \delta S = 2 J_0 \int d^\rd x \sqrt{|g|}\big(\mathbb{D}^{ab} \delta\bcP_{ab} + \tilde{\mathbb{E}}_c^{\ ab} \delta \tC^c_{\ ab} \cr
    & \hmten + \hmtwo \tfrac{1}{4 J_0\kappa_0} \mathcal{H}_{mn} \delta g^{mn} \hmtwo + \hmtwo\tfrac{1}{\rd+1} (\bnabla_a\mathbb{D}^{ab}) \delta \hat{\Gamma}^m_{\ bm}  \hmtwo + \hmtwo \tilde{\mathbb{E}}_c^{\ ab} \delta\hat{\Gamma}^c_{\ ab}\big).~~~~ 
\end{align}
Using Eqs.~\eqref{e:delGammatodelg} and~\eqref{e:delGammaTracetodelg} we can rewrite the second line in terms of the variation of $\delta g^{mn}$
\begin{align}
    &\hmfive\delta S \hmtwo =  \hmtwo 2 J_0  \hmfour \int \hmfour d^\rd x \sqrt{|g|}\Big[\hmone\mathbb{D}^{ab} \hmtwo \delta\bcP_{ab}  \hmtwo  +  \hmtwo  \tilde{\mathbb{E}}_c^{\ ab} \delta \tC^c_{\ ab} \hmtwo  +  \hmtwo\tfrac{1}{4 J_0\kappa_0} \mathcal{H}_{mn}  \delta g^{mn} \nonumber\\
    & \qquad  +  \left(\tfrac{1}{2(\rd+1)} g_{mn} \bnabla_a \bnabla_b \mathbb{D}^{ab} \hmtwo+ \hmtwo\tfrac{1}{2}\hat{\nabla}_a \tilde{\mathbb{E}}^{* \ \ a}_{mn} \right) \hmtwo\delta g^{mn} \Big].
\end{align}
Switching from $\hat{\nabla}_a$ to $\bnabla_a$ via Eq.~\eqref{e:GammabarC}, we find that the resulting coefficients of all $\delta g^{mn}$ terms combine to form $\mathbb{H}_{mn}$ as defined in Eq.~\eqref{e:mtapiH}. This leaves us with our final result:
\begin{align}\label{e:varSnewfields}
    & \delta S \hmtwo = \hmtwo 2 J_0 \hmfour \int \hmfour  d^\rd x \sqrt{|g|}\hmtwo \left[\mathbb{D}^{ab} \hmone\delta\bcP_{ab}\hmthree  + \hmthree \tilde{\mathbb{E}}_c^{\ ab} \hmone \delta \tC^c{}_{ab}\hmthree  +\hmthree  \tfrac{1}{4 J_0\kappa_0} \mathbb{H}_{mn} \delta g^{mn} \hmtwo \right]\hmtwo. 
\end{align}

All individual equations of motion and variations in Eq.~\eqref{e:varSnewfields} in terms of the new fields are tensorial. This is not true for Eq.~\eqref{e:varSoldfields} in terms of the old fields. It must be true, however, that the variation of the action is general coordinate invariant either way. What saves Eq.~\eqref{e:varSoldfields}, maintaining its general coordinate invariance, is that the non-tensorial parts of $\delta\cD_{ab}$ and $\tilde{\mathcal{E}}_c{}^{ab}$ conspire to cancel in Eq.~\eqref{e:varSoldfields} while all other fields and variations are tensorial. Specifically, performing a coordinate transformation on Eqs.~\eqref{e:varD} and~\eqref{e:cEtilde} we find
\begin{align}
    &\hmseven \delta\cD'_{ab} = \frac{\partial x^m}{\partial x'^a}\frac{\partial x^n}{\partial x'^b}\left[ \delta \cD_{mn} +j_c \delta \Pi^c{}_{mn} \right]\\
%%%%%%%%%%%%%%%%%%%%%%%%%%%%%%%%%%%%%%
    &\hmseven \tilde{\mathcal{E}}'_c{}^{ab} \hmthree = \hmthree \frac{\partial x^p}{\partial x'^c}\hmtwo\frac{\partial x'^a}{\partial x^m}\hmtwo\frac{\partial x'^b}{\partial x^n} \hmtwo\left[\tilde{\mathcal{E}}_p^{\ mn} \hmthree - \hmthree j_p \mathbb{D}^{mn} \hmthree  + \hmthree \tfrac{1}{\rd+1} \delta_p^{\ (m} j_d\mathbb{D}^{n) d} \right]\\
%%%%%%%%%%%%%%%%%%%%%%%%%%%%%%%%%%%%%%
    &\hmseven j_c = \tfrac{1}{\rd +1} \partial_c \log J= \tfrac{1}{\rd+1} \frac{\partial x'^q}{\partial x^p} \frac{\partial^2 x^p}{\partial x^c \partial x'^q} \\
%%%%%%%%%%%%%%%%%%%%%%%%%%%%%%%%%%%%%%
     &\hmseven J = |\partial x^p/\partial x'^q|
\end{align}
where the non-tensorial parts of these transformations stem from their dependence on $g_a$ as all other parts are tensorial (though a connection itself is not tensorial, its variation is tensorial). Note that the above result for $\delta \mathcal{D}'_{ab}$ is consistent with taking the functional variation of Eq.~\eqref{e:Dtrans} as $\delta j_m = 0$ necessarily. 

With the above results for $\delta \cD'_{ab}$ and $\tilde{\mathcal{E}}'_c{}^{ab}$, we find that the following combination found in Eq.~\eqref{e:varSoldfields} indeed transforms as a scalar
\begin{align}
    \mathbb{D}'^{ab} \delta\cD'_{ab} + \tilde{\mathcal{E}}'_c{}^{ab} \delta \Pi'^c{}_{ab} =\mathbb{D}^{ab} \delta\cD_{ab} + \tilde{\mathcal{E}}_c{}^{ab} \delta \Pi^c{}_{ab} 
\end{align}
where each term individually is not coordinate invariant, but their sum is coordinate invariant.
These nuances between the different formulations, in particular the tensorial and non-tensorial nature of the different fields and variations, may have implications as to certain features within a quantum theory. In particular, these features may prove vital in developing a Faddeev-Popov-type of gauge fixing procedure.

%%%%%%%%%%%%%%%%%%%%%%%%%%%%%%%%%%
%%%%%%%%%%%%%%%%%%%%%%%%%%%%%%%%%%
\subsection{Summary of Field Equations for Each Formulation}
We see that regardless of which set of fields with respect to we vary the action, we get the same field equations on-shell classically. That is, with one set of fields, the classical equations of motion are linear combinations of the fields for the other set. The equations of motion derived from variation of the projectively invariant and tensorial fields $\bcP_{ab}$, $\tC^c{}_{ab}$, and $g^{mn}$ are
\begin{align}
%%%%%%%%%%%%%%%%%PEOm
    &0=\mathbb{D}^{ab} = \lambda_0^2 \bar{\nabla}_c K^{(ab)c} -(\rd - 1) \mathbb{K}^{ab} -2\mathcal{K}^{ab}
 \\
%%%%%%%%%%%%%%%%%%%EEOM
& 0 = \tbbE_c{}^{ab} \equiv  \bbE_c{}^{ab} - \tfrac{1}{\rd+1} \delta_c{}^{(a} \bbE_m{}^{b)m}  \\
%%%%%%%%%%%%%%%%%%%HEOM
\label{e:bbHmnSummary}
 & 0=\mathbb{H}_{mn} = \mathcal{H}_{mn} + \tfrac{2 J_0\kappa_0 }{\rd+1}g_{mn}\bnabla_a\bnabla_b \bbD^{ab}  
   \cr
    & \qquad\qquad\hspace*{5 pt}  + 2 J_0 \kappa_0  \left[ \bnabla_a \tbbE^{* \ \ a}_{mn} \hmtwo  + \hmtwo  \tC^p_{\ a(m}\tbbE^{* \ \  a}_{n)p}  \right]~~~~~~~~  
\end{align}
where
\begin{align}
%%%%%%%%%%%%%%%%%%%%%E
& \mathbb{E}_c^{\ ab} =  \bnabla_c \mathbb{K}^{ab}  -  \bnabla_d  \cK_c^{\ (ab)d}  - \lambda_0^2 \bcP_{dc} K^{(ab)d}   \\
%%%%%%%%%%%%%%%%%%%%%%%%%bbKdef
&\mathbb{K}^{ab} \equiv g^{ab}   \left(\cK + \tfrac{1}{4 J_0 \kappa_0}\right) - 4 \mathcal{K}^{ab}\\
%%%%%%%%%%%%%%%%%%%%%E*def
    &\tbbE^{* \ \ a}_{mn} \equiv  \tbbE_{(mn)}^{\ \ \ \ \ a} - \tbbE^a_{\ mn} \\
%%%%%%%%%%%%%%%%%%%%%%cH
 &\mathcal{H}_{mn} = -\mathcal{K}_{mn} +  \kappa_0 \left( 2 \mathbb{L}_{mn} -  g_{mn} \mathcal{L} \right) \cr
    & \qquad\hspace*{12 pt} -  g_{mn} \tfrac{2 J_0  \kappa_0  \lambda_0^2}{\rd + 1}\bnabla_a \left(\cK^a{}_{bcd}K^{bcd}\right)  \\
%%%%%%%%%%%%%%%%%%%%%LTW
	\label{e:LTWSummary}
    &\mathcal{L} = -\frac{1}{2 \kappa_0}(\mathcal{K}+ 2 \Lambda_0)   + \mathbb{L} \\
%%%%%%%%%%%%%%%%%%%%%%LPGB
    &\mathbb{L} = J_0 \left[\lambda_0^2  K_{bcd} K^{bcd} \hspace*{-3 pt} -  \hspace*{-2 pt}\mathcal{K}_{abcd} \mathcal{K}^{abcd} \hspace*{-2 pt}+ \hspace*{-2 pt}4 \mathcal{K}_{bd} \mathcal{K}^{bd}\hspace*{-3 pt} -\hspace*{-2 pt} \mathcal{K}^2 \right]\hspace*{-2 pt} ~~~ \\
%%%%%%%%%%%%%%%%%%%%%%LPGBmn
      & \mathbb{L}_{mn} \hmtwo =\hmtwo J_0 \lambda_0^2  (K_{mcd}K_n^{\ cd}\hmtwo +\hmtwo 2 K_{bcm}K^{bc}_{\ \ n}) \cr
    & \qquad \quad +\hmtwo J_0  \hmtwo (\cK_{mbcd}\cK_n^{\ bcd}\hmtwo -\hmtwo\cK^a_{\ mcd}\cK_{an}^{\ \ cd} \hmtwo-\hmtwo 2 \cK^{abc}_{\ \ \ m} \cK_{abcn}) \nonumber\\
    & \qquad \quad +\hmtwo J_0  \left( 8 \cK_{mb}\cK^b_{\ n}\hmtwo -\hmtwo 2 \cK \cK_{mn}\right).
\end{align}

The equations of motion derived from variation of the original fields $\cD_{ab}$, $\Pi^c{}_{ab}$, and $g^{ab}$ are
\begin{align}\label{e:DEOMHEOM}
	\mathbb{D}^{ab} =& 0~~~,~~~\mathcal{H}_{mn} = 0 \\
	\label{e:EtildeOnShell}
\tilde{\mathcal{E}}_c{}^{ab} =& \tilde{\mathbb{E}}_c{}^{ab} + \left(g_c \mathbb{D}^{ab} - \tfrac{1}{\rd+1} \delta_c{}^{(a}g_{m} \mathbb{D}^{b)m} \right)=0~. 
\end{align}
Upon enforcing the equations of motion (going on-shell), these two sets of equations simultaneously reduce to
\begin{align}\label{e:TWEOMSummary}
	\mathbb{D}^{ab} =& 0~,~ \tilde{\mathcal{E}}_c^{\ ab} = \tilde{\mathbb{E}}_c^{\ ab} =0~,~\mathbb{H}_{mn} = \mathcal{H}_{mn} =  0~.
\end{align}

Finally, it is useful to express the equations of motion for $g^{mn}$ in the form of Einstein equations in a covariant and manifestly projective invariant form. On-shell, this becomes for either formulation 
\begin{align}\label{e:EECPI}
	&\hat{R}_{mn} - \tfrac{1}{2} g_{mn} \hat{R} = \kappa_0 \hat{T}_{mn}  + g_{mn} \Lambda_0\\
	\label{e:EECPITmn}
    &\hat{T}_{mn} \hmtwo = \hmtwo
   -\kappa_0^{-1} \hmtwo\left[\tilde{Q}_{mn}\hmtwo -\hmtwo \tfrac{1}{2}g_{mn} \tilde{Q}\hmtwo + \hmtwo(\rd-1) \left(\bcP_{mn} \hmtwo  -\hmtwo \tfrac{1}{2}g_{mn} \bcP \right)\right]\nonumber \\ 
  & \qquad\hspace*{6 pt}  - \hmtwo g_{mn} \tfrac{2J_0 \lambda_0^2 }{\rd+1}\bnabla_a (K_{bcd} \mathcal{K}^{abcd}) \hmtwo + \hmtwo 2 \mathbb{L}_{mn}\hmtwo - \hmtwo g_{mn} \mathbb{L} 
\end{align}
where 
\begin{align}
	\tilde{Q}_{mn}  = &\bnabla_{a} \tC^a_{\ mn} + \tC^b_{\ ma}\tC^a_{\ nb} \\
	\tilde{Q} = &\bnabla_{a} \tC^{am}_{\ \ \ m} - \tC^b_{\ ma}\tC^{am}_{\ \ \ b}~.
\end{align}
The most straightforward way to arrive at the on-shell form of the stress-energy tensor $\hat{T}_{mn}$ in Eq.~\eqref{e:EECPITmn} is to put $\mathbb{T}_{mn}$ in Eq.~\eqref{e:EECPITmnFull} on-shell. That is, set $\mathbb{D}^{ab}$ and $\tilde{\mathbb{E}}_c^{\ ab}$ to vanish. Alternatively, to arrive at Eq.~\eqref{e:EECPITmn} from either Eq.~\eqref{e:EE} or Eq.~\eqref{e:bbHmnSummary}, Eqs.~\eqref{e:LTWSummary}, \eqref{e:KRicciQExpandAllForms}, and~\eqref{e:KScalarQExpandAllForms} prove useful.

Recall that in the form above the Einstein equations for TW gravity are covariant and manifestly projective invariant. Clearly this is true for the stress energy tensor $\hat{T}_{mn}$, though it may seem at first glance that projective invariance is violated on the left hand side of the Einstein equations. This is not so, indeed the left-hand side is projectively invariant: the Levi-Civita Ricci tensor $\hat{R}_{mn}$ and Levi-Civita Ricci scalar $\hat{R}$  know not of projective transformations as they depend only on the metric and its derivatives. The projective transformation associated with $R_{ab}$ has been fully hidden in the projectively invariant Palatini field $\tC^c{}_{ab}$, leaving $\hat{R}_{mn}$ and $\hat{R}$ projectively invariant. Thus the Einstein equations in the above form are indeed covariant and manifestly projective invariant.
%%%%%%%%%%%%%%%%%%%%%%%%%%%%%%
%%%%%%%%%%%%%%%%%%%%%%%%%%%%%%
\subsection{Solutions}
In this section, we revisit the de Sitter space solution originally presented in~\cite{Brensinger:2019mnx} in constant volume coordinates $g_a = 0$. We demonstrate here that the de Sitter space solution is still valid even with non-constant volume coordinates 
$g_a \ne 0$.   We start with a brief overview of de Sitter space. A more thorough review can be found in the literature~\cite{deSitter:1917zz,Strominger:2001pn,Spradlin:2001pw,Kim:2002uz} in the east coast metric sign convention (mostly plus) or~\cite{Aldrovandi:2006vr} in the west coast metric sign convention (mostly minus). De Sitter space in $\rd$-dimensions is Minkowski space in $\rd+1$-dimensions constrained on a spacetime surface of constant de Sitter radius $L$ 
\begin{align}
	\label{e:dSlineelement}
	g_{ab} dx^a dx^b =& \eta_{ab} dx^a dx^b - dz^2 \\
	\label{e:dSconstraint}
	\eta_{ab} x^a x^b - z^2 =& - L^2 \\
	\label{e:eta}
	\eta_{ab} =& \text{diag} (1,-1,-1, \cdots , -1)~.
\end{align} 
Here we are using the west coast metric sign convention as we do throughout the entire paper.
Note that the coordinate $z$ is \emph{not} the projective coordinate $\lambda$. This is simply an extra coordinate typically used to succinctly encode $\rd$-dimensional de Sitter space as above and has nothing to do with the $\rd+1$-dimensional Thomas cone. It is straightforward to remove the $z$-dependence by enforcing the constraint Eq.~\eqref{e:dSconstraint}, and then solve for the inverse metric, Christoffel symbols, the Riemann and Ricci curvature tensors, and the Ricci scalar. These results are~\footnote{De Sitter space is often described as a space of constant \emph{positive} spacetime curvature. This physically corresponds to a \emph{positive} cosmological constant in any convention of which we are aware, though the sign of the Ricci scalar is convention dependent. In our conventions, the Ricci scalar  is negative in de Sitter space. The expressions for the Riemann and Ricci tensors each also differ by minus signs across east coast and west coast conventions.}
\begin{align}
	g_{ab} =& \eta_{ab} - \frac{\eta_{ac}\eta_{bd} x^c x^d}{L^2 + \eta_{ef} x^e x^f} ~,~
	g^{ab} = \eta^{ab} + \frac{x^a x^b}{L^2} \\\hat{\Gamma}^c{}_{ab} =& - \frac{x^c}{L^2} g_{ab} ~,~
	\hat{R}_{abcd} = -\frac{1}{L^2} g_{a[c}g_{d]b}\\
	\hat{R}_{ab} =& - \frac{\rd-1}{L^2} g_{ab}~,~
	\hat{R}= - \frac{\rd(\rd-1)}{L^2}~.
\end{align}
As the Ricci tensor is proportional to the metric, so too is the Einstein tensor
\begin{align}\label{e:Einsteintensor}
	\hat{G}_{ab} \equiv& \hat{R}_{ab} - \frac{1}{2} g_{ab} \hat{R} = \Lambda g_{ab} \\
	\label{e:CosmologicalConstant}
	\Lambda =& \frac{(\rd-1)(\rd - 2)}{2 L^2}
\end{align}
with the proportionality factor $\Lambda$ the cosmological constant. Here we see the relationship between the cosmological constant $\Lambda$ and the de Sitter radius $L$. Note that in $\rd=1,2$ the cosmological constant $\Lambda$ vanishes, and thus so too does the Einstein tensor $\hat{G}_{ab}$.

The solution to the equations of motion of TW-gravity in de Sitter space is
\begin{align}
\tilde{C}^a{}_{bc}&=0~~~,~~~\bar{\cP}_{ab}  = \mu g_{ab}~~~,~~~\bar{R}_{abcd} =\hat{R}_{abcd} 
\end{align}
with $\mu$ an arbitrary constant and $\hat{R}_{abcd}$ as above. The $\Pi^a{}_{bc}$ equations of motion are trivially satisfied. For the diffeomorphism field and Einstein equations, we restore the coefficients $J_0$ and $\kappa_0^{-1}$, respectively, as these were factored out of the variation, Eq.~\eqref{e:varSnewfields} 
\begin{widetext}
\begin{align}
	\label{e:DiffEOMdeSitter}
	& 0  =  J_0 \mathbb{D}^{ab}  = (\rd  -  1) \left[(\rd  -  2)(\rd  -  3)J_0 \left(L^{-2} -  \mu\right) -  \frac{1}{4\kappa_0}\right] g^{ab} \\
	\label{e:EEdeSitter}
	&\frac{1}{\kappa_0} \hat{G}_{mn} = \left\{\frac{ \Lambda_0}{\kappa_0}+ (\rd-1)(\rd-2)\left[\frac{ \mu}{2\kappa_0} + (\rd-3)(\rd-4)J_0\left(L^{-2}-\mu\right)^2\right] \right\}g_{mn}.
\end{align}
\end{widetext}
%two column version of equation
%\begin{align}
%	\label{e:DiffEOMdeSitter}
%	& 0 \hmtwo = \hmtwo J_0 \mathbb{D}^{ab} \hmtwo = (\rd \hmtwo - \hmtwo 1)\hmtwo \left[(\rd \hmtwo - \hmtwo 2)(\rd \hmtwo - \hmtwo 3)J_0 \left(L^{-2}\hmtwo - \hmtwo \mu\right)\hmtwo - \hmtwo \frac{1}{4\kappa_0}\right]\hmtwo g^{ab} \\
%	\label{e:EEdeSitter}
%	&\frac{1}{\kappa_0} \hat{G}_{mn} = \left\{\frac{ \Lambda_0}{\kappa_0}+ (\rd-1)(\rd-2)\left[\frac{ \mu}{2\kappa_0}\right.\right.  \cr & \qquad\qquad\hspace*{5 pt} \left.\left.  + (\rd-3)(\rd-4)J_0\left(L^{-2}-\mu\right)^2\right] \right\}g_{mn}.
%\end{align}
For $\rd = 1$, the diffeomorphism field equations $\mathbb{D}^{ab}$ are trivially satisfied and the Einstein equations are approximately satisfied in the regime $\Lambda_0 << \kappa_0$.  For $\rd = 2, 3$, the equations are satisfied only for $\kappa_0 = \infty$: these models must be purely  projective Gauss-Bonnet.

For $\rd \ge 4$, we use Eqs.~\eqref{e:Einsteintensor} and~\eqref{e:CosmologicalConstant} to simplify Eqs.~\eqref{e:DiffEOMdeSitter} and~\eqref{e:EEdeSitter} to the forms
%%Old Forms
%\begin{align}
%	\Lambda = \tfrac{(\rd-1)(\rd-2)}{2}\mu + \tfrac{\rd-1}{8(\rd-3)J_0\kappa_0}~~~,~~~\Lambda_0 = \tfrac{\rd(\rd-1)}{16(\rd-2)(\rd-3)J_0\kappa_0} 
%\end{align}
\begin{align}
	\Lambda =& \frac{(\rd-1)(\rd-2)}{2}\mu + \frac{2}{\rd}(\rd-2)\Lambda_0 \\
	\Lambda_0 =& \frac{\rd(\rd-1)}{16(\rd-2)(\rd-3)J_0\kappa_0} 
\end{align}
which for $\rd = 4$ become
\begin{align}\label{e:LambdaDE}
	\Lambda = 3 \mu + \Lambda_0~~~,~~~\Lambda_0 = \frac{3}{8J_0 \kappa_0}~.
\end{align}
The solution presented in~\cite{Brensinger:2019mnx} corresponds to the $\kappa_0 \to \infty$ limit. As done in that work, a further constraint can be applied to $\Lambda$ upon plugging the solution back into the action, and matching the coefficient of the Einstein-Hilbert term to Newton's constant. A very recent work~\cite{Fiedorowicz:2024hip} investigates perturbative solutions around Minkowski space, $\Lambda_0 = - 3 \mu$ in Eq.~\eqref{e:LambdaDE}, where perturbations of $\bcP_{ab}$ act as a form of gravitational radiation.

\section{Conclusion}
In this paper, a covariant and manifestly projective invariant formulation of TW gravity was presented in $\rd$-dimensions. The importance of this formulation is that TW gravity is now written in the way general relativity is typically written, in a covariant form, while being manifestly projective invariant at the same time. A key feature available in this new formulation is the ability to utilize partial integration at the Lagrangian level when deriving field equations, as was demonstrated in this paper. Furthermore, the resulting covariant and manifestly projective invariant field equations are much easier to work with in this form than in their original form.

The TW action, built out of the projective Einstein-Hilbert and projective Gauss-Bonnet actions, ties projective geometry to string theory and gravity. The projective connection and the metric are independent in the spirit of the Palatini formalism, giving rise to non-metricity degrees of freedom. The TW field equations then have degrees of freedom from the metric, the diffeomorphism sector, and the Palatini field, which may be encapsulated in different combinations of fields that we have shown are equivalent up to linear combinations. In this way projective geometry can influence general relativity by providing sources in the energy-momentum tensor, giving a possible geometric origin to dark energy, dark matter, inflation, and other physical phenomenon. 

This paper analyzed pure TW gravity, with no matter or other energy sources: all of the aforementioned applications \emph{are of TW gravity itself}. Interactions with matter and energy, from standard model sources for instance, are straightforward to include: source terms would appear on the right hand sides of Eqs.~\eqref{e:TWEOMSummary}. In the form of the Einstein equations~\eqref{e:EECPI}, interactions with other sources would appear as additions to the stress-energy tensor in the usual way. Indeed, this was already discussed in the previous paper~\cite{Brensinger:2020gcv}. We look forward to investigating the form of such interactions, in this covariant and manifestly projective invariant formulation, in a future work. Indeed, one of the authors (TG) has a work in progress on applications to black hole thermodynamics that considers perturbative solutions around the de Sitter space solution that include traceless parts of $\bcP_{ab}$.

As the equations of motion are different off-shell as derived with respect to different sets of fields, this will have interesting implications at the quantum level in future works. If for instance different gauge fixing procedures produce different results, this could perhaps be related to the trace anomaly in string theory. To iron out such details, the development of a Fadeev-Popov-type of gauge fixing procedure for TW gravity is under consideration. TW gravity itself may be an effective action of a deeper, underlying theory like string theory. Logarithmic running of the coupling constants is already present in the classical theory as demonstrated in this and previous works. Now having a formulation that is covariant and manifestly projective invariant will provide a platform for these further investigations.

% If in two-column mode, this environment will change to single-column
% format so that long equations can be displayed. Use
% sparingly.
%\begin{widetext}
% put long equation here
%\end{widetext}

%%%%%%%%%%%%%%%%%%%%%%%%
%%%%%%%%%%%%%%%%%%%%%%%%%%
%%%%%%%%%%%%%%%%%%%%%%%%%%%%%
% If you have acknowledgments, this puts in the proper section head.
\begin{acknowledgments}
The research of K. S. was supported in part by the
endowment of the Ford Foundation Professorship of
Physics at Brown University. The authors would like to thank the Diffeomorphisms and Geometry Research Group and the High Energy Physics Group at the University of Iowa for constructive conversations throughout this work. In particular, V.G.J. Rodgers was the source of much inspiration and elucidation. 
\end{acknowledgments}

%%%%%%%%%%%%%%%%%%%%%%%
%%%%%%%%%%%%%%%%%%%%%%%
%%%%%%%%%%%%%%%%%%%%%%%%%
%%%%%%%%%%%%%%%%%%%%%%%%%%%
\appendix

%%%%%%%%%%%%%%%%%%%%%%%%%%%%%%%%%%%
%%%%%%%%%%%%%%%%%%%%%%%%%%%%%%%%%%%%

\section{\label{s:Conventions}Conventions and Definitions}

Through most of the paper, we work in units where $c=1$. Specifically, we do this starting in Sec.~\ref{s:MPIEOM}. The action we use is that of TW gravity as presented in~\cite{Brensinger:2020gcv}, with addition of cosmological constant as in~\cite{InflationNewPaper:2022}. Our index convention is as in~\cite{Brensinger:2020gcv}: Latin indices $a,b,c,\dots$ are over the $\rd$-dimensional manifold $\mathcal{M}$ and Greek indices $\alpha, \beta, \gamma, \dots$ excluding $\lambda$ are over the $\rd+1$-dimensional manifold $\mathcal{N}$, also called the Thomas cone. The connection and curvature tensors on the Thomas cone are succinctly listed in Secs.~\ref{s:TWConnections} and~\ref{s:TWCurvatureTensors}. The various objects on $\mathcal{M}$ are located throughout the paper, so we succinctly list them all in this section. Parentheses and brackets denote symmetrization and antisymmetrization, respectively. For example, for some arbitrary rank 2 tensor $T_{ab}$ we define 

\begin{align}\label{e:SymmetrizationAndAntisymmetrization}
    T_{(ab)} = T_{ab} + T_{ba}~~~,~~~ T_{[ab]} = T_{ab} - T_{ba}
\end{align}
where we do \emph{not} include a factor of $\frac{1}{2}$. We use the west coast metric sign convention, which for diagonal metrics has a positive time component with all spatial components negative. Minkowski space in $\rd =4$ is for instance
\begin{align}
	\eta_{mn} = \text{diag}(1, -1, -1, -1)~.
\end{align}

The different connections used throughout are related as
\begin{align}\label{e:GammabarApp}
	\bar{\Gamma}^a{}_{bc} & = \Gamma^a{}_{bc} - \delta^a{}_{(b} \Vin_{c)} = \Pi^a_{\ bc} - \delta^a_{\ (b}g_{c)}~.  
\end{align}
The piecemeal parts used to construct fields are 
\begin{align}
    \alpha_m = & - \tfrac{1}{d+1} \Gamma^e_{\ em}\\
    \alpha_\mu = & (\alpha_m, \lambda^{-1}) \\
    g_m = & - \tfrac{1}{d+1} \partial_m \log \sqrt{|g|}=  - \tfrac{1}{d+1} \hat{\Gamma}^e{}_{me} \\ 
    g_\mu = & (g_m, \lambda^{-1}) \\
    \label{e:VincentApp}
    \Vin_m = & \ g_m - \alpha_m 
\end{align}
with $|g| = \det g_{ab}$ where $g_{ab}$ is the metric on the manifold $\mathcal{M}$.
Each connection can be expanded in terms of the Levi-Civita connection $\hat{\Gamma}^a{}_{bc}$ and Palatini tensor $C^{a}{}_{bc}$  as
\begin{align}
	\bar{\Gamma}^a{}_{bc} =& \hat{\Gamma}^a{}_{bc} + \tilde{C}^a{}_{bc} \\
	\Gamma^a{}_{bc} =& \hat{\Gamma}^a{}_{bc} + C^a{}_{bc} \\
	\Pi^a{}_{bc} =& \hat{\Gamma}^a{}_{bc} + \delta^a{}_{(b} g_{c)} + \tilde{C}^a{}_{bc}~. 
\end{align}
All connections as well as the Palatini tensor are symmetric in their lower two indices as for instance
\begin{align}
	C^a{}_{bc} = C^a{}_{cb}~.
\end{align}
The Levi-Civita connection's components are the canonical Christoffel symbols and $\tilde{C}^a{}_{bc}$ is the traceless part of the arbitrary Palatini tensor
\begin{align}
	\hat{\Gamma}^c{}_{ab} =& \frac{1}{2}g^{cm}\left(\partial_{(a}g_{b)m} - \partial_m g_{ab}\right) \\
	\tilde{C}^a_{\ bc} =&C^a_{\ bc}  - \tfrac{1}{\rd+1} \delta^a_{\ (b} C^{m}_{\ c)m} = C^a_{\ bc} - \delta^a_{\ (b} \Vin_{c)} ~. 
\end{align}

In matrix form, the metric and its inverse on the Thomas cone are 
\begin{align}
       G_{\alpha\beta} = & 
      \left[ \begin{array}{c|c}
            g_{ab}-\lambda_0^{\ 2} g_a g_b & -\frac{\lambda_0^{\ 2}}{\lambda} g_a \\
            \midrule
            -\frac{\lambda_0^{\ 2}}{\lambda} g_b & -\frac{\lambda_0^{\ 2}}{\lambda^2}
        \end{array} 
        \right] \\ 
%%%%%%%%%%%%%%%%%%%%%%%%%%%
    G^{\alpha\beta} = & 
       \left[ \begin{array}{c|c}
            g^{ab} & -\lambda g^{am} g_m \\
            \midrule
            -\lambda g^{bm}g_m & \frac{\lambda^2}{\lambda_0^{\ 2}}\left(-1 + g^{mn}\lambda_0^{\ 2}g_m g_n \right)
        \end{array} 
        \right] ~. 
\end{align}
More succinctly, we have
\begin{align}
        &G_{\alpha\beta} = \delta^a_{\,\,\alpha} \delta^b_{\,\,\beta} \,g_{ab} - \lambda_0^2 g_\alpha g_\beta \\
        &G^{\alpha\beta} = g^{ab} (\delta^\alpha_{\ a} - g_a \Upsilon^\alpha)(\delta^\beta_{\ b} - g_b \Upsilon^\alpha) - \lambda_0^{-2} \Upsilon^\alpha \Upsilon^\beta \\
        &\Upsilon^\alpha = (0,0,\dots,0,\lambda)~. 
\end{align} 

On $\mathcal{M}$, the curvature tensors are 
\begin{align}
        \hat{R}^a{}_{bcd} = & \hat{\Gamma}^a{}_{b[d,c]} + \hat{\Gamma}^m{}_{b[d} \hat{\Gamma}^a{}_{c]m} \\
        Q^a_{\ bcd} = & \hat{\nabla}_{[c} C^a_{\ d]b}  -  C^m_{\ \ b[c}C^a_{\ d]m} \\ 
        \tilde{Q}^a_{\ bcd} = & \hat{\nabla}_{[c} \tC^a_{\ d]b}  -  \tC^m_{\ \ b[c}\tC^a_{\ d]m}\cr
        =& \bar{\nabla}_{[c} \tC^a_{\ d]b}  +  \tC^m_{\ \ b[c}\tC^a_{\ d]m}\\ 
	   \mathcal{R}^a{}_{bcd} =& \Pi^a{}_{b[d,c]} + \Pi^m{}_{b[d} \Pi^a{}_{c]m}  \\
	   R^a{}_{bcd} = & \Gamma^a{}_{b[d,c]} + \Gamma^m{}_{b[d} \Gamma^a{}_{c]m}  = \hat{R}^a_{\ bcd} + Q^a_{\ bcd}\\ 
	   \bar{R}^a{}_{bcd} = & \bar{\Gamma}^a{}_{b[d,c]} + \bar{\Gamma}^m{}_{b[d} \bar{\Gamma}^a{}_{c]m} = \hat{R}^a{}_{bcd} + \tilde{Q}^a_{\ bcd}  \\
	   \label{e:KRiemannApp}
        \mathcal{K}^a{}_{bcd}  = & \mathcal{R}^{a}_{\ bcd} + \delta^a_{\ [c}\mathcal{D}_{d]b} \cr 
            = & R^a{}_{bcd} + \delta^{a}_{\ [ \, c}{} \mathcal{P}_{d]b}- \delta^a{}_b \mathcal{P}_{[cd]}\cr
            =& \bar{R}^a{}_{bcd} + \delta^a{}_{[c} \bar{\mathcal{P}}_{d]b} ~. 
\end{align}
The projective Cotton-York tensor is
\begin{align}
       K_{bcd} = & \breve{\nabla}_{[c} \mathcal{D}_{d]b} + g_a \mathcal{K}^a{}_{bcd} \nonumber \\ 
    = & \nabla_{[c} \mathcal{P}_{d]b} + \Vin_a \mathcal{K}^a{}_{bcd} \nonumber \\
    = & \bar{\nabla}_{[c} \bar{\mathcal{P}}_{d]b}~. 
\end{align}
Ricci tensors are always constructed by contracting over the first and third indices
\begin{align}%%%%%%%%%%%%%%%%%%%%%%%%%%%%%%%%%%%%%%%
    \mathcal{R}_{bd} \equiv \mathcal{R}^a_{\ bad} &   \\
    %%%%%%%%%%%%%%%%%%%%%%%%%%%%
        \label{e:RicciApp}
        R_{bd}  \equiv R^a_{\ bad}   &=  \hat{R}_{bd} + Q_{bd}  \equiv  \hat{R}^a_{\ bad} + Q^a_{\ bad}\\
    %%%%%%%%%%%%%%%%%%%%%%%%%%%%%%%%%
	\bar{R}_{bd}  \equiv \bar{R}^a{}_{bad}&=  \hat{R}_{bd} + \tilde{Q}_{bd}  \equiv \hat{R}^a_{\ bad} + \tilde{Q}^a_{\ bad}   \\ 
    %%%%%%%%%%%%%%%%%%%%%%%%%%%%%
        \label{e:KRicciApp}
    \mathcal{K}_{bd} \equiv \mathcal{K}^{a}_{\ bad} &= \mathcal{R}_{bd} + (\rd-1) \mathcal{D}_{bd} \nonumber\\
    &= R_{bd} + (\rd -1) \mathcal{P}_{bd} - \rd \mathcal{P}_{[bd]} \nonumber \\
    &= \bar{R}_{mn} + (\rd -1) \bar{\mathcal{P}}_{mn} ~. 
\end{align}

Ricci scalars are the trace of their associated Ricci tensors as seen below 
\begin{align}
    %%%%%%%%%%%%%%%%%%%%%%%%%%%%%%%%%%%%%
        \mathcal{R} \equiv g^{bd} \mathcal{R}_{bd}& \\
        %%%%%%%%%%%%%%%%%%%%%%%%%%%%%%%%%%%
        \label{e:RicciScalarApp}
    R \equiv  g^{bd} R_{bd}& = \hat{R} + Q \equiv g^{bd} \hat{R}_{bd} + g^{bd} Q_{bd} \\ 
        %%%%%%%%%%%%%%%%%%%%%%%%%%%%%%%%%%%%%
    \bar{R}  \equiv  g^{bd} \bar{R}_{bd} &= \hat{R} + \tilde{Q} \equiv  g^{bd} \hat{R}_{bd} + g^{bd} \tilde{Q}_{bd} \\
         %%%%%%%%%%%%%%%%%%%%%%%%%%%%%%%%%%%
        \label{e:KRicciScalarApp}
    \mathcal{K} \equiv  g^{ab} \mathcal{K}_{ab} &=\mathcal{R} + (\rd - 1) \mathcal{D} \cr
    &= R + (\rd - 1) \mathcal{P} \nonumber \\
    &=\bar{R} + (\rd -1) \bar{\mathcal{P}}              \\ %%%%%%%%%%%%%%%%%%%%%%%%%%%%%%%%%%%%%%%%%
    \label{e:scalarsApp}
        \mathcal{P} =  g^{ab} \mathcal{P}_{ab}& ~,~
        \mathcal{D} =  g^{ab} \mathcal{D}_{ab} ~,~Q = g^{ab}Q_{ab} ~. 
\end{align}

The diffeomorphism field and its related tensors are
\begin{align}
    \label{eq:PBarDefApp}
    \bar{\mathcal{P}}_{mn} =& \mathcal{D}_{mn} - \breve{\nabla}_m g_n - g_m g_n \cr
    =& \mathcal{P}_{mn} - \nabla_m \Vin_n - \Vin_m \Vin_n  \\
    \label{e:PbarScalarApp}
	\bar{\mathcal{P}} =  g^{mn} \bar{\mathcal{P}}_{mn} =& \mathcal{D} - g^{mn}\breve{\nabla}_m g_n - g^{mn}g_m g_n  \cr
	=& \mathcal{P} - \nabla^m \Vin_m - \Vin_m \Vin^m~.  
\end{align}
The following are used to write the  equations of motion of Thomas-Whitehead gravity in their original form as in~\cite{Brensinger:2020gcv}
\begin{align} 
    \label{e:khatApp}
    &\hmseven\hat{\mathcal{K}}_d^{\  a b c} =  \mathcal{K}^{ \bar d}_{\ \bar a \bar b \bar c} \mathcal{G}_{d \bar d}^{\ \ a \bar a [b |\bar b| c] \bar c} \\
    \label{e:cGApp}
    &\hmseven\mathcal{G}_{d \bar{d}}^{\ \ \ a \bar{a} b \bar{b} c \bar{c}} \nonumber\\
    &\hspace*{3 pt}= G_{d \bar{d}} g^{a \bar{a}} g^{b \bar{b}} g^{c \bar{c}} - 4 \delta^b_{\ d} \delta^{\bar{b}}_{\ \bar{d}} g^{a \bar{a}} g^{c \bar{c}} + \delta^b_{\ d} \delta^{\bar{b}}_{\ \bar{d}} g^{ac} g^{\bar{a} \bar{c}}\\
    \label{e:bcKApp}
    &\hmseven\breve{\mathcal{K}}^{abc} = g^{ap}g^{bm}g^{cn}\breve{\nabla}_{[m}\cD_{n]p} ~.
\end{align}
Note that the object $\hat{\cK}_d^{\ abc}$ simply has $G_{d\bar{d}}$ in its definition as on $\mathcal{M}$ indices are raised or lowered via $g^{ma}$ or $g_{ma}$ as
\begin{align}
\label{e:cKraiselower1App}
    \cK_{abcd} \equiv & g_{ma} \cK^m{}_{bcd}\\
\label{e:cKraiselower2App}
    \cK^{abcd} = & g^{bp} g^{cm} g^{dn} \cK^a{}_{pmn}\\
\label{e:cKraiselower3App}
    K^{bcd} \equiv & g^{bp}g^{cm}g^{dn}K_{pmn}~.
\end{align}

Important relationships between the projective curvatures and the Levi-Civita curvatures are
\begin{align}
%%%%%%%%%%%%%%%%%%%%%%%%%%%
\label{e:KRiemannExpQApp}
\mathcal{K}^a{}_{bcd} = & \hat{R}^a_{\ bcd} + \tilde{Q}^a_{\ bcd}  + \delta^a{}_{[c} \bar{\mathcal{P}}_{d]b} \cr
    = & \hat{R}^a_{\ bcd} + Q^a_{\ bcd} + \delta^a{}_{[c} \mathcal{P}_{d]b} - \delta^a{}_b \mathcal{P}_{[cd]} \\
%%%%%%%%%%%%%%%%%%%%%%%
\label{e:KRicciExpQApp}
\mathcal{K}_{mn} = &  \hat{R}_{mn} + \tilde{Q}_{mn} + (\rd -1) \bar{\mathcal{P}}_{mn} \cr
        = & \hat{R}_{mn} + Q_{mn} + (\rd-1) \mathcal{P}_{mn} - \rd \mathcal{P}_{[mn]} \\
%%%%%%%%%%%%%%%%%%%%%%%%%%
\label{e:KScalarExpQApp}
        \mathcal{K}  = & \hat{R} + \tilde{Q} + (\rd -1) \bar{\mathcal{P}} \cr
   = & \hat{R} + Q + (\rd-1) \mathcal{P} ~.
\end{align}
These allow for straightforward transitions to the form of Einstein equations as well as transitions between formulations.

%%%%%%%%%%%%%%%%%%%%%%%%%%%%%%%%%%%%
%%%%%%%%%%%%%%%%%%%%%%%%%%%%%%%%%%%%
%%%%%%%%%%%%%%%%%%%%%%%%%%%%%%%%%%%%
%%%%%%%%%%%%%%%%%%%%%%%%%%%%%%%%%%%%%
\section{Recasting the Original TW Field Equations in a Covariant Form}\label{a:EOMPItoTdetails}
To rewrite the Field equations in terms of $g_a$ instead of explicit factors of $\sqrt{|g|}$, the following property will be useful
\begin{align}\label{e:BreveExposeg}
    &\hmseven \tfrac{1}{\sqrt{|g|}}\breve{\nabla}_a (\sqrt{|g|}~ \text{anything}) \hmtwo = \hmtwo (\breve{\nabla}_a \hmtwo - \hmtwo (\rd\hmtwo + \hmtwo1) g_a) \text{anything.}
\end{align}

\subsection{Recasting \texorpdfstring{$\mathbb{D}^{ab}$}{Dab} in a Covariant Form}
Equation~\eqref{e:BreveExposeg} allows us to expand terms such as
\begin{align}\label{e:bnablaCYT}
    &\tfrac{1}{\sqrt{|g|}}\breve{\nabla}_c (\sqrt{|g|} K^{(ab)c}) = (\breve{\nabla}_c - (\rd+1) g_c) K^{(ab)c} \cr
    &\qquad\qquad\qquad = (\nabla_c - (\rd+2) \Vin_c) K^{(ab)c} + g_c K^{(ab)c}~~~~~~~
\end{align}
where we have used Eqs.~\eqref{e:GammabarApp} and~\eqref{e:VincentApp} to produce the last line.
It will also be of use to expand Eqs.~\eqref{e:khatApp} and~\eqref{e:bcKApp}
\begin{align}\label{e:hatKAppExpanded}
    &\hat{\cK}_c^{\ (ab)d} = 2 \cK_c^{\ (ab)d} - 2 \lambda_0^2 g_c g_m \cK^{m(ab)d} \nonumber\\
    &\qquad + \delta^d_{\ c}(8 \cK^{ab} - 2 g^{ab} \cK) + \delta^{(a}_{\ \ c} (g^{b)d}\cK - 4 \cK^{b)d}) \\
    \label{e:bcKAppExpanded}
    &\breve{\cK}^{(ab)c} = K^{(ab)c} - g_m \cK^{m(ab)c}~. 
\end{align}
Note that $\hat{\cK}_d^{\ abc}$, that includes $G_{d\bar{d}}$ in its definition, should \emph{not} be confused with $\cK_d^{\ abc} = g_{dm} \cK^{mabc}$ that,  here on $\mathcal{M}$, has its indices raised or lowered via $g^{dm}$ or $g_{dm}$  as in Eq.~\eqref{e:cKraiselower2App}. In expanding $\breve{\cK}^{(ab)c}$ above, we have made use of the definition of the projective Cotton-York tensor from Eq.~\eqref{e:CottonYorkDef}. 

Contracting indices in Eq.~\ref{e:hatKAppExpanded} and simplifying yields the following useful result
\begin{align}\label{e:hatKAppExpandedContracted}
    \tfrac{1}{2} \hat{\cK}_c{}^{(ab)c} =& - \lambda_0^2 g_c g_m \cK^{m(ab)c} + 2 (2 \rd - 3) \cK^{ab} \cr
    & \qquad\qquad\qquad\qquad - (\rd - 1) g^{ab} \cK~.
\end{align}
Recall the field equation $\mathbb{D}^{ab}$ in its manifestly projective invariant form
\begin{align}
     0= \mathbb{D}^{ab} \equiv & - \tfrac{(\rd - 1)}{4 J_0\kappa_0}  g^{ab} + \tfrac{1}{2}\hat{\mathcal{K}}_{c}^{\ (ab)c} \cr
    & +  \tfrac{\lambda_0^2}{\sqrt{|g|}} \breve{\nabla}_c ( \sqrt{|g|} K^{(ab)c}) -  \lambda_0^2 g_c \breve{\mathcal{K}}^{(ab)c} .\tag{\ref{e:DEOM}}
\end{align}
Utilizing the expansions in Eqs.~\eqref{e:bnablaCYT}, \eqref{e:hatKAppExpandedContracted}, and~\eqref{e:bcKAppExpanded} and simplifying casts Eq.~$\mathbb{D}^{ab}$ into its covariant form
\begin{align}
    & \hmsix 0 \hmtwo =\hmtwo\mathbb{D}^{ab}\hmtwo = 
    \hmtwo \lambda_0^2 (\nabla_c \hmtwo- \hmtwo(\rd\hmtwo+\hmtwo 2)\Vin_c) K^{(ab)c} \hmtwo -\hmtwo(\rd \hmtwo -\hmtwo 1) \mathbb{K}^{ab}\hmtwo -\hmtwo 2 \mathcal{K}^{ab} \tag{\ref{e:DEOMtensorial}}\\
    \label{e:bbKApp}
    &\hspace*{6 pt} \mathbb{K}^{ab} \equiv g^{ab}   \left(\cK + \tfrac{1}{4 J_0 \kappa_0}\right) - 4 \mathcal{K}^{ab}~. 
\end{align}
Via use of Eq.~\eqref{e:bnablatoOthernablas}, it is straightforward to show the first term collapses to an overall $\bar{\nabla}_c$ derivative. This reduces the equation to the covariant and manifestly projective invariant form
\begin{align}
0=\mathbb{D}^{ab} =& \lambda_0^2 \bar{\nabla}_c K^{(ab)c} -(\rd - 1) \mathbb{K}^{ab} -2\mathcal{K}^{ab} ~.\tag{\ref{e:PbarEOM}}
\end{align}
\subsection{Recasting \texorpdfstring{$\mathcal{H}_{mn}$}{Hab} in a Covariant Form}
Using Eqs.~\eqref{e:BreveExposeg}, \eqref{e:GammabarApp}, and~\eqref{e:VincentApp} we find the following identity that we will find useful in this section
\begin{align}\label{e:bnablavector}
    &\hmten \tfrac{1}{\sqrt{|g|}} \breve{\nabla}_a (\sqrt{|g|} \cK^a_{\ bcd}K^{bcd}) \hmtwo = \hmtwo (\nabla_a  \hmtwo  - \hmtwo (\rd \hmtwo + \hmtwo 1) \Vin_a)(\cK^a_{\ bcd}K^{bcd}).
\end{align}
Beginning with the equation of motion $\mathcal{H}_{mn}$
\begin{align}
0 = \mathcal{H}_{mn} \equiv &  -\mathcal{K}_{mn} +  \kappa_0 \left( 2 \mathbb{L}_{mn} -  g_{mn} \mathcal{L} \right)   \cr 
	&  -  g_{mn} \tfrac{2 J_0 \kappa_0 \l_0^2}{(\rd+1)\sqrt{|g|}} \breve{\nabla}_a (\sqrt{|g|}\cK^{a}{}_{bcd}K^{bcd}) \tag{\ref{e:cHmn}} \\
%%%%%%%%%%%%%%%%%%%%
\mathcal{L} = & -\frac{1}{2 \kappa_0}(\mathcal{K}+ 2 \Lambda_0)   + \mathbb{L} \tag{\ref{e:LTW}}
\end{align}
we use Eq.~\eqref{e:bnablavector} along with Eqs.~\eqref{e:KRicciExpQApp} and \eqref{e:KScalarExpQApp}. Simplifying yields the covariant form of Einstein's equations
\begin{align}
	\hat{R}_{mn} - & \tfrac{1}{2} g_{mn} \hat{R} = \kappa_0 \hat{T}_{mn}  + g_{mn} \Lambda_0 \tag{\ref{e:EE}}\\
 \hat{T}_{mn} =&
   - \kappa_0^{-1}\left( Q_{mn} - \tfrac{1}{2}  g_{mn} Q \right) + 2 \mathbb{L}_{mn} - g_{mn} \mathbb{L} \cr 
   & - \kappa_0^{-1} \left((\rd -1) \mathcal{P}_{mn} - \rd \cP_{[mn]} - \tfrac{\rd - 1}{2} g_{mn} \mathcal{P} \right)\cr 
  & - g_{mn} \tfrac{2 J_0 \lambda_0^2}{\rd+1} (\nabla_a -(\rd+1) \Vin_a)(K_{bcd} \mathcal{K}^{abcd} )~. \tag{\ref{e:Thatmn}}
\end{align}
	
\subsection{Recasting \texorpdfstring{$\tilde{\mathcal{E}}_c{}^{ab}$}{Ecab} in terms of Tensorial Objects to Demonstrate it is not a Tensor}
Using Eqs.~\eqref{e:BreveExposeg} and~\eqref{e:bnablaCYT} it is straightforward to prove the following identities that we will find useful in this section
\begin{align}
\label{e:bnablaT}
\tfrac{1}{\sqrt{|g|}} \breve{\nabla}_c (\sqrt{|g|} T^{ab}) =& (\nabla_c - 2 \Vin_c) T^{ab} - \delta^{(a}_{\ \ c}\Vin_m T^{b)m} \cr
&- (\rd-1) g_c T^{ab} +\delta^{(a}_{\ \ c} g_m T^{b)m}
\end{align}
\begin{align}
    \label{e:bnablacK4}
    \tfrac{1}{\sqrt{|g|}} \breve{\nabla}_d (\sqrt{|g|}K_c{}^{(ab)d}) = & (\nabla_d - (\rd+1) \Vin_d) \cK_c{}^{(ab)d} \cr
    & -2 \Vin_c \cK^{ab} + 2 g_c \cK^{ab}\\
    \label{e:bnablagCT}
    \tfrac{1}{\sqrt{|g|}} \breve{\nabla}_d (\sqrt{|g|} g_c K^{(ab)d}) =& K^{(ab)d}(\breve{\nabla}_c g_d  + g_c g_d) \cr
    & \hmten + \hmtwo g_c (\nabla_d \hmtwo - \hmtwo (\rd \hmtwo + \hmtwo 2) \Vin_d)K^{(ab)d}.
\end{align} 
In the above, $T^{ab} = T^{ba}$ represents an arbitrary symmetric tensor. 

Starting with the original form of the equations of motion
\begin{align}
0=\tilde{\mathcal{E}}_c{}^{ab} \equiv & \mathcal{E}_c^{\ ab} - \tfrac{1}{d+1} \delta_c^{\ ( a}\mathcal{E}_{d}^{\ b)d} \tag{\ref{e:EOMGamma}} \\
%%%%%%%%%%%%%%%%%%%%%%%%%%%%%%%%%%
\mathcal{E}_c^{\ ab}  \equiv & \tfrac{1}{4 J_0\kappa_0 \sqrt{|g|}} \breve{\nabla}_c( \sqrt{|g|}g^{ab})\hspace*{-2 pt} - \hspace*{-2 pt} \tfrac{1}{2\sqrt{|g|}}\breve{\nabla}_d (\sqrt{|g|}\hat{\mathcal{K}}_{c}^{\ (ab) d}) \nonumber\\
    & + \tfrac{\lambda_0^2}{\sqrt{|g|}} \breve{\nabla}_d (\sqrt{|g|}g_c \breve{\mathcal{K}}^{(ab)d})  \hmtwo - \hmtwo \lambda_0^2  K^{(ab)d} \mathcal{D}_{dc} \tag{\ref{e:CurlyE}}
\end{align}
we use Eqs.~\eqref{e:hatKAppExpanded} and~\eqref{e:bcKAppExpanded} to substitute for $\hat{\cK}_c{}^{(ab)d}$ and $\breve{\cK}^{(ab)d}$. After some simplification this results in
\begin{align}
\mathcal{E}_c^{\ ab} =&  \tfrac{1}{\sqrt{|g|}} \breve{\nabla}_c( \sqrt{|g|}\mathbb{K}^{ab}) - \tfrac{1}{\sqrt{|g|}}\breve{\nabla}_d (\sqrt{|g|}\mathcal{K}_{c}^{\ (ab) d})  \quad \nonumber\\
    &+ \lambda_0^2 \left[ \tfrac{1}{\sqrt{|g|}} \breve{\nabla}_d (\sqrt{|g|} g_c K^{(ab)d}) - \cD_{cd} K^{(ab)d}\right] \nonumber\\
    &- \tfrac{1}{2} \delta^{(a}{}_{c} \tfrac{1}{\sqrt{|g|}} \breve{\nabla}_d \left[ \sqrt{|g|} \left(g^{b)d} \cK - 4 \cK^{b)d}\right)\right]
\end{align}
where $\mathbb{K}^{ab}$ is as in Eq.~\eqref{e:bbKApp}.
Next, we substitute using Eqs.~\eqref{e:bnablaT}, \eqref{e:bnablacK4}, and~\eqref{e:bnablagCT} and simplify to the final form
\begin{align}
    \mathcal{E}_c{}^{ab} =&\mathbb{E}_c{}^{ab} + g_c \mathbb{D}^{ab} + \delta^{(a}{}_{c}\mathfrak{g}^{b)}\tag{\ref{e:GammaEOMNablaP}}\\
    %%%%%%%%%%%%%%%%%%%%%%%%%%%%%%%%%%%
    \mathbb{E}_c{}^{ab} \equiv& (\nabla_c - 2 \Vin_c) \mathbb{K}^{ab} - \delta^{(a}{}_{c} \Vin_d \mathbb{K}^{b)d} \cr
    &- [(\nabla_d - (\rd+1)\Vin_d) \mathcal{K}_c{}^{(ab)d} - 2 \Vin_c \mathcal{K}^{ab}] \cr
    &- \lambda_0^2 (\mathcal{P}_{cd} - \nabla_c \Vin_d - \Vin_c \Vin_d) K^{(ab)d} \tag{\ref{e:GammaEOMEscrK}}\\
    %%%%%%%%%%%%%%%%%%%%%%%%%%%%%%%%%%%
    \mathfrak{g}^b \equiv & \tfrac{1}{4 J_0 \kappa_0}g_d g^{bd} \hspace*{-2 pt} - \hspace*{-2 pt}  \tfrac{1}{2}(\nabla_d \hspace*{-2 pt} -\hspace*{-2 pt} (\rd+3) \Vin_d)(g^{bd} \mathcal{K}\hspace*{-2 pt} -\hspace*{-2 pt} 4 \mathcal{K}^{bd}).~~~~~~ \tag{\ref{e:mathfrakg}}
\end{align}
where in the last line of the definition of $\mathbb{E}_c{}^{ab}$, we have made use of Eq.~\eqref{eq:PBarDefApp} to switch between $\cD_{cd}$ and $\cP_{cd}$. Subtracting the trace results in a form of the equation that is clearly non-tensorial
\begin{align}
     0=\tilde{\mathcal{E}}_c^{\ ab} =& \tbbE_c{}^{ab}+\hspace*{-2 pt} g_c \mathbb{D}^{ab} \hspace*{-2 pt} - \hspace*{-2 pt} \tfrac{1}{\rd+1} \delta^{(a}_{\ \  c}g_m \mathbb{D}^{b)m}    \tag{\ref{e:cEtilde}} \\
       \tbbE_c{}^{ab} \equiv & \bbE_c{}^{ab} - \tfrac{1}{\rd+1} \delta_c{}^{(a} \bbE_m{}^{b)m}~. \tag{\ref{e:tbbE}}
\end{align}

%%%%%%%%%%%%%%%%%%%%%%%%%%%%%%%%%%%%
%%%%%%%%%%%%%%%%%%%%%%%%%%%%%%%%%%%%
%%%%%%%%%%%%%%%%%%%%%%%%%%%%%%%%%%%%
%%%%%%%%%%%%%%%%%%%%%%%%%%%%%%%%%%%%%
\section{Proofs}\label{a:proofs}
\subsection{Proof of Eq.~\texorpdfstring{\eqref{e:Va}}{1111}}\label{a:Va}
We start by calculating the trace $\bbE_b{}^{ab}$ via Eq.~\eqref{e:bbE}. Inserting this into the definition of $V^a$ in Eq.~\eqref{e:Va} and simplifying results in
\begin{align}
   &\hmseven V^a \hmtwo = \hmthree\tfrac{-1}{\rd+1}\hmthree\left\{\hmtwo\bnabla_d\hmtwo \left[(\rd\hmtwo-\hmtwo 1)\mathbb{K}^{ad} \hmtwo + \hmtwo2 \cK_b^{\ (ab)d}\right]
\hmthree     +\hmtwo 2 \lambda_0^2 \bcP_{db}K^{(ab)d}   \right\}
\end{align}
with $\mathbb{K}^{ab}$ defined in Eq.~\eqref{e:bbKApp}.
Using the fact that $\cK_b^{\ bad}$ vanishes trivially from its definition, the second term results in an additional projective Ricci term. Owing to the symmetry of $\bcP_{db}$ and antisymmetry of the last two indices of $K^{abd}$, the only non-vanishing contribution from the third term is $\bcP_{db}K^{bad}$. Making these simplifications, relabeling indices, and combining like terms we have then
\begin{align}
       \hmsix V^a \hmtwo =& -\tfrac{1}{\rd+1}\left\{\bnabla_b[\hmone (\rd \hmtwo -\hmtwo 1)\mathbb{K}^{ab} \hmtwo  + \hmtwo 2 \cK^{ab}]\hmtwo+\hmtwo 2 \lambda_0^2 \bcP_{db}K^{bad}   \right\}.
\end{align}
By inspection of Eq.~\eqref{e:PbarEOM}, we see that the term in brackets is equivalent to $\lambda_0^2 \bnabla_c K^{(ab)c}-\mathbb{D}^{ab} $. Substituting this and simplifying results in the second line of Eq.~\eqref{e:Va}
\begin{align}\label{e:VaSimple}
    V^a = & \tfrac{1}{\rd+1}\bnabla_b \bbD^{ab} \hmtwo - \hmtwo \tfrac{\lambda_0^2}{\rd + 1}\left(\bnabla_b \bnabla_c K^{(ab)c} \hmtwo + \hmtwo 2 \bcP_{db}K^{bad}\right).
\end{align}

%%%%%%%%%%%%%%%%%%%%%%%%%%%%%%%%%%%%%
%%%%%%%%%%%%%%%%%%%%%%%%%%%%%%%%%%%%%
\subsection{Proof of Eqs.~\texorpdfstring{\eqref{e:delGammatodelg}}{1111} and~\texorpdfstring{\eqref{e:delGammaTracetodelg}}{1111}}\label{a:delGammatodelg}
We expand the variation of the Levi-Civita connection in terms of the metric
\begin{align}
    \delta\hat{\Gamma}^c{}_{ab} =& \tfrac{1}{2}\delta \left[ g^{cm}(g_{m(a,b)} - g_{ab,m}) \right]
\end{align}
where a comma denotes a partial derivative. Using Leibniz's rule for the variation leads to 
\begin{align}
    &\delta\hat{\Gamma}^c{}_{ab} \hmtwo = \hmtwo \tfrac{1}{2}\left[\delta g^{cm}(g_{m(a,b)}\hmtwo -\hmtwo g_{ab,m})\hmtwo + \hmtwo g^{cm}(\delta g_{m(a,b)}\hmtwo - \hmtwo \delta g_{ab,m}) \right].
\end{align}
Switching to Levi-Civita covariant derivatives in the second set of parenthesis yields
\begin{align}
    \delta\hat{\Gamma}^c{}_{ab} \hmtwo &= \hmtwo \tfrac{1}{2}\big[ \delta g^{cm}(g_{m(a,b)}\hmtwo - \hmtwo g_{ab,m})\hmtwo + \hmtwo g^{cm}(\hat{\nabla}_{(a}\delta g_{b)m} \hmtwo - \hmtwo \hat{\nabla}_m\delta g_{ab}) \cr
    & + g^{cm} \hat{\Gamma}^p_{\ (ab)} \delta g_{pm} \hmtwo + \hmtwo g^{cm} \hat{\Gamma}^p_{\ m(a} \delta g_{b)p} \hmtwo - \hmtwo g^{cm} \hat{\Gamma}^p_{\ m(a} \delta g_{b)p}   \big].
\end{align}
Canceling the last two terms, expanding the Levi-Civita connection in terms of the metric in the first term on the second line, and rearranging we have
\begin{align}
     \delta\hat{\Gamma}^c{}_{ab} =& \tfrac{1}{2}  g^{cm}(\hat{\nabla}_{(a}\delta g_{b)m} - \hat{\nabla}_m\delta g_{ab}) \cr
     &  + \tfrac{1}{2} \delta g^{cm}(g_{m(a,b)}  -  g_{ab,m})   \cr
     & +  \tfrac{1}{2} \delta g_{pm}g^{cm} g^{pn}(g_{n(a,b)}  -  g_{ab,n}) 
\end{align}
raising the indices on the variation in the last term, at the cost of a minus sign, yields
\begin{align}
    \delta\hat{\Gamma}^c{}_{ab} =& \tfrac{1}{2} g^{cm}(\hat{\nabla}_{(a}\delta g_{b)m} - \hat{\nabla}_m\delta g_{ab}) \cr
     &+ \tfrac{1}{2} \delta g^{cm}(g_{m(a,b)} - g_{ab,m})  \cr
     &- \tfrac{1}{2} \delta g^{cn}(g_{n(a,b)} - g_{ab,n}) ~.
\end{align}
Relabeling indices, the second line vanishes, resulting in
\begin{align}\label{e:varhG}
    \delta\hat{\Gamma}^c{}_{ab} =& \tfrac{1}{2} g^{cm}(\hat{\nabla}_{(a}\delta g_{b)m} - \hat{\nabla}_m\delta g_{ab})~.
\end{align}
Contracting indices and simplifying, we find for the variation of the trace
\begin{align}\label{e:varhGTrace}
     \delta\hat{\Gamma}^m{}_{am} =& -\tfrac{1}{2} g_{mn} \hat{\nabla}_a \delta g^{mn}~.
\end{align}
Next, we take a Lagrangian variation of the form
\begin{align}
    \tilde{\mathbb{E}}_c{}^{ab} \delta \hat{\Gamma}^c{}_{ab} ~.
\end{align}
Substituting Eq.~\eqref{e:varhG} we have
\begin{align}
    \tilde{\mathbb{E}}_c{}^{ab} \delta \hat{\Gamma}^c{}_{ab} =& \tilde{\mathbb{E}}_c{}^{ab} \tfrac{1}{2} g^{cm}(\hat{\nabla}_{(a}\delta g_{b)m} - \hat{\nabla}_m\delta g_{ab}).
\end{align}
Up to a total derivative, this is equivalent to the following
\begin{align}
    \hmfive \tilde{\mathbb{E}}_c{}^{ab} \delta \hat{\Gamma}^c{}_{ab} =& \tfrac{1}{2} g^{cm}( \delta g_{ab}  \hat{\nabla}_m \tilde{\mathbb{E}}_c{}^{ab} - 2\delta g_{bm}\hat{\nabla}_{a}\tilde{\mathbb{E}}_c{}^{ab} ).
\end{align}
Contracting with the metric, relabeling indices, and simplifying, we have
\begin{align}
    \tilde{\mathbb{E}}_c{}^{ab} \delta \hat{\Gamma}^c{}_{ab} =& \tfrac{1}{2} \left[\hat{\nabla}_a(\tilde{\mathbb{E}}^{amn} - 2 \tilde{\mathbb{E}}^{mna}) \right]\delta g_{mn} ~.
\end{align}
Raising indices on the variation of the metric at the cost of a minus sign and using symmetry of the metric indices to symmetrize indices on $\tilde{\mathbb{E}}_{mn}{}^a$ yields the final result
\begin{align}
          \tilde{\mathbb{E}}_c{}^{ab} \delta\hat{\Gamma}^c{}_{ab} =&  \tfrac{1}{2} \delta g^{mn} \hat{\nabla}_a \tilde{\mathbb{E}}^{*\ \ a}_{mn} \tag{\ref{e:delGammatodelg}}\\
        \tilde{\mathbb{E}}^{*\ \ a}_{mn} \equiv & \tilde{\mathbb{E}}_{(mn)}^{\ \ \ \ \ a} - \tilde{\mathbb{E}}^a_{\ mn} \tag{\ref{e:Stardef}}
\end{align}
Doing the corresponding calculation for the variation of the trace $\delta\hat{\Gamma}^m{}_{am}$ contracted with  $V^a$ quickly yields the final result
\begin{align}
            V^a \delta\hat{\Gamma}^m{}_{am} =& V^a(-\tfrac{1}{2} g_{mn} \hat{\nabla}_a \delta g^{mn}) \cr
            =&\tfrac{1}{2}\delta g^{mn} g_{mn} \hat{\nabla}_a V^a  \cr
            =&  \tfrac{1}{2}\delta g^{mn} g_{mn} \bar{\nabla}_a V^a  ~. \tag{\ref{e:delGammaTracetodelg}}
\end{align}
The second equality holds, again, up to a total derivative. In going from the second to third lines, we have used the interchangeability of the divergences of $\hat{\nabla}_a$ and $\bnabla_a$, as proved in Eq.~\eqref{e:divbnablaProof}.
%%%%%%%%%%%%%%%%%%%%%%%%%%%%%%%%%%%%
%%%%%%%%%%%%%%%%%%%%%%%%%%%%%%%%%%%%
\subsection{Proof of Eq.~\texorpdfstring{\eqref{e:DivV}}{1111}}\label{a:DivV}
First, we note some useful identities. Curvature as defined in Eqs.~\eqref{e:CurvatureContravariant} and~\eqref{e:CurvatureCovariant} can be extended to an arbitrary number of indices, and to any curvature tensor associated with its covariant derivative. Of particular use here will be the following such curvature identities
\begin{align}
	\label{e:CurvatureIdentity1}
	&\hmsix[\bnabla_b , \bnabla_a] (\bnabla_c K^{abc}) \hmtwo = \hmtwo \bar{R}^a_{\ bda} \bnabla_c K^{dbc} \hmtwo+\hmtwo \bar{R}^b_{\ dba} \bnabla_c K^{adc} \\
	\label{e:CurvatureIdentity2}
	&\hmsix[\bnabla_b , \bnabla_c] K^{abc} \hmtwo=\hmtwo \bar{R}^a_{\ dbc} K^{dbc} \hmtwo+\hmtwo \bar{R}^b_{\ dbc} K^{adc} \hmtwo+\hmtwo \bar{R}^c_{\ dbc}K^{abd}.
\end{align}

We start by taking the divergence of Eq.~\eqref{e:VaSimple}, yielding
\begin{align}\label{e:DivVStart}
    &\bnabla_a V^a = \tfrac{1}{\rd+1} \bnabla_a \bnabla_b \mathbb{D}^{ab} \cr
    &\qquad - \tfrac{\lambda_0^2}{\rd+1} \left[\bnabla_a\bnabla_b \bnabla_c K^{(ab)c} + 2\bnabla_a( \bcP_{db}K^{bad}) \right].~~~~~~
\end{align}
We focus on the triple derivative, simplifying as follows
\begin{align}
    \bnabla_a\bnabla_b \bnabla_c K^{(ab)c} =& \bnabla_a\bnabla_b \bnabla_c K^{abc} + \bnabla_b \bnabla_a \bnabla_c K^{abc} \cr
    =& 2\bnabla_a\bnabla_b \bnabla_c K^{abc} + \bR^a{}_{dba}\bnabla_cK^{dbc} \cr
    & + \bR^b{}_{dba}\bnabla_cK^{adc} 
\end{align}
where in going from the first to the second equality we have the curvature identity in Eq.~\eqref{e:CurvatureIdentity1}. The second and third terms cancel owing to the antisymmetry and symmetry properties $\bR^a{}_{dba} = - \bR^a{}_{dab} = -\bR_{db} = - \bR_{bd}$ and a relabeling of indices. Using the antisymmetry of $K^{abc}$ on the remaining first term, we apply the curvature identity in Eq.~\eqref{e:CurvatureIdentity2} to reduce this to 
\begin{align}
    \bnabla_a\bnabla_b \bnabla_c K^{(ab)c} 
    &\hmtwo =\hmtwo  \bnabla_a(\bR^a_{\ dbc}K^{dbc}\hmtwo  + \hmtwo \bR^b_{\ dbc}K^{adc} \hmtwo + \hmtwo \bR^c_{\ dbc}K^{abd}) \nonumber\\
    &\hmtwo= \bnabla_a(\bR^a_{\ dbc}K^{dbc} + \bR_{dc}K^{adc} - \bR_{db}K^{abd}) \nonumber\\
    &\hmtwo= \bnabla_a (\bR^a_{\ dbc}K^{dbc})
\end{align}
where the second and third terms have vanished owing to the symmetric $\bar{R}_{dc}$ contracted over the antisymmetric indices of $K^{adc}$. Relabeling indices and inserting into Eq.~\eqref{e:DivVStart} we have
\begin{align}
   &\bnabla_a V^a \hmtwo =  \tfrac{1}{\rd+1} \hmtwo \bnabla_a \hmtwo  \bnabla_b \mathbb{D}^{ab} \hmtwo  - \hmtwo \tfrac{\lambda_0^2}{\rd+1} \bnabla_a \hmtwo \left(\bR^a_{\ bcd}K^{bcd}\hmtwo  + \hmtwo 2\bcP_{db}K^{bad} \right) \nonumber\\
   &\hspace*{4 pt} = \tfrac{1}{\rd+1} \hmtwo \bnabla_a \hmtwo  \bnabla_b \mathbb{D}^{ab} \hmtwo  - \hmtwo \tfrac{\lambda_0^2}{\rd+1} \bnabla_a \hmtwo \left([\bR^a_{\ bcd}\hmtwo  + \hmtwo \delta^a_{\ [c}\bcP_{d]b}]K^{bcd} \right)
\end{align}
where in going from the first to last equality we have relabeled some indices, introduced a Kronecker delta symbol, and utilized the antisymmetry of the last two indices of $K^{bcd}$.
Comparing with Eq.~\eqref{e:KRiemannApp}, we identity the terms in brackets as $\cK^a{}_{bcd}$, thus the expression simplifies to
\begin{align}
   \bnabla_a V^a = \tfrac{1}{\rd+1} \bnabla_a\bnabla_b \bbD^{ab} - \tfrac{\lambda_0^2}{\rd+1} \bnabla_a\left(\cK^a{}_{bcd} K^{bcd}\right)~. \tag{\ref{e:DivV}}
\end{align}

% Create the reference section using BibTeX:
\bibliography{RosettaBibliography}

%apsrev4-2.bst 2019-01-14 (MD) hand-edited version of apsrev4-1.bst
%Control: key (0)
%Control: author (8) initials jnrlst
%Control: editor formatted (1) identically to author
%Control: production of article title (0) allowed
%Control: page (0) single
%Control: year (1) truncated
%Control: production of eprint (0) enabled
\begin{thebibliography}{52}%
\makeatletter
\providecommand \@ifxundefined [1]{%
 \@ifx{#1\undefined}
}%
\providecommand \@ifnum [1]{%
 \ifnum #1\expandafter \@firstoftwo
 \else \expandafter \@secondoftwo
 \fi
}%
\providecommand \@ifx [1]{%
 \ifx #1\expandafter \@firstoftwo
 \else \expandafter \@secondoftwo
 \fi
}%
\providecommand \natexlab [1]{#1}%
\providecommand \enquote  [1]{``#1''}%
\providecommand \bibnamefont  [1]{#1}%
\providecommand \bibfnamefont [1]{#1}%
\providecommand \citenamefont [1]{#1}%
\providecommand \href@noop [0]{\@secondoftwo}%
\providecommand \href [0]{\begingroup \@sanitize@url \@href}%
\providecommand \@href[1]{\@@startlink{#1}\@@href}%
\providecommand \@@href[1]{\endgroup#1\@@endlink}%
\providecommand \@sanitize@url [0]{\catcode `\\12\catcode `\$12\catcode
  `\&12\catcode `\#12\catcode `\^12\catcode `\_12\catcode `\%12\relax}%
\providecommand \@@startlink[1]{}%
\providecommand \@@endlink[0]{}%
\providecommand \url  [0]{\begingroup\@sanitize@url \@url }%
\providecommand \@url [1]{\endgroup\@href {#1}{\urlprefix }}%
\providecommand \urlprefix  [0]{URL }%
\providecommand \Eprint [0]{\href }%
\providecommand \doibase [0]{https://doi.org/}%
\providecommand \selectlanguage [0]{\@gobble}%
\providecommand \bibinfo  [0]{\@secondoftwo}%
\providecommand \bibfield  [0]{\@secondoftwo}%
\providecommand \translation [1]{[#1]}%
\providecommand \BibitemOpen [0]{}%
\providecommand \bibitemStop [0]{}%
\providecommand \bibitemNoStop [0]{.\EOS\space}%
\providecommand \EOS [0]{\spacefactor3000\relax}%
\providecommand \BibitemShut  [1]{\csname bibitem#1\endcsname}%
\let\auto@bib@innerbib\@empty
%</preamble>
\bibitem [{\citenamefont {Thomas}(1925{\natexlab{a}})}]{Thomas:1925a}%
  \BibitemOpen
  \bibfield  {author} {\bibinfo {author} {\bibfnamefont {T.~Y.}\ \bibnamefont
  {Thomas}},\ }\bibfield  {title} {\bibinfo {title} {{Announcement of a
  Projective Theory of Affinely Connected Manifolds}},\ }\href@noop {}
  {\bibfield  {journal} {\bibinfo  {journal} {Proc. of the Nat. Acad. of
  Sciences of the USA}\ }\textbf {\bibinfo {volume} {11}},\ \bibinfo {pages}
  {588 } (\bibinfo {year} {1925}{\natexlab{a}})}\BibitemShut {NoStop}%
\bibitem [{\citenamefont {Thomas}(1925{\natexlab{b}})}]{Thomas:1925b}%
  \BibitemOpen
  \bibfield  {author} {\bibinfo {author} {\bibfnamefont {T.~Y.}\ \bibnamefont
  {Thomas}},\ }\bibfield  {title} {\bibinfo {title} {{On the Projective and
  Equi-Projective Geometries of Paths}},\ }\href@noop {} {\bibfield  {journal}
  {\bibinfo  {journal} {Proc. of the Nat. Acad. of Sciences of the USA}\
  }\textbf {\bibinfo {volume} {11}},\ \bibinfo {pages} {199} (\bibinfo {year}
  {1925}{\natexlab{b}})}\BibitemShut {NoStop}%
\bibitem [{\citenamefont {Veblen}\ and\ \citenamefont
  {Hoffmann}(1930)}]{Veblen1930}%
  \BibitemOpen
  \bibfield  {author} {\bibinfo {author} {\bibfnamefont {O.}~\bibnamefont
  {Veblen}}\ and\ \bibinfo {author} {\bibfnamefont {B.}~\bibnamefont
  {Hoffmann}},\ }\bibfield  {title} {\bibinfo {title} {Projective relativity},\
  }\href {https://doi.org/10.1103/PhysRev.36.810} {\bibfield  {journal}
  {\bibinfo  {journal} {Phys. Rev.}\ }\textbf {\bibinfo {volume} {36}},\
  \bibinfo {pages} {810} (\bibinfo {year} {1930})}\BibitemShut {NoStop}%
\bibitem [{\citenamefont {Kaluza}(1921)}]{Kaluza:1921tu}%
  \BibitemOpen
  \bibfield  {author} {\bibinfo {author} {\bibfnamefont {T.}~\bibnamefont
  {Kaluza}},\ }\bibfield  {title} {\bibinfo {title} {{Zum Unit\"atsproblem der
  Physik}},\ }\href {https://doi.org/10.1142/S0218271818700017} {\bibfield
  {journal} {\bibinfo  {journal} {Sitzungsber. Preuss. Akad. Wiss. Berlin
  (Math. Phys. )}\ }\textbf {\bibinfo {volume} {1921}},\ \bibinfo {pages} {966}
  (\bibinfo {year} {1921})},\ \Eprint {https://arxiv.org/abs/1803.08616}
  {arXiv:1803.08616 [physics.hist-ph]} \BibitemShut {NoStop}%
\bibitem [{\citenamefont {Klein}(1926)}]{Klein:1926tv}%
  \BibitemOpen
  \bibfield  {author} {\bibinfo {author} {\bibfnamefont {O.}~\bibnamefont
  {Klein}},\ }\bibfield  {title} {\bibinfo {title} {{Quantum Theory and
  Five-Dimensional Theory of Relativity. (In German and English)}},\ }\href
  {https://doi.org/10.1007/BF01397481} {\bibfield  {journal} {\bibinfo
  {journal} {Z. Phys.}\ }\textbf {\bibinfo {volume} {37}},\ \bibinfo {pages}
  {895} (\bibinfo {year} {1926})}\BibitemShut {NoStop}%
\bibitem [{\citenamefont {Whitehead}(1931)}]{Whitehead}%
  \BibitemOpen
  \bibfield  {author} {\bibinfo {author} {\bibfnamefont {J.}~\bibnamefont
  {Whitehead}},\ }\bibfield  {title} {\bibinfo {title} {{The representation of
  projective spaces}},\ }\href@noop {} {\bibfield  {journal} {\bibinfo
  {journal} {Annals of Mathematics}\ }\textbf {\bibinfo {volume} {32}},\
  \bibinfo {pages} {327} (\bibinfo {year} {1931})}\BibitemShut {NoStop}%
\bibitem [{\citenamefont {Bailey}\ \emph {et~al.}(1994)\citenamefont {Bailey},
  \citenamefont {Eastwood},\ and\ \citenamefont {Gover}}]{BaileyT.N.1994TSBf}%
  \BibitemOpen
  \bibfield  {author} {\bibinfo {author} {\bibfnamefont {T.}~\bibnamefont
  {Bailey}}, \bibinfo {author} {\bibfnamefont {M.}~\bibnamefont {Eastwood}},\
  and\ \bibinfo {author} {\bibfnamefont {A.}~\bibnamefont {Gover}},\ }\bibfield
   {title} {\bibinfo {title} {Thomas's structure bundle for conformal,
  projective and related structures},\ }\href@noop {} {\bibfield  {journal}
  {\bibinfo  {journal} {Rocky Mountain J. Math.}\ }\textbf {\bibinfo {volume}
  {24}},\ \bibinfo {pages} {1191} (\bibinfo {year} {1994})}\BibitemShut
  {NoStop}%
\bibitem [{\citenamefont {Roberts}(1995)}]{Roberts}%
  \BibitemOpen
  \bibfield  {author} {\bibinfo {author} {\bibfnamefont {C.}~\bibnamefont
  {Roberts}},\ }\bibfield  {title} {\bibinfo {title} {{The Projective
  Connections of T.Y. Thomas and J.H.C. Whitehead Applied to Invariant
  Connections}},\ }\href@noop {} {\bibfield  {journal} {\bibinfo  {journal}
  {Differential Geometry and its Applications}\ }\textbf {\bibinfo {volume}
  {5}},\ \bibinfo {pages} {237} (\bibinfo {year} {1995})}\BibitemShut {NoStop}%
\bibitem [{\citenamefont {Eastwood}\ and\ \citenamefont
  {Matveev}(2007)}]{Eastwood}%
  \BibitemOpen
  \bibfield  {author} {\bibinfo {author} {\bibfnamefont {M.}~\bibnamefont
  {Eastwood}}\ and\ \bibinfo {author} {\bibfnamefont {V.~S.}\ \bibnamefont
  {Matveev}},\ }\bibfield  {title} {\bibinfo {title} {{Metric connections in
  projective differential geometry in ``Symmetries and Overdetermined Systems
  of Partial Differential Equations"}},\ }\href@noop {} {\bibfield  {journal}
  {\bibinfo  {journal} {Math. Appl.}\ }\textbf {\bibinfo {volume} {144}},\
  \bibinfo {pages} {339} (\bibinfo {year} {2007})},\ \Eprint
  {https://arxiv.org/abs/0806.3998} {arXiv:0806.3998 [math-dg]} \BibitemShut
  {NoStop}%
\bibitem [{\citenamefont {Eastwood}(2007)}]{Eastwood2}%
  \BibitemOpen
  \bibfield  {author} {\bibinfo {author} {\bibfnamefont {M.}~\bibnamefont
  {Eastwood}},\ }\bibfield  {title} {\bibinfo {title} {{Notes on projective
  differential geometry in ``Symmetries and Overdetermined Systems of Partial
  Differential Equations"}},\ }\href@noop {} {\bibfield  {journal} {\bibinfo
  {journal} {Math. Appl.}\ }\textbf {\bibinfo {volume} {144}},\ \bibinfo
  {pages} {41} (\bibinfo {year} {2007})}\BibitemShut {NoStop}%
\bibitem [{\citenamefont {Crampin}\ and\ \citenamefont
  {Saunders}(2007)}]{Crampin}%
  \BibitemOpen
  \bibfield  {author} {\bibinfo {author} {\bibfnamefont {M.}~\bibnamefont
  {Crampin}}\ and\ \bibinfo {author} {\bibfnamefont {D.}~\bibnamefont
  {Saunders}},\ }\bibfield  {title} {\bibinfo {title} {{Projective
  Connections}},\ }\href@noop {} {\bibfield  {journal} {\bibinfo  {journal}
  {Journal of Geometry and Physics}\ }\textbf {\bibinfo {volume} {57}},\
  \bibinfo {pages} {691} (\bibinfo {year} {2007})}\BibitemShut {NoStop}%
\bibitem [{\citenamefont {Hall}\ and\ \citenamefont
  {Lonie}(2007)}]{Hall:2007wp}%
  \BibitemOpen
  \bibfield  {author} {\bibinfo {author} {\bibfnamefont {G.~S.}\ \bibnamefont
  {Hall}}\ and\ \bibinfo {author} {\bibfnamefont {D.~P.}\ \bibnamefont
  {Lonie}},\ }\bibfield  {title} {\bibinfo {title} {{The Principle of
  equivalence and projective structure in space-times}},\ }\href
  {https://doi.org/10.1088/0264-9381/24/14/005} {\bibfield  {journal} {\bibinfo
   {journal} {Class. Quant. Grav.}\ }\textbf {\bibinfo {volume} {24}},\
  \bibinfo {pages} {3617} (\bibinfo {year} {2007})},\ \Eprint
  {https://arxiv.org/abs/gr-qc/0703104} {arXiv:gr-qc/0703104} \BibitemShut
  {NoStop}%
\bibitem [{\citenamefont {Hall}\ and\ \citenamefont {Lonie}(2008)}]{Hall:2008}%
  \BibitemOpen
  \bibfield  {author} {\bibinfo {author} {\bibfnamefont {G.~S.}\ \bibnamefont
  {Hall}}\ and\ \bibinfo {author} {\bibfnamefont {D.~P.}\ \bibnamefont
  {Lonie}},\ }\bibfield  {title} {\bibinfo {title} {{The Principle of
  equivalence and cosmological metrics}},\ }\href
  {https://doi.org/10.1063/1.2837431} {\bibfield  {journal} {\bibinfo
  {journal} {J. Math. Phys.}\ }\textbf {\bibinfo {volume} {49}},\ \bibinfo
  {pages} {022502} (\bibinfo {year} {2008})}\BibitemShut {NoStop}%
\bibitem [{\citenamefont {Hall}\ and\ \citenamefont
  {Lonie}(2009)}]{Hall:2009zza}%
  \BibitemOpen
  \bibfield  {author} {\bibinfo {author} {\bibfnamefont {G.~S.}\ \bibnamefont
  {Hall}}\ and\ \bibinfo {author} {\bibfnamefont {D.~P.}\ \bibnamefont
  {Lonie}},\ }\bibfield  {title} {\bibinfo {title} {{Projective equivalence of
  Einstein spaces in general relativity}},\ }\href
  {https://doi.org/10.1088/0264-9381/26/12/125009} {\bibfield  {journal}
  {\bibinfo  {journal} {Class. Quant. Grav.}\ }\textbf {\bibinfo {volume}
  {26}},\ \bibinfo {pages} {125009} (\bibinfo {year} {2009})}\BibitemShut
  {NoStop}%
\bibitem [{\citenamefont {Hall}\ and\ \citenamefont
  {Lonie}(2011)}]{Hall:2011zza}%
  \BibitemOpen
  \bibfield  {author} {\bibinfo {author} {\bibfnamefont {G.~S.}\ \bibnamefont
  {Hall}}\ and\ \bibinfo {author} {\bibfnamefont {D.~P.}\ \bibnamefont
  {Lonie}},\ }\bibfield  {title} {\bibinfo {title} {{Projective structure and
  holonomy in general relativity}},\ }\href
  {https://doi.org/10.1088/0264-9381/28/8/083101} {\bibfield  {journal}
  {\bibinfo  {journal} {Class. Quant. Grav.}\ }\textbf {\bibinfo {volume}
  {28}},\ \bibinfo {pages} {083101} (\bibinfo {year} {2011})}\BibitemShut
  {NoStop}%
\bibitem [{\citenamefont {Gover}\ and\ \citenamefont
  {Macbeth}(2014)}]{GoverA.Rod2012DEgE}%
  \BibitemOpen
  \bibfield  {author} {\bibinfo {author} {\bibfnamefont {A.~R.}\ \bibnamefont
  {Gover}}\ and\ \bibinfo {author} {\bibfnamefont {H.~R.}\ \bibnamefont
  {Macbeth}},\ }\bibfield  {title} {\bibinfo {title} {{Detecting Einstein
  geodesics: Einstein metrics in projective and conformal geometry}},\ }\href
  {https://doi.org/10.1016/j.difgeo.2013.10.011} {\bibfield  {journal}
  {\bibinfo  {journal} {Differ. Geom. Appl.}\ }\textbf {\bibinfo {volume}
  {33}},\ \bibinfo {pages} {44} (\bibinfo {year} {2014})},\ \Eprint
  {https://arxiv.org/abs/1212.6286} {arXiv:1212.6286 [math.DG]} \BibitemShut
  {NoStop}%
\bibitem [{\citenamefont {Cap}\ \emph {et~al.}(2014)\citenamefont {Cap},
  \citenamefont {Gover},\ and\ \citenamefont {Macbeth}}]{CapA.2014Emip}%
  \BibitemOpen
  \bibfield  {author} {\bibinfo {author} {\bibfnamefont {A.}~\bibnamefont
  {Cap}}, \bibinfo {author} {\bibfnamefont {A.~R.}\ \bibnamefont {Gover}},\
  and\ \bibinfo {author} {\bibfnamefont {H.~R.}\ \bibnamefont {Macbeth}},\
  }\bibfield  {title} {\bibinfo {title} {Einstein metrics in projective
  geometry.(report)},\ }\href@noop {} {\bibfield  {journal} {\bibinfo
  {journal} {Geometriae Dedicata}\ }\textbf {\bibinfo {volume} {168}} (\bibinfo
  {year} {2014})}\BibitemShut {NoStop}%
\bibitem [{\citenamefont {Brensinger}\ and\ \citenamefont
  {Rodgers}(2019)}]{Brensinger:2017gtb}%
  \BibitemOpen
  \bibfield  {author} {\bibinfo {author} {\bibfnamefont {S.}~\bibnamefont
  {Brensinger}}\ and\ \bibinfo {author} {\bibfnamefont {V.~G.~J.}\ \bibnamefont
  {Rodgers}},\ }\bibfield  {title} {\bibinfo {title} {{Dynamical Projective
  Curvature in Gravitation}},\ }\href
  {https://doi.org/10.1142/S0217751X18502238} {\bibfield  {journal} {\bibinfo
  {journal} {Int. J. Mod. Phys. A}\ }\textbf {\bibinfo {volume} {33}},\
  \bibinfo {pages} {1850223} (\bibinfo {year} {2019})},\ \Eprint
  {https://arxiv.org/abs/1712.05394} {arXiv:1712.05394 [hep-th]} \BibitemShut
  {NoStop}%
\bibitem [{\citenamefont {Brensinger}\ \emph {et~al.}(2020)\citenamefont
  {Brensinger}, \citenamefont {Heitritter}, \citenamefont {Rodgers},
  \citenamefont {Stiffler},\ and\ \citenamefont
  {Whiting}}]{Brensinger:2019mnx}%
  \BibitemOpen
  \bibfield  {author} {\bibinfo {author} {\bibfnamefont {S.}~\bibnamefont
  {Brensinger}}, \bibinfo {author} {\bibfnamefont {K.}~\bibnamefont
  {Heitritter}}, \bibinfo {author} {\bibfnamefont {V.~G.~J.}\ \bibnamefont
  {Rodgers}}, \bibinfo {author} {\bibfnamefont {K.}~\bibnamefont {Stiffler}},\
  and\ \bibinfo {author} {\bibfnamefont {C.~A.}\ \bibnamefont {Whiting}},\
  }\bibfield  {title} {\bibinfo {title} {{Dark Energy From Dynamical Projective
  Connections}},\ }\href {https://doi.org/10.1088/1361-6382/ab685d} {\bibfield
  {journal} {\bibinfo  {journal} {Class. Quant. Grav.}\ }\textbf {\bibinfo
  {volume} {37}},\ \bibinfo {pages} {055003} (\bibinfo {year} {2020})},\
  \Eprint {https://arxiv.org/abs/1907.05334} {arXiv:1907.05334 [hep-th]}
  \BibitemShut {NoStop}%
\bibitem [{\citenamefont {Brensinger}\ \emph {et~al.}(2021)\citenamefont
  {Brensinger}, \citenamefont {Heitritter}, \citenamefont {Rodgers},\ and\
  \citenamefont {Stiffler}}]{Brensinger:2020gcv}%
  \BibitemOpen
  \bibfield  {author} {\bibinfo {author} {\bibfnamefont {S.}~\bibnamefont
  {Brensinger}}, \bibinfo {author} {\bibfnamefont {K.}~\bibnamefont
  {Heitritter}}, \bibinfo {author} {\bibfnamefont {V.~G.~J.}\ \bibnamefont
  {Rodgers}},\ and\ \bibinfo {author} {\bibfnamefont {K.}~\bibnamefont
  {Stiffler}},\ }\bibfield  {title} {\bibinfo {title} {{General structure of
  Thomas$-$Whitehead gravity}},\ }\href
  {https://doi.org/10.1103/PhysRevD.103.044060} {\bibfield  {journal} {\bibinfo
   {journal} {Phys. Rev. D}\ }\textbf {\bibinfo {volume} {103}},\ \bibinfo
  {pages} {044060} (\bibinfo {year} {2021})},\ \Eprint
  {https://arxiv.org/abs/2009.06730} {arXiv:2009.06730 [hep-th]} \BibitemShut
  {NoStop}%
\bibitem [{\citenamefont {Brensinger}(2020)}]{SamBrensinger}%
  \BibitemOpen
  \bibfield  {author} {\bibinfo {author} {\bibfnamefont {S.~J.}\ \bibnamefont
  {Brensinger}},\ }\href@noop {} {\bibinfo {title} {Projective gauge
  gravity}},\ \bibinfo {howpublished}
  {\href{https://www.proquest.com/dissertations-theses/projective-gauge-gravity/docview/2448833356/se-2}{ProQuest
  Dissertations and Theses}} (\bibinfo {year} {2020})\BibitemShut {NoStop}%
\bibitem [{\citenamefont {Palatini}(1919)}]{palatini}%
  \BibitemOpen
  \bibfield  {author} {\bibinfo {author} {\bibfnamefont {A.}~\bibnamefont
  {Palatini}},\ }\bibfield  {title} {\bibinfo {title} {Deduzione invariantiva
  delle equazioni gravitazionali dal principio di hamilton},\ }\href@noop {}
  {\bibfield  {journal} {\bibinfo  {journal} {Rend. Circ. Mat. Palermo}\
  }\textbf {\bibinfo {volume} {43}},\ \bibinfo {pages} {203} (\bibinfo {year}
  {1919})}\BibitemShut {NoStop}%
\bibitem [{\citenamefont {Borunda}\ \emph {et~al.}(2008)\citenamefont
  {Borunda}, \citenamefont {Janssen},\ and\ \citenamefont
  {Bastero-Gil}}]{Borunda:2008}%
  \BibitemOpen
  \bibfield  {author} {\bibinfo {author} {\bibfnamefont {M.}~\bibnamefont
  {Borunda}}, \bibinfo {author} {\bibfnamefont {B.}~\bibnamefont {Janssen}},\
  and\ \bibinfo {author} {\bibfnamefont {M.}~\bibnamefont {Bastero-Gil}},\
  }\bibfield  {title} {\bibinfo {title} {{Palatini versus metric formulation in
  higher curvature gravity}},\ }\href
  {https://doi.org/10.1088/1475-7516/2008/11/008} {\bibfield  {journal}
  {\bibinfo  {journal} {JCAP}\ }\textbf {\bibinfo {volume} {11}},\ \bibinfo
  {pages} {008}},\ \Eprint {https://arxiv.org/abs/0804.4440} {arXiv:0804.4440
  [hep-th]} \BibitemShut {NoStop}%
\bibitem [{\citenamefont {Abdullah}\ \emph {et~al.}(2022)\citenamefont
  {Abdullah}, \citenamefont {Bavor}, \citenamefont {Chafamo}, \citenamefont
  {Jiang}, \citenamefont {Kalim}, \citenamefont {Stiffler},\ and\ \citenamefont
  {Whiting}}]{InflationNewPaper:2022}%
  \BibitemOpen
  \bibfield  {author} {\bibinfo {author} {\bibfnamefont {M.}~\bibnamefont
  {Abdullah}}, \bibinfo {author} {\bibfnamefont {C.}~\bibnamefont {Bavor}},
  \bibinfo {author} {\bibfnamefont {B.}~\bibnamefont {Chafamo}}, \bibinfo
  {author} {\bibfnamefont {X.}~\bibnamefont {Jiang}}, \bibinfo {author}
  {\bibfnamefont {M.~H.}\ \bibnamefont {Kalim}}, \bibinfo {author}
  {\bibfnamefont {K.}~\bibnamefont {Stiffler}},\ and\ \bibinfo {author}
  {\bibfnamefont {C.~A.}\ \bibnamefont {Whiting}},\ }\bibfield  {title}
  {\bibinfo {title} {Inflation from dynamical projective connections},\ }\href
  {https://doi.org/10.1103/PhysRevD.106.084049} {\bibfield  {journal} {\bibinfo
   {journal} {Phys. Rev. D}\ }\textbf {\bibinfo {volume} {106}},\ \bibinfo
  {pages} {084049} (\bibinfo {year} {2022})}\BibitemShut {NoStop}%
\bibitem [{\citenamefont {Kirillov}(1962)}]{Kirillov:1962}%
  \BibitemOpen
  \bibfield  {author} {\bibinfo {author} {\bibfnamefont {A.~A.}\ \bibnamefont
  {Kirillov}},\ }\bibfield  {title} {\bibinfo {title} {Unitary representations
  of nilpotent lie groups},\ }\href@noop {} {\bibfield  {journal} {\bibinfo
  {journal} {Russian Mathematical Surveys}\ }\textbf {\bibinfo {volume} {17}},\
  \bibinfo {pages} {53} (\bibinfo {year} {1962})}\BibitemShut {NoStop}%
\bibitem [{\citenamefont {Kirillov}(1982)}]{Kirillov:1982kav}%
  \BibitemOpen
  \bibfield  {author} {\bibinfo {author} {\bibfnamefont {A.~A.}\ \bibnamefont
  {Kirillov}},\ }\bibfield  {title} {\bibinfo {title} {{Infinite dimensional
  lie groups; their orbits, invariants and representations. The geometry of
  moments}},\ }\bibfield  {booktitle} {\emph {\bibinfo {booktitle}
  {{Proceedings, IV Bulgarian School on Elementary Particles and High-Energy
  Physics: Mathematical Problems of Quantum Field Theory: Primorsko, Bulgaria,
  September 16-25, 1980}}},\ }\href {https://doi.org/10.1007/BFb0066026}
  {\bibfield  {journal} {\bibinfo  {journal} {Lect. Notes Math.}\ }\textbf
  {\bibinfo {volume} {970}},\ \bibinfo {pages} {101} (\bibinfo {year}
  {1982})}\BibitemShut {NoStop}%
%%CITATION = LNMAA,970,101;%%
\bibitem [{\citenamefont {Zaccaria}\ \emph {et~al.}(1983)\citenamefont
  {Zaccaria}, \citenamefont {Sudarshan}, \citenamefont {Nilsson}, \citenamefont
  {Mukunda}, \citenamefont {Marmo},\ and\ \citenamefont
  {Balachandran}}]{Zaccaria:1981fi}%
  \BibitemOpen
  \bibfield  {author} {\bibinfo {author} {\bibfnamefont {F.}~\bibnamefont
  {Zaccaria}}, \bibinfo {author} {\bibfnamefont {E.~C.~G.}\ \bibnamefont
  {Sudarshan}}, \bibinfo {author} {\bibfnamefont {J.~S.}\ \bibnamefont
  {Nilsson}}, \bibinfo {author} {\bibfnamefont {N.}~\bibnamefont {Mukunda}},
  \bibinfo {author} {\bibfnamefont {G.}~\bibnamefont {Marmo}},\ and\ \bibinfo
  {author} {\bibfnamefont {A.~P.}\ \bibnamefont {Balachandran}},\ }\bibfield
  {title} {\bibinfo {title} {{Universal Unfolding of Hamiltonian Systems: From
  Symplectic Structure to Fiber Bundles}},\ }\href
  {https://doi.org/10.1103/PhysRevD.27.2327} {\bibfield  {journal} {\bibinfo
  {journal} {Phys. Rev. D}\ }\textbf {\bibinfo {volume} {27}},\ \bibinfo
  {pages} {2327} (\bibinfo {year} {1983})}\BibitemShut {NoStop}%
\bibitem [{\citenamefont {Balachandran}\ \emph {et~al.}(1983)\citenamefont
  {Balachandran}, \citenamefont {Marmo}, \citenamefont {Skagerstam},\ and\
  \citenamefont {Stern}}]{Balachandran:1983oit}%
  \BibitemOpen
  \bibfield  {author} {\bibinfo {author} {\bibfnamefont {A.~P.}\ \bibnamefont
  {Balachandran}}, \bibinfo {author} {\bibfnamefont {G.}~\bibnamefont {Marmo}},
  \bibinfo {author} {\bibfnamefont {B.~S.}\ \bibnamefont {Skagerstam}},\ and\
  \bibinfo {author} {\bibfnamefont {A.}~\bibnamefont {Stern}},\ }\href
  {https://doi.org/10.1007/3-540-12724-0\_1} {\emph {\bibinfo {title} {{Gauge
  Theories and Fibre Bundles - Applications to Particle Dynamics}}}},\ Vol.\
  \bibinfo {volume} {188}\ (\bibinfo  {publisher} {Springer},\ \bibinfo {year}
  {1983})\ \Eprint {https://arxiv.org/abs/1702.08910} {arXiv:1702.08910
  [quant-ph]} \BibitemShut {NoStop}%
\bibitem [{\citenamefont {Balachandran}\ \emph {et~al.}(1987)\citenamefont
  {Balachandran}, \citenamefont {Gomm},\ and\ \citenamefont
  {Sorkin}}]{Balachandran:1986hv}%
  \BibitemOpen
  \bibfield  {author} {\bibinfo {author} {\bibfnamefont {A.~P.}\ \bibnamefont
  {Balachandran}}, \bibinfo {author} {\bibfnamefont {H.}~\bibnamefont {Gomm}},\
  and\ \bibinfo {author} {\bibfnamefont {R.~D.}\ \bibnamefont {Sorkin}},\
  }\bibfield  {title} {\bibinfo {title} {{Quantum Symmetries From Quantum
  Phases: Fermions From Bosons, a $Z$(2) Anomaly and Galilean Invariance}},\
  }\href {https://doi.org/10.1016/0550-3213(87)90420-2} {\bibfield  {journal}
  {\bibinfo  {journal} {Nucl. Phys.}\ }\textbf {\bibinfo {volume} {B281}},\
  \bibinfo {pages} {573} (\bibinfo {year} {1987})}\BibitemShut {NoStop}%
%%CITATION = NUPHA,B281,573;%%
\bibitem [{\citenamefont {Balachandran}(1987)}]{Balachandran:1987st}%
  \BibitemOpen
  \bibfield  {author} {\bibinfo {author} {\bibfnamefont {A.~P.}\ \bibnamefont
  {Balachandran}},\ }\bibfield  {title} {\bibinfo {title} {{Wess-ZuminoTerms
  and Quantum Symmetries}},\ }in\ \href@noop {} {\emph {\bibinfo {booktitle}
  {{1st Asia Pacific Conference on High-energy Physics: Superstrings, Anomalies
  and Field Theory Singapore, Singapore, June 21-28, 1987}}}}\ (\bibinfo
  {publisher} {World Scientific Publication, Teaneck},\ \bibinfo {year}
  {1987})\ pp.\ \bibinfo {pages} {375--407}\BibitemShut {NoStop}%
%%CITATION = SU-4428-361;%%
\bibitem [{\citenamefont {Alekseev}\ and\ \citenamefont
  {Shatashvili}(1989)}]{Alekseev:1988ce}%
  \BibitemOpen
  \bibfield  {author} {\bibinfo {author} {\bibfnamefont {A.}~\bibnamefont
  {Alekseev}}\ and\ \bibinfo {author} {\bibfnamefont {S.~L.}\ \bibnamefont
  {Shatashvili}},\ }\bibfield  {title} {\bibinfo {title} {{Path Integral
  Quantization of the Coadjoint Orbits of the Virasoro Group and 2D Gravity}},\
  }\href {https://doi.org/10.1016/0550-3213(89)90130-2} {\bibfield  {journal}
  {\bibinfo  {journal} {Nucl. Phys.}\ }\textbf {\bibinfo {volume} {B323}},\
  \bibinfo {pages} {719} (\bibinfo {year} {1989})}\BibitemShut {NoStop}%
%%CITATION = NUPHA,B323,719;%%
\bibitem [{\citenamefont {Alekseev}\ \emph {et~al.}(1988)\citenamefont
  {Alekseev}, \citenamefont {Faddeev},\ and\ \citenamefont
  {Shatashvili}}]{Alekseev:1988vx}%
  \BibitemOpen
  \bibfield  {author} {\bibinfo {author} {\bibfnamefont {A.}~\bibnamefont
  {Alekseev}}, \bibinfo {author} {\bibfnamefont {L.~D.}\ \bibnamefont
  {Faddeev}},\ and\ \bibinfo {author} {\bibfnamefont {S.~L.}\ \bibnamefont
  {Shatashvili}},\ }\bibfield  {title} {\bibinfo {title} {{Quantization of
  symplectic orbits of compact Lie groups by means of the functional
  integral}},\ }\href {https://doi.org/10.1016/0393-0440(88)90031-9} {\bibfield
   {journal} {\bibinfo  {journal} {J. Geom. Phys.}\ }\textbf {\bibinfo {volume}
  {5}},\ \bibinfo {pages} {391} (\bibinfo {year} {1988})}\BibitemShut {NoStop}%
%%CITATION = JGPHE,5,391;%%
\bibitem [{\citenamefont {Delius}\ \emph {et~al.}(1990)\citenamefont {Delius},
  \citenamefont {van Nieuwenhuizen},\ and\ \citenamefont
  {Rodgers}}]{Delius:1990pt}%
  \BibitemOpen
  \bibfield  {author} {\bibinfo {author} {\bibfnamefont {G.~W.}\ \bibnamefont
  {Delius}}, \bibinfo {author} {\bibfnamefont {P.}~\bibnamefont {van
  Nieuwenhuizen}},\ and\ \bibinfo {author} {\bibfnamefont {V.~G.~J.}\
  \bibnamefont {Rodgers}},\ }\bibfield  {title} {\bibinfo {title} {{The Method
  of Coadjoint Orbits: An Algorithm for the Construction of Invariant
  Actions}},\ }\href {https://doi.org/10.1142/S0217751X90001690} {\bibfield
  {journal} {\bibinfo  {journal} {Int. J. Mod. Phys.}\ }\textbf {\bibinfo
  {volume} {A5}},\ \bibinfo {pages} {3943} (\bibinfo {year}
  {1990})}\BibitemShut {NoStop}%
%%CITATION = IMPAE,A5,3943;%%
\bibitem [{\citenamefont {Rai}\ and\ \citenamefont
  {Rodgers}(1990)}]{Rai:1989js}%
  \BibitemOpen
  \bibfield  {author} {\bibinfo {author} {\bibfnamefont {B.}~\bibnamefont
  {Rai}}\ and\ \bibinfo {author} {\bibfnamefont {V.~G.~J.}\ \bibnamefont
  {Rodgers}},\ }\bibfield  {title} {\bibinfo {title} {{From Coadjoint Orbits to
  Scale Invariant {WZNW} Type Actions and 2-$D$ Quantum Gravity Action}},\
  }\href {https://doi.org/10.1016/0550-3213(90)90264-E} {\bibfield  {journal}
  {\bibinfo  {journal} {Nucl. Phys.}\ }\textbf {\bibinfo {volume} {B341}},\
  \bibinfo {pages} {119} (\bibinfo {year} {1990})}\BibitemShut {NoStop}%
%%CITATION = NUPHA,B341,119;%%
\bibitem [{\citenamefont {Di~Vecchia}\ \emph {et~al.}(1984)\citenamefont
  {Di~Vecchia}, \citenamefont {Durhuus},\ and\ \citenamefont
  {Petersen}}]{DiVecchia:1984ksr}%
  \BibitemOpen
  \bibfield  {author} {\bibinfo {author} {\bibfnamefont {P.}~\bibnamefont
  {Di~Vecchia}}, \bibinfo {author} {\bibfnamefont {B.}~\bibnamefont
  {Durhuus}},\ and\ \bibinfo {author} {\bibfnamefont {J.~L.}\ \bibnamefont
  {Petersen}},\ }\bibfield  {title} {\bibinfo {title} {{The Wess-Zumino Action
  in Two-Dimensions and Nonabelian Bosonization}},\ }\href
  {https://doi.org/10.1016/0370-2693(84)91813-6} {\bibfield  {journal}
  {\bibinfo  {journal} {Phys. Lett. B}\ }\textbf {\bibinfo {volume} {144}},\
  \bibinfo {pages} {245} (\bibinfo {year} {1984})}\BibitemShut {NoStop}%
\bibitem [{\citenamefont {Witten}(1984)}]{Witten:1983ar}%
  \BibitemOpen
  \bibfield  {author} {\bibinfo {author} {\bibfnamefont {E.}~\bibnamefont
  {Witten}},\ }\bibfield  {title} {\bibinfo {title} {{Nonabelian Bosonization
  in Two-Dimensions}},\ }\href {https://doi.org/10.1007/BF01215276} {\bibfield
  {journal} {\bibinfo  {journal} {Commun. Math. Phys.}\ }\textbf {\bibinfo
  {volume} {92}},\ \bibinfo {pages} {455} (\bibinfo {year} {1984})}\BibitemShut
  {NoStop}%
\bibitem [{\citenamefont {Polyakov}(1987)}]{Polyakov:1987zb}%
  \BibitemOpen
  \bibfield  {author} {\bibinfo {author} {\bibfnamefont {A.~M.}\ \bibnamefont
  {Polyakov}},\ }\bibfield  {title} {\bibinfo {title} {{Quantum Gravity in
  Two-Dimensions}},\ }\href {https://doi.org/10.1142/S0217732387001130}
  {\bibfield  {journal} {\bibinfo  {journal} {Mod. Phys. Lett.}\ }\textbf
  {\bibinfo {volume} {A2}},\ \bibinfo {pages} {893} (\bibinfo {year}
  {1987})}\BibitemShut {NoStop}%
%%CITATION = MPLAE,A2,893;%%
\bibitem [{\citenamefont {Polyakov}(1981)}]{Polyakov:1981rd}%
  \BibitemOpen
  \bibfield  {author} {\bibinfo {author} {\bibfnamefont {A.~M.}\ \bibnamefont
  {Polyakov}},\ }\bibfield  {title} {\bibinfo {title} {{Quantum Geometry of
  Bosonic Strings}},\ }\href {https://doi.org/10.1016/0370-2693(81)90743-7}
  {\bibfield  {journal} {\bibinfo  {journal} {Phys. Lett.}\ }\textbf {\bibinfo
  {volume} {103B}},\ \bibinfo {pages} {207} (\bibinfo {year}
  {1981})}\BibitemShut {NoStop}%
%%CITATION = PHLTA,103B,207;%%
\bibitem [{\citenamefont {Rodgers}(1994)}]{Rodgers:1994ck}%
  \BibitemOpen
  \bibfield  {author} {\bibinfo {author} {\bibfnamefont {V.~G.~J.}\
  \bibnamefont {Rodgers}},\ }\bibfield  {title} {\bibinfo {title} {{A 2-D
  inspired 4-D theory of gravity}},\ }\href
  {https://doi.org/10.1016/0370-2693(94)90543-6} {\bibfield  {journal}
  {\bibinfo  {journal} {Phys. Lett. B}\ }\textbf {\bibinfo {volume} {336}},\
  \bibinfo {pages} {343} (\bibinfo {year} {1994})},\ \Eprint
  {https://arxiv.org/abs/hep-th/9407119} {arXiv:hep-th/9407119} \BibitemShut
  {NoStop}%
\bibitem [{\citenamefont {Lano}\ and\ \citenamefont
  {Rodgers}(1995)}]{Lano:1994gx}%
  \BibitemOpen
  \bibfield  {author} {\bibinfo {author} {\bibfnamefont {R.~P.}\ \bibnamefont
  {Lano}}\ and\ \bibinfo {author} {\bibfnamefont {V.~G.~J.}\ \bibnamefont
  {Rodgers}},\ }\bibfield  {title} {\bibinfo {title} {{A Study of fermions
  coupled to gauge and gravitational fields on a cylinder}},\ }\href
  {https://doi.org/10.1016/0550-3213(94)00544-O} {\bibfield  {journal}
  {\bibinfo  {journal} {Nucl. Phys.}\ }\textbf {\bibinfo {volume} {B437}},\
  \bibinfo {pages} {45} (\bibinfo {year} {1995})},\ \Eprint
  {https://arxiv.org/abs/hep-th/9401039} {arXiv:hep-th/9401039 [hep-th]}
  \BibitemShut {NoStop}%
%%CITATION = HEP-TH/9401039;%%
\bibitem [{\citenamefont {Branson}\ \emph {et~al.}(1997)\citenamefont
  {Branson}, \citenamefont {Lano},\ and\ \citenamefont
  {Rodgers}}]{Branson:1996pe}%
  \BibitemOpen
  \bibfield  {author} {\bibinfo {author} {\bibfnamefont {T.}~\bibnamefont
  {Branson}}, \bibinfo {author} {\bibfnamefont {R.~P.}\ \bibnamefont {Lano}},\
  and\ \bibinfo {author} {\bibfnamefont {V.~G.~J.}\ \bibnamefont {Rodgers}},\
  }\bibfield  {title} {\bibinfo {title} {{Yang-Mills, gravity, and 2-D string
  symmetries}},\ }\href {https://doi.org/10.1016/S0370-2693(97)01029-0}
  {\bibfield  {journal} {\bibinfo  {journal} {Phys. Lett.}\ }\textbf {\bibinfo
  {volume} {B412}},\ \bibinfo {pages} {253} (\bibinfo {year} {1997})},\ \Eprint
  {https://arxiv.org/abs/hep-th/9610023} {arXiv:hep-th/9610023 [hep-th]}
  \BibitemShut {NoStop}%
%%CITATION = HEP-TH/9610023;%%
\bibitem [{\citenamefont {Branson}\ \emph {et~al.}(2000)\citenamefont
  {Branson}, \citenamefont {Rodgers},\ and\ \citenamefont
  {Yasuda}}]{Branson:1998bc}%
  \BibitemOpen
  \bibfield  {author} {\bibinfo {author} {\bibfnamefont {T.~P.}\ \bibnamefont
  {Branson}}, \bibinfo {author} {\bibfnamefont {V.~G.~J.}\ \bibnamefont
  {Rodgers}},\ and\ \bibinfo {author} {\bibfnamefont {T.}~\bibnamefont
  {Yasuda}},\ }\bibfield  {title} {\bibinfo {title} {{Interaction of a string
  inspired graviton}},\ }\href {https://doi.org/10.1142/S0217751X0000121X,
  10.1142/S0217751X00001218} {\bibfield  {journal} {\bibinfo  {journal} {Int.
  J. Mod. Phys.}\ }\textbf {\bibinfo {volume} {A15}},\ \bibinfo {pages} {3549}
  (\bibinfo {year} {2000})},\ \Eprint {https://arxiv.org/abs/hep-th/9812098}
  {arXiv:hep-th/9812098 [hep-th]} \BibitemShut {NoStop}%
%%CITATION = HEP-TH/9812098;%%
\bibitem [{\citenamefont {Rodgers}(2022)}]{Rodgers:2022rvo}%
  \BibitemOpen
  \bibfield  {author} {\bibinfo {author} {\bibfnamefont {V.~G.~J.}\
  \bibnamefont {Rodgers}},\ }\bibfield  {title} {\bibinfo {title} {{From
  Virasoro Algebra to Cosmology}},\ }\href@noop {} {\  (\bibinfo {year}
  {2022})},\ \Eprint {https://arxiv.org/abs/2212.08715} {arXiv:2212.08715
  [hep-th]} \BibitemShut {NoStop}%
\bibitem [{\citenamefont {Brensinger}\ and\ \citenamefont
  {Vecera}(2024)}]{Brensinger:2024udu}%
  \BibitemOpen
  \bibfield  {author} {\bibinfo {author} {\bibfnamefont {S.~J.}\ \bibnamefont
  {Brensinger}}\ and\ \bibinfo {author} {\bibfnamefont {P.}~\bibnamefont
  {Vecera}},\ }\bibfield  {title} {\bibinfo {title} {{Geometrical Heavy
  Lifting: Yang-Mills, Spin, and Torsion in Dynamical Projective
  Gravitation}},\ }\href@noop {} {\  (\bibinfo {year} {2024})},\ \Eprint
  {https://arxiv.org/abs/2404.02243} {arXiv:2404.02243 [gr-qc]} \BibitemShut
  {NoStop}%
\bibitem [{\citenamefont {Lovelock}(1971)}]{Lovelock:1971yv}%
  \BibitemOpen
  \bibfield  {author} {\bibinfo {author} {\bibfnamefont {D.}~\bibnamefont
  {Lovelock}},\ }\bibfield  {title} {\bibinfo {title} {{The Einstein tensor and
  its generalizations}},\ }\href {https://doi.org/10.1063/1.1665613} {\bibfield
   {journal} {\bibinfo  {journal} {J. Math. Phys.}\ }\textbf {\bibinfo {volume}
  {12}},\ \bibinfo {pages} {498} (\bibinfo {year} {1971})}\BibitemShut
  {NoStop}%
\bibitem [{\citenamefont {Lanczos}(1938)}]{Lanczos:1938sf}%
  \BibitemOpen
  \bibfield  {author} {\bibinfo {author} {\bibfnamefont {C.}~\bibnamefont
  {Lanczos}},\ }\bibfield  {title} {\bibinfo {title} {{A Remarkable property of
  the Riemann-Christoffel tensor in four dimensions}},\ }\href
  {https://doi.org/10.2307/1968467} {\bibfield  {journal} {\bibinfo  {journal}
  {Annals Math.}\ }\textbf {\bibinfo {volume} {39}},\ \bibinfo {pages} {842}
  (\bibinfo {year} {1938})}\BibitemShut {NoStop}%
\bibitem [{\citenamefont {de~Sitter}(1917)}]{deSitter:1917zz}%
  \BibitemOpen
  \bibfield  {author} {\bibinfo {author} {\bibfnamefont {W.}~\bibnamefont
  {de~Sitter}},\ }\bibfield  {title} {\bibinfo {title} {{On Einstein's Theory
  of Gravitation and its Astronomical Consequences. Third Paper.}},\ }\href
  {https://doi.org/10.1093/mnras/78.1.3} {\bibfield  {journal} {\bibinfo
  {journal} {Mon. Not. Roy. Astron. Soc.}\ }\textbf {\bibinfo {volume} {78}},\
  \bibinfo {pages} {3} (\bibinfo {year} {1917})}\BibitemShut {NoStop}%
\bibitem [{\citenamefont {Strominger}(2001)}]{Strominger:2001pn}%
  \BibitemOpen
  \bibfield  {author} {\bibinfo {author} {\bibfnamefont {A.}~\bibnamefont
  {Strominger}},\ }\bibfield  {title} {\bibinfo {title} {{The dS / CFT
  correspondence}},\ }\href {https://doi.org/10.1088/1126-6708/2001/10/034}
  {\bibfield  {journal} {\bibinfo  {journal} {JHEP}\ }\textbf {\bibinfo
  {volume} {10}},\ \bibinfo {pages} {034}},\ \Eprint
  {https://arxiv.org/abs/hep-th/0106113} {arXiv:hep-th/0106113} \BibitemShut
  {NoStop}%
\bibitem [{\citenamefont {Spradlin}\ \emph {et~al.}(2001)\citenamefont
  {Spradlin}, \citenamefont {Strominger},\ and\ \citenamefont
  {Volovich}}]{Spradlin:2001pw}%
  \BibitemOpen
  \bibfield  {author} {\bibinfo {author} {\bibfnamefont {M.}~\bibnamefont
  {Spradlin}}, \bibinfo {author} {\bibfnamefont {A.}~\bibnamefont
  {Strominger}},\ and\ \bibinfo {author} {\bibfnamefont {A.}~\bibnamefont
  {Volovich}},\ }\bibfield  {title} {\bibinfo {title} {{Les Houches lectures on
  de Sitter space}},\ }in\ \href@noop {} {\emph {\bibinfo {booktitle} {{Les
  Houches Summer School: Session 76: Euro Summer School on Unity of Fundamental
  Physics: Gravity, Gauge Theory and Strings}}}}\ (\bibinfo {year} {2001})\
  pp.\ \bibinfo {pages} {423--453},\ \Eprint
  {https://arxiv.org/abs/hep-th/0110007} {arXiv:hep-th/0110007} \BibitemShut
  {NoStop}%
\bibitem [{\citenamefont {Kim}\ \emph {et~al.}(2002)\citenamefont {Kim},
  \citenamefont {Oh},\ and\ \citenamefont {Park}}]{Kim:2002uz}%
  \BibitemOpen
  \bibfield  {author} {\bibinfo {author} {\bibfnamefont {Y.-b.}\ \bibnamefont
  {Kim}}, \bibinfo {author} {\bibfnamefont {C.~Y.}\ \bibnamefont {Oh}},\ and\
  \bibinfo {author} {\bibfnamefont {N.}~\bibnamefont {Park}},\ }\bibfield
  {title} {\bibinfo {title} {{Classical geometry of de Sitter space-time: An
  Introductory review}},\ }\href@noop {} {\  (\bibinfo {year} {2002})},\
  \Eprint {https://arxiv.org/abs/hep-th/0212326} {arXiv:hep-th/0212326}
  \BibitemShut {NoStop}%
\bibitem [{\citenamefont {Aldrovandi}\ \emph {et~al.}(2007)\citenamefont
  {Aldrovandi}, \citenamefont {Beltran~Almeida},\ and\ \citenamefont
  {Pereira}}]{Aldrovandi:2006vr}%
  \BibitemOpen
  \bibfield  {author} {\bibinfo {author} {\bibfnamefont {R.}~\bibnamefont
  {Aldrovandi}}, \bibinfo {author} {\bibfnamefont {J.~P.}\ \bibnamefont
  {Beltran~Almeida}},\ and\ \bibinfo {author} {\bibfnamefont {J.~G.}\
  \bibnamefont {Pereira}},\ }\bibfield  {title} {\bibinfo {title} {{de Sitter
  special relativity}},\ }\href {https://doi.org/10.1088/0264-9381/24/6/002}
  {\bibfield  {journal} {\bibinfo  {journal} {Class. Quant. Grav.}\ }\textbf
  {\bibinfo {volume} {24}},\ \bibinfo {pages} {1385} (\bibinfo {year}
  {2007})},\ \Eprint {https://arxiv.org/abs/gr-qc/0606122}
  {arXiv:gr-qc/0606122} \BibitemShut {NoStop}%
\bibitem [{\citenamefont {Fiedorowicz}\ \emph {et~al.}(2024)\citenamefont
  {Fiedorowicz}, \citenamefont {Grover}, \citenamefont {Rodgers},\ and\
  \citenamefont {Zenger}}]{Fiedorowicz:2024hip}%
  \BibitemOpen
  \bibfield  {author} {\bibinfo {author} {\bibfnamefont {O.}~\bibnamefont
  {Fiedorowicz}}, \bibinfo {author} {\bibfnamefont {T.~C.}\ \bibnamefont
  {Grover}}, \bibinfo {author} {\bibfnamefont {V.~G.~J.}\ \bibnamefont
  {Rodgers}},\ and\ \bibinfo {author} {\bibfnamefont {H.~D.}\ \bibnamefont
  {Zenger}},\ }\bibfield  {title} {\bibinfo {title} {{Diffeomorphism Radiative
  Degrees of Freedom of Thomas-Whitehead Gravity}},\ }\href@noop {} {\
  (\bibinfo {year} {2024})},\ \Eprint {https://arxiv.org/abs/2405.11101}
  {arXiv:2405.11101 [gr-qc]} \BibitemShut {NoStop}%
\end{thebibliography}%

\end{document}